\documentclass[a4paper,12pt,english]{article}

\usepackage{amsfonts,bm,amssymb,euscript,array,babel,cite}
\usepackage{mathtools}
\usepackage{tikz-cd}
\usepackage{amsmath}
\usepackage{tensor,accents}
\usepackage{hhline}
\usepackage[amsmath]{empheq}
\usepackage[pdfencoding=auto, psdextra]{hyperref}
\usepackage{cancel}
\usepackage{multirow}
\usepackage{mathrsfs}
\usepackage{pdfsync}
\usepackage{multirow}
\usepackage{tikz}
\usepackage[most]{tcolorbox}

\usepackage{framed}

\newtcbox{\mymath}[1][]{%
    nobeforeafter, math upper, tcbox raise base,
    enhanced, colframe=blue!30!black,
    colback=blue!30, boxrule=1pt,
    #1}

\newsavebox{\mysaveboxM}
\newsavebox{\mysaveboxT}

\makeatletter

\newcommand{\dd}{\mathrm{d}}
\newcommand{\DD}{\mathrm{D}}

\newcommand{\w}{\wedge}

\newcommand{\be}{\begin{equation}}
\newcommand{\ee}{\end{equation}}
\newcommand{\sfrac}[2]{{\textstyle\frac{#1}{#2}}}
\def\nn{\nonumber}
\def \bea{\begin{eqnarray}} 
\def\eea{\end{eqnarray}}
\newcommand{\mf}{\mathfrak}
\def\mc{\mathcal}

\def\bi{\begin{itemize}} 
\def\ei{\end{itemize}}

\def\E{\textit{\tiny{E}}} 
\def\TM{\textit{\tiny{TM}}}
\def\Eh{\widehat{\textit{\tiny{E}}}} 
\newcommand{\sbullet}{%
  \hbox{\fontfamily{lmr}\fontsize{.8\dimexpr(\f@size pt)}{0}\selectfont\textbullet}}
\DeclareRobustCommand{\mathbullet}{\accentset{\sbullet}}

\makeatother

\newtheorem{corollary}[equation]{Corollary}

\newtheorem{prop}[equation]{Proposition}
\newtheorem{defn}[equation]{Definition}
\newtheorem{rmk}[equation]{Remark}

\def\a{\alpha} \def\b{\beta} \def\g{\gamma} \def\G{\Gamma} \def\d{\delta} \def\D{\Delta}
\def\e{\epsilon} 
   \def\k{\kappa}
\def\l{\lambda}  \def\m{\mu}
\def\n{\nu} \def\o{\omega}   \def\r{\rho}
\def\s{\sigma} \def\S{\Sigma}

 \def\Z{{\mathbb Z}} 

\def\one{\mbox{1 \kern-.59em {\rm l}}}

\numberwithin{equation}{section}

\begin{document}

\makeatother
\parindent=0cm
\renewcommand{\title}[1]{\vspace{10mm}\noindent{\Large{\bf #1}}\vspace{8mm}} \newcommand{\authors}[1]{\noindent{\large #1}\vspace{5mm}} \newcommand{\address}[1]{{\itshape #1\vspace{2mm}}}

\begin{titlepage}

\begin{flushright} 
RBI-ThPhys-2023-4
\end{flushright}

\begin{center}

\title{ {\Large {Basic curvature \&  the Atiyah cocycle in gauge theory }}}

 \vskip 3mm

 \authors{\large Athanasios Chatzistavrakidis$^{\a}$ and Larisa Jonke$^{\a,\b}$
  }

 \vskip 2mm
 
  \address{ $^{\a}$ Division of Theoretical Physics, Rudjer Bo\v skovi\'c Institute \\ Bijeni\v cka 54, 10000 Zagreb, Croatia \\[5pt]
 
 $^{\beta}$ School of Theoretical Physics,  Dublin Institute for Advanced Studies \\
 10 Burlington Road,  Dublin 4,  Ireland
  }

\vskip 2cm

\begin{abstract}
    Motivated from target space covariant formulations of topological sigma models and from a graded-geometric approach to higher gauge theory, we study connections on Lie and Courant algebroids and on their description as differential graded (dg) manifolds. We revisit the notion of basic curvature for a connection on a Lie algebroid, which measures its compatibility with the Lie bracket and it appears in the BV-BRST differential of 2D gauge theories such as (twisted) Poisson and Dirac sigma models. We define a basic curvature tensor for connections on Courant algebroids and we show that in the description of a Courant algebroid as a QP2 manifold it appears naturally as part of the homological vector field together with the Gualtieri torsion of an induced generalised connection. Furthermore, we consider connections on dg manifolds and revisit the structure of gauge transformations in the graded-geometric approach to higher gauge theories from a manifestly covariant perspective. We argue that it is governed by the brackets of a Kapranov L$_{\infty}[1]$ algebra, whose binary bracket is given by the Atiyah cocycle that measures the compatibility of the connection with the homological vector field. We also revisit some aspects of the derived bracket construction and show how to extend it to connections and tensors. 
\end{abstract}

\end{center}

\vskip 2cm

\end{titlepage}

\setcounter{footnote}{0}
\tableofcontents

\newpage

\section{Introduction} 
\label{sec1}

\subsection{Motivation} 

The Batalin-Vilkovisky (BV) quantization of general gauge field theories \cite{Batalin:1981jr,Batalin:1983ggl,HTbook} has revealed a variety of connections between physics and higher structures. Among them, the AKSZ construction of topological sigma models in various dimensions established that solutions to the classical master equation, which neatly encodes the gauge structure of the theory, are in correspondence to differential graded (dg) manifolds admitting a compatible graded symplectic form \cite{Alexandrov:1995kv}, also known as QP manifolds. Prototypical examples where this construction applies are the topological A- and B-models \cite{Witten:1988xj} and Chern-Simons theory, as already discussed in \cite{Alexandrov:1995kv} and the AKSZ formulation of the Poisson sigma model as presented in \cite{Cattaneo:2001ys}. Further examples include the generalization of Chern-Simons theory to the Courant sigma model \cite{Ikeda:2002wh,Roytenberg:2006qz} and the AKSZ formulation of Rozansky-Witten theory \cite{Rozansky:1996bq} in \cite{Qiu:2009zv}. 

On the other hand,  QP manifolds with symplectic form of degree 1 and 2 exhibit a relation to the differential geometric notions of Lie and Courant algebroids respectively \cite{Roytenberg:2002nu}. In particular, QP1 manifolds are in one to one correspondence with the Lie algebroid on the cotangent bundle of a smooth Poisson manifold $M$. More generally such a correspondence exists for any Lie algebroid on a vector bundle $E$ and the graded (Q, not P) manifold $E[1]$ endowed with a suitable homological vector field \cite{Vaintrob}. At the simplest level, this includes a correspondence between a Lie algebra $\mf g$ (a Lie algebroid over the point) and the degree-shifted $\mf g[1]$ with the Chevalley-Eilenberg differential as homological vector field.  

The topological quantum field theories in 2D and 3D mentioned above are related to Lie and Courant algebroids through this correspondence to QP1 and QP2 manifolds. For their formulation there is no need to introduce any additional, noncanonical geometrical data. However, if one wishes to express the gauge transformations, the field equations and ultimately the BV action in a manifestly target space covariant form, one is led to consider (auxiliary) connections \cite{Baulieu:2001fi,Cattaneo:2000hz,Ikeda:2019czt}. Once a connection is introduced, one can attempt to write down its torsion and curvature tensors. This is straightforward for Lie algebroids, but more complicated for Courant algebroids. On the other hand, for a Lie algebroid there exists a tensor which measures the compatibility of an ordinary vector bundle connection on it with the Lie bracket on its space of sections. This tensor was essentially introduced in \cite{Blaom} in the study of Cartan connections on Lie algebroids and it was called ``basic curvature'' in Ref. \cite{Crainic} in the context of representations up to homotopy (see also Ref. \cite{Kotov:2016lpx}). Regardless of the motivation presented here, studying connections on such structures is important more broadly, in view of other applications such as in gravity models \cite{Jurco:2016emw}. 

For dg manifolds there is a notion of affine connection for graded vector bundles and Lie algebroids thereof \cite{AtiyahDG} and the graded-geometric versions of torsion and curvature tensors. Additionally, there exists a tensor that measures the compatibility of the connection with the homological vector field. This is called the Atiyah 1-cocycle of a dg vector bundle \cite{AtiyahDG98,Lyakhovich:2009qq,AtiyahDG}; recall that the Atiyah class of a holomorphic vector bundle \cite{Atiyah} constitutes an obstruction to the existence of a global holomorphic connection.  Interestingly, the Atiyah class appears in Rozansky-Witten theory and in the topological field theory for the Witten genus as explained in \cite{Kapranov} and \cite{Costello:2011nq} respectively.  

Let us explain in more detail what is the motivation to study these notions from a field-theoretical perspective. First recall that Witten's  A- and B-models were constructed by topologically twisting the ${\cal N}=(2,2)$ supersymmetric sigma model in 2D. Alternatively, one can directly define them  as the BV-BRST quantization of classical bosonic sigma models, see for example the various approaches collected in the  review \cite{Birmingham:1991ty}. In any case the 4-fermion terms in the action functional are of the form 
    \be \label{4F one}
 S_{\text{4-fermion}} = \int\dd^2\sigma \, R_{\m\n\rho\sigma}\bar\psi^{\m}\bar\psi^{\n}\psi^{\rho}\psi^{\sigma}\,,
    \ee 
    where $R_{\m\n\rho\sigma}$ are the components of the Riemann curvature tensor of a metric connection on the target space. This is  not a feature specific to 2D theories, since it reflects the openness of the gauge algebra; the Riemann tensor multiplies the 4-fermion terms also in 1D supersymmetric quantum mechanics{\footnote{The relation to the AKSZ construction in 1D is explained in \cite{Grigoriev:1999qz}.}} and in 3D Rozansky-Witten theory, which is also obtained from a supersymmetric sigma model by topologically twisting, albeit this time starting with ${\cal N}=2$ supersymmetry in 6D and dimensionally reducing first to 3D. 
    
    The topological string models are obtained as AKSZ models via a gauge-fixing of the Poisson sigma model \cite{SchallerStrobl,Ikeda}, see for example \cite{Bonechi:2016wqz,Bonechi:2007ar,Kokenyesi:2018xgj} for more details on this procedure. The Poisson sigma model is a topological field theory of scalar fields and 1-forms in 2D and its target space is a Poisson manifold with Poisson bivector $\Pi^{\m\n}$. In its BV action, one may isolate the 4-fermion terms, 
    \be \label{S4fb}
S'_{\text{4-fermion}}=\int\dd^2\sigma \, \partial_{\m}\partial_{\n}\Pi^{\rho\sigma}A_+^{\m}A_+^{\n}\e_{\rho}\e_{\sigma}\,
    \ee 
    where $A_+$ is the antifield of the 1-form $A$  and $\epsilon$ is the BRST ghost of its gauge symmetry. Unlike \eqref{4F one}, the coefficient in \eqref{S4fb} is not manifestly covariant. One may ask what is the covariant tensor underlying this term. According to \cite{Ikeda:2019czt},  it is the  basic curvature tensor for a Lie algebroid connection on the Poisson Lie algebroid. 
    Remarkably, the above statement is true also for the extension of the Poisson sigma model by a 3-form Wess-Zumino-Witten term. Even though in this case the BV action cannot be determined from the vanilla AKSZ construction due to an obstruction to the existence of a QP structure on the target space, nevertheless it was found that the 4-fermion term is still controlled by the basic curvature \cite{Ikeda:2019czt} (of a different connection though, this time with torsion.)  A similar result holds for topological Dirac sigma models in 2D whose target spaces are Dirac manifolds; the 4-fermion terms in the BV action appear together with the basic curvature of two  Lie algebroid connections on a Dirac structure \cite{Chatzistavrakidis:2022wdd}.   

    These arguments are not only relevant for 2D sigma models, but also in higher dimensions. In 3D, the Courant sigma model has a BV action whose 4-fermion term  corresponds when covariantized to a basic curvature tensor, this time for connections on Courant algebroids and their twisted version \cite{Chatzistavrakidis:2022hlu,CIJ}. Here we give a general definition of such a tensor,{\footnote{This definition was also given earlier in \cite{Jotz} in the context of representations up to homotopy for Lie n-algebroids, along with higher analogs thereof.}} 
    in other words from a field-theoretical perspective we answer the question:  {\it{Is there a unifying framework where the curvature and basic curvature tensors that multiply  4-fermion terms find a common origin through a suitable curvature tensor on a Courant algebroid? }} 

    From a geometric standpoint, the basic curvature is the tensor measuring the compatibility of a connection on a Lie algebroid and the Lie bracket on it. The usual definition of an $E$-curvature for a Lie algebroid connection does not give the correct criterion for compatibility; indeed it gives a too weak criterion  \cite{Blaom,Kotov:2016lpx}. 
    It is however useful to mention that for the Lie algebroid  on the tangent bundle, the basic curvature is equal to (minus, in our conventions) the usual curvature for affine connections without torsion. It is for general Lie algebroids that the two notions depart from each other. 
    
    For generalised connections on Courant algebroids, the usual, ``naive'' definition for a torsion tensor is generically not linear in all its arguments. This issue can be solved by the Gualtieri torsion introduced in \cite{Gualtieri:2007bq}.{\footnote{Another interesting proposal in terms of graded Poisson structures was put forward in Ref. \cite{Boffo:2019zus}.}} Moreover, the naive definition of a  curvature as commutator of generalized covariant derivatives is not a tensor either unless one uses the canonical induced connection by the anchor---this already tells us that it behaves differently, since there exists a connection for which the naive definition is a tensor already. When the naive definition for the basic curvature on a Courant algebroid is employed though, the result is never a tensor and must be corrected, as happens with the torsion. It is important to note already that when the dg manifold picture for Courant algebroids is considered, it will turn out that the homological vector field encodes the Gualtieri torsion and the Courant algebroid basic curvature; on the contrary, an analog of the usual curvature plays no role. This in turn becomes manifest in the context of the Courant sigma model and the corresponding tensorial structures that multiply the various terms in the BV action of the theory.
   
Moving to the graded-geometric side, the Atiyah cocycle and its cohomology class can be thought of as avatars of the curvature; replacing the curvature by the Atiyah class, replaces the (differential) Bianchi identity by a Jacobi-like identity \cite{Kapranov}. In the simplest case of connections on a dg manifold, there is an L$_{\infty}$[1] algebra underlying this, with the Atiyah cocycle as binary bracket \cite{Chen:2012yb,Seol:2021tol}. In recent years, there has been a renewal of activity in the physics literature with regard to the relation between field theory and L$_{\infty}$ algebras \cite{Hohm:2017pnh,Jurco:2018sby,Arvanitakis:2020rrk}, motivated by the original ties between these ideas found in closed string field theory and developed in \cite{Zwiebach:1992ie,Lada:1992wc}; see also the closely related unfolded dynamics approach reviewed in \cite{Vasiliev:2005zu} and \cite{Barnich:2005ru}. Another related perspective includes the description of higher gauge theory with towers of differential forms in terms of dg manifolds and dg vector bundles \cite{Grutzmann:2014hkn}. In this approach, for higher gauge theories with 1- and 2-forms (non-Abelian gerbes) \cite{Strobl:2016aph}, the following situation is encountered: (i) gauge transformations are controlled by the first derivative of the components of the homological vector field and (ii) the transformation rules of the field strengths are controlled by the second derivatives of the components of the homological vector field. This situation is generic up to possible dependencies on the field strengths, at least at a local, noncovariant level. One may then observe that for the trivial connection on a dg manifold, the components of the Atiyah cocycle are precisely the second derivatives of the homological vector field on it, while the first derivative is its tangent lift. This motivates us to seek for a relation between higher gauge theory and the Atiyah cocycle. One of our goals is then to make precise the correspondence between field theoretical data and the Atiyah cocycle on a dg manifold, at least is some restricted but still fairly general context.  

\subsection{Structure and results of the paper} 

Let us summarize the main questions we ask and the results of the paper, and describe its structure. In the introductory Section \ref{sec2} we give a quick review of ordinary and $E$-connections on Lie algebroids $E$ and of their tensors. To make this physics-friendly, we provide explicit formulas for basic examples of Lie algebroids, the tangent and the Poisson ones. The main new result of this section is Eq. \eqref{REvsSE}, which gives a relation between the $E$-curvature of an arbitrary $E$-connection and the basic curvature of an ordinary connection, more precisely it tells us that the $E$-curvature is uniquely determined from the $E$-torsion, the basic curvature and the endomorphism that corresponds to the difference of the given $E$-connection from the canonically induced one on $E$. 

In Section \ref{sec3} we define the basic curvature tensor for an ordinary connection on a Courant algebroid $\widehat{E}$  and determine its relation to the Gualtieri torsion and the ordinary curvature.
We give a definition using both the Dorfman bracket and the skew-symmetric Courant bracket. Moreover, we compute it in two basic examples, the standard Courant algebroid with anchor being the projection to the tangent bundle and for the contravariant Courant algebroid with anchor being the extension of the Poisson Lie algebroid on $T^{\ast}M$ by a trivial (totally intransitive) Lie algebroid structure on $TM$. As a side result, we also propose an indirect way to define an $\widehat{E}$-curvature for Courant algebroids which is fully tensorial by construction; the idea is to generalise the relation \eqref{REvsSE} between the $E$-curvature and the basic curvature of Lie algebroids to the Courant algebroid realm, instead of generalizing the usual definition (the commutator of covariant derivatives) that does not yield a tensorial quantity in any case other than for the canonically induced connection. Since this is not central to the physical motivation here, we discuss it in Appendix \ref{appb}.

Our next goal is to understand how the basic curvature on Lie and Courant algebroids can be related to structures on the dg manifold side in a similar spirit to the derived bracket construction \cite{derived1,derived2}. First we recall the well known relation between Lie algebroids $E$ and dg manifolds $E[1]$ due to Vaintrob \cite{Vaintrob} and the way that the anchor and Lie bracket can be obtained from the homological vector field $Q$ and the bracket of vector fields on $E[1]$. Then, introducing connections, torsion and curvature tensors on the dg manifold $E[1]$, we ask: 
 \textit{For a choice of connection $\grave\nabla$ on $\mc M=E[1]$ together with its torsion $\grave{T}$ and curvature $\grave{R}$ tensors, how are a connection $\nabla$ on a Lie algebroid $E$ and its induced $E$-connections $\nabla^{\E}$ obtained, together with their $E$-torsion $T^{\E}$, basic curvature $S^{\E}$ and $E$-curvature $R^{\E}$ tensors?}
To answer this question we invoke the Atiyah 1-cocycle, reviewing its definition and main properties, but also computing it in component form. This reveals a neat relation between the Atiyah cocycle and the second (covariant) derivatives on the components of the homological vector field together with the torsion and curvature tensors on $E[1]$, Eq. \eqref{Atiyahcomponents}. We also briefly recall one of the main results of \cite{AtiyahDG} on the construction of Kapranov L$_{\infty}[1]$ algebras, where the unary bracket is the tangent lift of the homological vector field and the binary bracket is given by the (opposite of the) Atiyah cocycle.  

 The main results of Section \ref{sec4} appear in Propositions \ref{derivedtorsionLie} and \ref{derivedbasicLie} together with Eq. \eqref{iotaRE1}. These state that if we consider the dg manifold with a torsion-free connection $(E[1],Q,\grave\nabla)$ and the contraction map $\iota:C^{\infty}(M,E)\to \G(TE[1])$, then we obtain a connection $\nabla$ on $E$ and the $E$-connection $\mathbullet{\nabla}^{\E}$ that is canonically induced from it through the anchor, for which the following formulas hold for the $E$-torsion and $E$-curvature of $\mathbullet{\nabla}^{\E}$ and for the basic curvature of $\nabla$:
\bea 
 \iota(\mathbullet T^{\E}(e,e'))&=&-{\rm At}(\iota(e),\iota(e'))\,, \nn\\[4pt]
 \iota(S^{\E}(e,e')X)&=&\grave R({\mc X_0},\mc Q\iota(e))\iota(e')-\grave R({\mc X_0},\mc Q\iota(e'))\iota(e)-\grave\nabla_{\mc X_0}\text{At}(\iota(e'),\iota(e))\,, \nn \\[4pt] 
\iota(\mathbullet{R}^{\E}(e,e')e'')&=&\grave R(\mc Q\iota(e),\mc Q\iota(e'))\iota(e'')\,. \nn
\eea
for $e\in \G(E)$, $X\in\mf X(M)$ and $\mc X_0\in \mf X_0(E[1])$, where $\grave R$ is the curvature of $\grave\nabla$ and $\text{At}$ its Atiyah cocycle.
Furthermore, utilizing the dg manifold description of a Courant algebroid, we first show how the Gualtieri torsion and the Courant algebroid basic curvature naturally appear in its homological vector field (see Eq. \eqref{Qcourcov}) and we propose how to obtain them from the tensors associated to a connection on the QP2 manifold. 

In the next two sections we return to field theory. In Section \ref{sec5}, we recall 2D topological sigma models with Poisson and Dirac manifolds as targets and revisit their structure from the angle of our approach in this paper. Starting with the BRST transformation, which is only nilpotent when the field equations of the model are taken into account (on-shell), we recall the construction of its deformation to the BV-BRST differential which is nilpotent off-shell and show that in a manifestly covariant form it is related to the $E$-torsion and basic curvature tensors as in Eqs. \eqref{qbvpsm1}-\eqref{qbvpsm4}; notably, for the 1-form field of the Poisson sigma model
\be  
Q_{\text{\tiny{BV}}}A^{\mathring\nabla}=\mathring{\DD}\e-T^{\ast}(A^{\mathring\nabla},\e)-S^{\ast}(\e\w\e)A^{\dagger}\,, \nn
\ee  
which expresses the BV-BRST differential in terms of $E$-tensors of the Lie algebroid on $T^{\ast}M$, the $E$-torsion $T^{\ast}$ and the basic curvature $S^{\ast}$. Although noncanonical, since it includes a choice of connection that is only auxiliary in this context, this expression remains true for twisted Poisson manifolds, for which a canonical AKSZ construction anyway does not exist. In the dg manifold approach, which is anyway the natural one for the AKSZ construction, one works with maps from $T[1]\S$ to $T^{\ast}[1]M$. In that case, we find an interesting relation of the gauge transformations of the fields and field strengths to the lowest brackets of the Kapranov L$_{\infty}[1]$ algebra. This is found in Proposition \ref{psmkapranov}.

In Section \ref{sec6}, we use the Poisson and Dirac sigma models as a springboard to the approach of Ref. \cite{Grutzmann:2014hkn} to gauge theory for higher degree $p$-forms and ask whether the suggested role of the Atiyah cocycle is also reflected in this higher gauge theory.
This contains a tower of differential form fields that satisfy some minimal requirements with regard to their gauge symmetries, the form of their field strengths and the Bianchi identities \cite{Grutzmann:2014hkn}. The answer to the above question is given in Proposition \ref{hgtatiyah}, where it is shown that the unary and binary brackets of a Kapranov L$_{\infty}[1]$ algebra give the gauge transformations of all the fields and of their strengths, while the commutator of gauge transformations  leads to the ternary bracket in the L$_{\infty}$[1] algebra. 
Finally, we conclude with a discussion on our results and an outlook to open problems in Section \ref{sec7}. Four appendices contain some additional material that can be useful in different parts of this paper, as indicated.

\section{Connections \&  Geometry I: Lie algebroids}
\label{sec2}

\subsection{Connections on Lie algebroids}
\label{sec21}

Recall that a linear connection on a smooth vector bundle $V\overset{\pi}\longrightarrow  M$ over $M$ is a map{\footnote{For terminological clarity, this should better be called a covariant derivative associated to a connection thought of as a map $\nabla:\G(V)\to \G(T^{\ast}M\otimes V)$ which satisfies the Leibniz rule $\nabla(f v)=f\nabla v+\dd f\otimes v$. We shall not insist on this since it should be clear from context, unless otherwise stated.}}
\bea 
\nabla: \G(TM\otimes V)&\to& \G(V) \nn\\[4pt] 
(X,v)&\mapsto& \nabla_{X}v\,,
\eea 
which satisfies the homogeneity and Leibniz rules
\bea 
\nabla_{f X}v=f\, \nabla_{X}v\,, \quad 
\nabla_{X}(fv)=(\nabla_{X}f) \, v+f\,\nabla_{X}v\,,
\eea 
 where $\nabla_{X}f=X(f)$ in standard notation and $f\in C^{\infty}(M)$, and linearity in both arguments. 
 If we consider a local basis $\mf v_m$ of $V$, in which case sections are expanded as $v=v^{m} \mf v_{m}$, 
\be 
\nabla_{X}v=X^{\mu}\left(\partial_{\m}v^{m}+v^{n}\o^{m}_{\m n}\right)\mf v_m\,,
\ee 
where connection coefficients are defined as 
\be 
\nabla_{\m}\mf v_m=\o^{n}_{\m m}\mf v_n\,.
\ee 
When $V=TM$, we have an affine connection on $M$ and use the more usual notation $\G_{\m\n}^{\rho}$.

We can generalize the notion of a vector bundle connection on $V$ to the case when the first entry is a section of some other vector bundle $E\longrightarrow M$ over the same manifold $M$. We denote sections of $E$ as $e\in \G(E)$. In addition, we assume that it is an anchored vector bundle $(E,\rho)$, see \cite{LGLS}, one which is equipped with a vector bundle morphism $\rho:E\to TM$. The role of the anchored vector bundle is that we can now use it to define a suitable generalization of a connection on $V$ as follows: 
Given a vector bundle $V$ and an anchored vector bundle $(E,\rho)$, an $E$-connection on $V$ is a map
\bea 
\nabla^{\E}:\G(E\otimes V)&\to &\G(V) \nn\\[4pt] 
(e,v)&\mapsto& \nabla^\E_e\, v\,,
\eea 
which satisfies the following properties,
\bea 
\nabla^{\E}_{fe}v=f\,\nabla^{\E}_e\, v\,, \quad 
\nabla^{\E}_{e}(fv)= (\nabla^{\E}_{e}f)\, v+f\, \nabla^{\E}_{e}\,v\,,
\eea 
and where in the Leibniz rule we have 
\be 
\nabla^{\E}_{e}f=\rho(e)(f)\,,
\ee 
namely we use the anchor to turn the section $e$ into a vector field and subsequently act with it on the function $f$. 
In addition, it was proven in \cite{LGLS} that under very mild assumptions every anchored bundle can be extended to an almost-Lie algebroid, where $E$ is additionally equipped with a skew-symmetric bracket $[\cdot,\cdot]_{\E}$ on its sections which satisfies the Leibniz rule 
\be 
[e,fe']_{\E}=f[e,e']_{\E}+\rho(e)(f)\, e' 
\ee 
and the homomorphism condition 
\be 
\rho([e,e']_\E)=[\rho(e),\rho(e')]\,,
\ee 
for arbitrary sections $e,e'\in \G(E)$. The bracket on $E$ does not necessarily satisfy the Jacobi identity, which is the reason the structure is called almost-Lie. We will not delve into details of how the kernels associated to ``almost'' are taken care of, since in the following we will primarily be interested in genuine Lie algebroids, where the Jacobi identity holds.  

Given a $TM$-connection $\nabla$ on $V$, hence we refer to it as an ordinary connection, there always exists a canonical induced $E$-connection on $V$ via the anchor, which we denote with a bullet, 
\be \label{trivialcon}
\mathbullet\nabla^{\E}_{e}\, v=\nabla_{\rho(e)}v\,.
\ee 
We will refer to it as the canonical $E$-connection.
This is obviously not the most general $E$-connection one can consider. In local bases, for a general $E$-connection on $V$ and for $e=e^a\mf{e}_a$ and $v=v^{m}\mf{v}_{m}$, 
\be 
\nabla^{\E}_e\, v= e^{a}\left(\rho_a{}^{\m}\partial_{\m}v^{m}+v^n\o^{m}_{an}\right)\mf{v}_{m}\,,
\ee 
whereas for the canonical $E$-connection this reduces to 
\be 
\mathbullet\nabla^{\E}_e\, v= \rho_a{}^{\m} e^{a}\left(\partial_{\m}v^{m}+v^n\o^{m}_{\m n}\right)\mf{v}_{m}\,.
\ee 
The connection coefficients of the canonical $E$-connection are obtained through the anchor map, which is not the case in general. As a final bit of notation in this respect, we define the difference 
\be \label{phidef}
\phi(e,v):=(\nabla^{\E}_e-  \mathbullet\nabla^\E_e)v\,,
\ee 
for an arbitrary $E$-connection on $V$.
Then $\phi$ is an endomorphism valued 1-form and transforms homogeneously under changes of coordinates. 
In a given local basis, its components are 
\be \label{omega is rho omega}
\phi^{m}_{an}=\o^{m}_{an}-\rho_{a}{}^{\m}\,\o^{m}_{\m n}\,.
\ee
Eq. \eqref{phidef} essentially means that we will be thinking of general $E$-connections on $V$ given as the sum of the canonical $E$-connection and an endomorphism 1-form: $\nabla^{\E}=\mathbullet{\nabla}^{\E}+\phi$. Note that 
the vector bundles $E$ and $V$ can also be identical. In that case we speak of an $E$-connection on $E$ (or a $V$-connection on $V$, of course, the name follows the notation here.) When this happens, in a local basis the Latin indices $a, b,\dots$ and $m,n,\dots$ become indistinguishable. 
For instance, if $E=T^{\ast}M$ then 
the arbitrary connection coefficients are of the form $\omega^{\m\n}_{\rho}$ and the endomorphism $\phi$ is in general 
\be 
\phi^{\m\n}_{\rho}=\o^{\m\n}_{\rho}+\Pi^{\m\s}\G^{\n}_{\s\rho}\,
\ee 
where recall that $\G$ are the coefficients of an affine connection on the base manifold $M$ over which the Lie algebroid with anchor given by the Poisson structure $\Pi$ is defined.{\footnote{One should be cautious of the fact that when $\omega$ is written for the cotangent bundle, it is identified with $\G^{\ast}$, namely the dual connection coefficients, which are in turn the opposite of $\G$.}} 

\subsection{Torsion and curvature tensors}
\label{sec22} 

Let us consider $E=V$ and an $E$-on-$E$ connection $\nabla^{\E}$. The tensors for this connection, hence referred as $E$-tensors, are defined as follows.{\footnote{More precisely, we are defining tensor maps here.}} The $E$-torsion is the map 
\bea 
&&T^{\E}:\G(E\otimes E)\to \G(E) \nn\\[4pt] 
&& T^{\E}(e,e')= \nabla^{\E}_{e}e'- \nabla^{\E}_{e'} e - [e,e']_{\E}\,,\label{etorsion}\eea 
given by the same formula as the ordinary torsion but this time with respect to the $E$-connection and the (almost-)Lie bracket on $E$.  
The $E$-curvature is defined as 
\bea \label{rlie}
&& R^{\E}:\Gamma(E\otimes E\otimes E) \to \G(E) \nn\\[4pt] 
&& R^{\E}(e,e')e'' =[\nabla^\E_{e}, \, \nabla^{\E}_{e'}]e''- \nabla^\E_{[e,e']_{\E}}e''\,,
\eea 
once more as a generalization of the usual formula with the commutator of covariant derivatives. 
Like their ordinary counterparts, they satisfy the antisymmetry properties 
\be 
T^\E(e,e')=-T^\E(e',e) \quad \text{and} \quad R^\E(e,e')e''=- R^\E(e',e)e''\,,
\ee 
which are obvious from their definitions, 
as well as the corresponding linearity properties, 
\bea 
T^{\E}(fX,gY)=fg\, T^{\E}(X,Y)\,, \quad R^{\E}(fX,gY)(hZ)=fgh\, R^{\E}(X,Y)Z\,.
\eea 
as can be easily proven using   $\rho(fe)=f\rho(e)$ and that $\rho$ is a homomorphism. In a local basis, the component form of these tensors becomes---note that we refrain from including the $E$ superscript when the nature of the object is clear from its index structure--- 
\bea 
T_{ab}^c&=&2\omega^{c}_{[ab]}-C^c_{ab}\,,
\\[4pt] 
R^{d}_{cab}&=&\partial_a\o_{bc}^d-\partial_b\o_{ac}^d+\o^e_{bc}\o^d_{ae}-\o^e_{ac}\o^d_{be}-C_{ab}^{e}\o^d_{ec}\,,
\eea 
where $C_{ab}^c$ the structure functions of the Lie algebroid in this basis, namely 
\be 
[\mf e_a,\mf e_b]_{\E}=C_{ab}^c\, \mf e_c\,.
\ee 
In the following, we shall denote the torsion and  curvature $E$-tensors that correspond to the canonical $E$-connection $\mathbullet{\nabla}^{\E}$ by $\mathbullet{T}^{\E}$ and $\mathbullet{R}^{\E}$. It is useful to note that the antisymmetric part of the endomorphism $\phi$ defined in the previous section satisfies 
\be \label{phiantisym}
2\, \phi(e_{[1},e_{2]}):=\phi(e_1,e_2)-\phi(e_2,e_1)=T^{\E}(e_1,e_2)-\mathbullet{T}^{\E}(e_1,e_2)\,,
\ee 
namely it is the difference of the $E$-torsion for the chosen $E$-connection from the one of the canonical $E$-connection.  

As case studies, consider the simplest nontrivial examples, first the tangent Lie algebroid with $E=TM$, $\rho=\text{id}$ and bracket the ordinary Lie bracket of vector fields and second the Poisson Lie algebroid for a Poisson manifold $M$ with $E=T^{\ast}M$, anchor $\rho=\Pi^{\sharp}$ given by the map induced by the Poisson structure $\Pi$ and bracket the Koszul-Schouten bracket of 1-forms---see Appendix \ref{appa} for the necessary definitions and conventions. Connection-wise, in the first case we have an ordinary affine connection $\nabla$ on $M$, possibly with torsion, and in the second a $T^{\ast}M$-connection on $T^{\ast}M$. For the sake of examples, in the latter case we shall consider the  $T^{\ast}M$-connection induced by a torsionless connection $\mathring{\nabla}$ on $M$ along the anchored 1-form, as in Eq. \eqref{trivialcon}.  Moreover, we denote all tensors in the tangent case by the corresponding symbols such as $T, R, S$ etc., whereas we use the notation $T^{\ast}, R^{\ast}, S^{\ast}$ for the tensors on the Poisson Lie algebroid. Then we have, respectively, the connection coefficients 
\bea 
\o_{\m\n}^{\rho}= \G_{\m\n}^{\rho}\,, \qquad
\o^{\ast}{}_{\m}^{\n\rho}=-\Pi^{\n\s}\mathring{\G}_{\s\m}^{\rho}\,, 
\eea 
and we can compute the $E$-torsion tensors 
\bea  
T_{\m\n}^{\rho}&=&2\G_{[\m\n]}^{\rho}\,, \\[4pt] 
\mathbullet{T}^{\ast}{}^{\m\n}_{\rho}&=& -\partial_{\rho}\Pi^{\m\n}+2\omega^{\ast}{}_{\rho}^{[\m\n]}=-\mathring{\nabla}_{\rho}\Pi^{\m\n}\,.
\eea  
 For the  $E$-curvature we find 
\bea 
R_{\rho\m\n}^{\s}&=& \partial_{\m}\G_{\n\rho}^{\s}-\partial_{\n}\G^{\s}_{\m\rho}+\G^{\s}_{\m\l}\G^{\l}_{\n\rho}- \G^{\s}_{\n\l}\G^{\l}_{\m\rho}\,, \\[4pt] 
\mathbullet{R}^{\ast}{}^{\rho\m\n}_{\s}&=& \Pi^{\m\k}\Pi^{\n\l}\mathring{R}^{\rho}_{\s\k\l}\,,\label{RETstarM}
\eea 
with $\mathring{R}$ the ordinary curvature of the torsionless connection $\mathring{\nabla}$ on $M$. As emphasized in the introduction, \eqref{RETstarM} shows that the $E$-curvature is unrelated to the coefficient of the  4-fermion term in the BV-BRST formalism of the Poisson sigma model. 

In the present case, however, there exists an additional important tensor. This is called the basic curvature in \cite{Crainic,Kotov:2016lpx}, initially defined in \cite{Blaom} as a compatibility condition between an ordinary connection on $E$ and its bracket.{\footnote{The reason it is a curvature is that it corresponds to the same integrability condition as the usual curvature tensor, but for a splitting of the short exact sequence of $E$ and its 1st jet bundle $J^1E$. Specifically, since to every vector bundle $E$ we can associate the short exact sequence 
\be 
0\longrightarrow T^{\ast}M\otimes E \longrightarrow J^{1}(E) \longrightarrow E \longrightarrow 0\,,
\ee 
a connection on $E$ corresponds to a splitting $\s:E\to J^{1}(E)$ of the sequence as $\s(e)=j_1(e)+\nabla e$, in terms of the first jet prolongation of $e$. Then $S(e,e')=[\s(e),\s(e')]-\s([e,e'])$, for the Lie  bracket on the jet bundle \cite{Kotov:2016lpx}. } 
 Given the anchored bundle $(E,\rho)$ and its bracket, consider a $TM$-connection on $E$ denoted as $\nabla$. 
 Then we define the following induced $E$-connections on $E$ and $TM$ respectively,
\bea 
\overline{\nabla}^{\E}_ee'&:=&\nabla_{\rho(e')}e+[e,e']_{\E}\,, \\[4pt] 
\overline{\nabla}^{\E}_eX&:=&\rho(\nabla_{X}e)+[\rho(e),X]\,. \label{EonTM}
\eea 
These are called basic connections in Ref. \cite{Crainic}.  Moreover, it holds that \cite{Blaom} 
\be 
\rho\circ \, \overline{\nabla}^\E= \overline{\nabla}^\E\circ \rho\,,
\ee 
which explains the united notation; in fact the two connections can be considered as a single one on the chain complex $E\overset{\rho}\rightarrow TM$ \cite{Crainic}.
Then the basic curvature is a map 
\bea 
&& S^\E:\G(E\otimes E\otimes TM)\to \G(E) \nn\\[4pt]
&& S^\E(e,e')X=\nabla_{X}[e,e']_{\E}-[\nabla_Xe,e']_{\E}-[e,\nabla_Xe']_{\E}-\nabla_{\overline{\nabla}^\E_{e'}X}e+\nabla_{\overline{\nabla}^\E_eX}e'\,. 
\label{basiccurvature}
\eea 
We observe that it contains the ordinary $TM$-connection on $E$ and the induced $E$-connection on $TM$, i.e. the roles of $E$ and $TM$ are interchanged in these two connections. For instance, when $E=TM$ with the standard algebroid structure, then 
\be  
\overline{\nabla}^{\E=\TM}_YX = \nabla_{X}Y+[Y,X]=\nabla_{Y}X-T(Y,X)\,,
\ee  
and this is the opposite affine connection. Note that the basic curvature refers to an ordinary and not to an $E$-connection on $E$; we still use the notation with the superscript $^\E$ to remember that we are working with connections on this bundle.
Note also that the last two terms in the definition are included so that the tensorial character of the quantity is guaranteed---without these terms it would fail to be linear in all its entries. 

In component form, the basic curvature reads 
\bea 
S^{c}{}_{\m ab}&=& \partial_{\m}C_{ab}^c+C_{ab}^d\o_{\m d}^c-C^c_{db}\o^d_{\m a}-C^{c}_{ad}\o^d_{\m b} \nn\\[4pt] 
&& -\, 2\rho_{[a}{}^{\n}\partial_{\n}\o^c_{\m b]}-2(\partial_{\m}\rho_{[a}{}^{\n}-\rho_d{}^{\n}\o^d_{\m [a})\o^c_{\n b]}\,.
\eea 
With reference to the two basic examples on the tangent and cotangent Lie algebroids, we find that the basic curvature in each case becomes 
\bea 
\label{STM} S^{\s}_{\rho\m\n}&=&-R^{\s}_{\rho\m\n}+2\nabla_{[\m}T^{\s}_{\n]\r}-T^{\k}_{\m\n}T^{\s}_{\r\k}-2T^{\k}_{\r[\m}T^{\s}_{\n]\k}\,, \\[4pt] 
\label{ST*M} {S}^{\ast}{}^{\m\n}_{\r\s}&=& \mathring\nabla_{\r}\mathring\nabla_{\s}\Pi^{\m\n}+2\Pi^{\k[\m}\mathring{R}^{\n]}{}_{\s\r\k}\,.
\eea 
Observe that the basic curvature on the tangent Lie algebroid with a torsion-free affine connection is simply the ordinary curvature (in fact its opposite in our conventions). Alternatively and in the spirit of teleparallelism, the basic curvature of the Weitzenb\"ock connection is the covariant derivative of the torsion tensor plus quadratic terms on it.  We also observe that in the Poisson case, the basic curvature is precisely the type of tensor that appears in the BV action of the Poisson sigma model, as for example in the target space covariant formulation of \cite{Baulieu:2001fi}, Eq. (3.30).

It is interesting to explain further the relation between the basic curvature on one hand and the $E$-torsion and  $E$-curvature on the other. To keep things general, we choose a $TM$-connection $\nabla$ on $E$, which induces the canonical $E$-connection, and we also consider an arbitrary $E$-connection on $E$, i.e. $\nabla^{\E}=\mathbullet{\nabla}^{\E}+\phi$, and determine the relation of its $E$-tensors to the ordinary and basic curvature tensors of $\nabla$.  This will also explain in basis-independent terms the structure of the basic curvature in the examples \eqref{STM} and \eqref{ST*M}. First, we introduce the notation 
\be 
\text{Cycl}\big(R^\E(e_1,e_2)e_3\big):= R^\E(e_1,e_2)e_3+ R^\E(e_3,e_1)e_2+ R^\E(e_2,e_3)e_1
\ee 
and similarly for any other quantity with 3 or more entries that can be permuted in a cyclic fashion. For a Lie algebroid, the algebraic Bianchi identity for the  $E$-curvature tensor reads{\footnote{Note that the first term on the right-hand side is, save the superscript notation, the tensor $\nabla T$ evaluated on $e_2$ and $e_3$ and it should not be confused with $\nabla(T(e_2,e_3))$.}}
 \be  \text{Cycl}\big(R^\E(e_1,e_2)e_3\big)=\text{Cycl}\bigg(\nabla^\E_{e_1}T^\E(e_2,e_3)+\, T^\E\left(T^\E(e_1,e_2),e_3\right)\bigg)\,.
 \ee 
This is easily proven using the definition of the  $E$-curvature, then turning all second derivative terms into first derivatives of the $E$-torsion via the definition of the latter and finally using the Jacobi identity for the Lie algebroid bracket.{\footnote{Note that for an almost-Lie algebroid the correct expression is a modification of this.}}} The  proof is found in Appendix \ref{appc}. 

Next, it is useful to rewrite the definition of the basic curvature in an equivalent form, in terms of $E$-tensors. It is a matter of simple algebra and use of the definition for the $E$-torsion to show that 
\be \label{screl}
S^\E(e_1,e_2)X=- \nabla_{X}T^\E(e_1,e_2)-{\cal C}(e_1,X)e_2+{\cal C}(e_2,X)e_1\,,
\ee 
where we have defined the tensor 
\be \label{ctensor}
{\cal C}(e_1,X)e_2:=[\nabla^\E_{e_1},\nabla_X]e_2-\nabla_{[\rho(e_1),X]}e_2+\phi(\nabla_Xe_1,e_2)\,,
\ee 
and we recall that $\phi$ is the difference of the $E$-connection we work with from the canonical $E$-connection.
It is straightforward to show that ${\cal C}$ is linear in all its arguments. Notably, this is the result of the combined effort of the the first pair of terms and the final one. The first pair looks like a would-be curvature, but it is not a tensor by itself in general.  The same is true for the last term which is not a tensor in $e_1$ (although it is in $X$ and $e_2$). 

Suppose now that we consider a vector field $X$ that is obtained through the anchoring of a section of $E$, say $\rho(e_3)$. Then, applying the above formulas we first find that the tensor ${\cal C}$ can be related to the $E$-curvature $R^\E$ as
\be 
{\cal C}(e_1,\rho(e_3))e_2= R^\E(e_1,e_3)e_2-\nabla^\E_{e_1}\phi(e_3,e_2)-\phi(T^\E(e_1,e_3),e_2)-\phi(\phi(e_3,e_1),e_2)\,,
\ee 
and to the ordinary curvature $R$  as
\be 
{\cal C}(e_1,\rho(e_3))e_2= R(\rho(e_1),\rho(e_3))e_2-\nabla_{\rho(e_3)}\phi(e_1,e_2)\,.
\ee 
It is now obvious that when the $E$-connection is chosen to be the canonical one, in which case $\phi$ is identically zero, ${\cal C}$ evaluated on $\rho(e_3)$ is identical to the  $E$-curvature and to the ordinary curvature along anchored sections, which are anyway one and the same thing in that case. This is of course not true for a general $E$-connection. 
 Using now the algebraic Bianchi identity, we find
\bea 
S^\E(e_1,e_2)\rho(e_3)= - \mathbullet\nabla^\E_{e_3} \mathbullet{T}^\E(e_1,e_2)-R(\rho(e_1),\rho(e_3))e_2+R(\rho(e_2),\rho(e_3))e_1\,,
\label{STErel}
\eea 
in terms of the ordinary endomorphism valued curvature 2-form $R\in\Omega^2(M,\text{End}(E))$ of $\nabla$ and the torsion $\mathbullet{T}^\E$. Alternatively, defining the difference 
\be  
\widehat{\phi}:=\phi-T^{\E}\,,
\ee 
we find the following relation to the $E$-curvature of the general $\nabla^{\E}$:
 \begin{tcolorbox}[ams nodisplayskip, ams align] 
R^{\textit{\tiny{E}}}(e_1,e_2)e_3+  S^{\textit{\tiny{E}}}(e_1,e_2)\rho(e_3)=2\nabla^{\E}_{e_{[1}}\widehat\phi(e_{|3|},e_{2]})  
+  \widehat\phi(e_3,T^{\E}(e_{1},e_{2}))+2\widehat\phi(\widehat{\phi}(e_3,e_{[1}),e_{2]})\,.\quad 
\label{REvsSE}\end{tcolorbox}
We have thus expressed the $E$-curvature of an arbitrary $E$-connection in terms of other known tensors of a Lie algebroid equipped with a $TM$-connection. Eq. \eqref{REvsSE} can be used as a definition of the  $E$-curvature instead of the usual one in terms of the commutator of covariant derivatives. We also observe that for $\widehat\phi=0$  the $E$-curvature is the opposite of the basic curvature on an anchored section. This includes the case of $\phi=0=T^{\E}$ (canonical $E$-connection of vanishing $E$-torsion), which is reflected in the local expressions for the two basic examples we considered above, Eqs. \eqref{STM} and \eqref{ST*M}. More generally the case of not necessarily vanishing $\phi=T^{\E}$ corresponds to the basic connection $\overline{\nabla}^{\E}$. That this is so can be seen through Eq. \eqref{phiantisym}, which for $\phi=T^{\E}$ implies that $\phi=-\mathbullet{T}^{\E}=T^{\E}$. This means that the connection at hand is 
\bea  
\nabla^{\E}_ee'=\mathbullet{\nabla}^{\E}_ee'-\mathbullet{T}^{\E}(e,e')=\nabla_{\rho(e)}e'-(\nabla_{\rho(e)}e'-\nabla_{\rho(e')}e-[e,e']_{\E})=\overline{\nabla}^{\E}_e e'\,.
\eea 
This was found in a more direct way in Ref. \cite{Crainic}. We can now summarize our findings: 
\begin{prop}
    Given a connection $\nabla$ and an $E$-connection $\nabla^{\E}$ on a Lie algebroid $E$ with anchor map $\rho$, the  $E$-curvature $R^{\E}$ of $\nabla^{\E}$ is uniquely determined from the $E$-torsion $T^{\E}$ of $\nabla^{\E}$, the basic curvature $S^{\E}$ of $\nabla$ and the endomorphism valued 1-form $\phi=\nabla^{\E}-\mathbullet{\nabla}^{\E}$ through Eq. \eqref{REvsSE}.   
\end{prop}
\begin{corollary}
    When $\phi=T^{\E}$, in which case the $E$-connection on $E$ is the basic connection, the $E$-curvature is the opposite of the basic curvature evaluated on an anchored section of $E$, as in 
    \be  {R}^{\E}(e_1,e_2)e_3=-{S}^{\E}(e_1,e_2)\rho(e_3)\,.
    \ee  
\end{corollary}
This corollary corresponds to the first part of Proposition 2.11 in Ref. \cite{Crainic}. For completeness, we report also the second leg of that proposition, 
\be 
R^{\E}(e_1,e_2)X=-\rho(S^{\E}(e_1,e_2)X)\,,
\ee 
where the curvature on the right hand side is computed for the basic $E$-connection on $TM$ given in \eqref{EonTM}.

\section{Connections \& Geometry II: Courant algebroids}
\label{sec3}

\subsection{Torsion for generalised connections}
\label{sec31} 

Let us now consider Courant algebroids, whose vector bundle is hence denoted by $\widehat{E}$ to differentiate it from Lie algebroids. A Courant algebroid features an anchor map $\rho: \widehat{E}\to TM$, a binary operation $\circ$ (Dorfman bracket) satisfying 
\be 
e_1\circ(e_2\circ e_3)=(e_1\circ e_2)\circ e_3+ e_2\circ (e_1\circ e_3)\,,
\ee 
for $e_1, e_2, e_3\in \G(\widehat{E})$ are generalized vector fields, making $(\Gamma(\widehat{E}),\circ)$ a Leibniz algebra, and a non-degenerate symmetric bilinear form $\langle\cdot,\cdot\rangle$ (a metric) on $\widehat{E}$, such that 
\be 
\langle e_1,e_2\circ e_2\rangle=\frac 12 \rho(e_1)\langle e_2,e_2\rangle= \langle e_1\circ e_2,e_2\rangle\,.
\ee 
Note that the Dorfman bracket has a symmetric part. Alternatively,
the structure may be defined in terms of the skew-symmetric Courant bracket
\be
[e_1,e_2]_{\Eh}=\sfrac 12(e_1\circ e_2-e_2\circ e_1),\nn
\ee
with suitable properties, see \cite{Liu:1995lsa}. The notion of an $\widehat{E}$-connection on $\widehat{E}$, also known as a generalized connection, still exists in the same way as before.  We require in addition that it is metric compatible, 
\be  \nabla^{\Eh}_{e}\langle e',e''\rangle=\langle \nabla^{\Eh}_{e}e',e''\rangle+\langle e,\nabla^{\Eh}_e e''\rangle\,.\ee 
A proposal for a $\widehat{E}$-torsion tensor was given in \cite{Gualtieri:2007bq}, hence referred as Gualtieri torsion. It is a section ${\cal T}^{\Eh} \in \G(\w^2\widehat{E}\otimes \widehat{E})$ given by the formula 
\bea \label{GTE}
\mc T^{\Eh}(e_1,e_2,e_3)=\langle \nabla^{\Eh}_{e_1}e_2- \nabla^{\Eh}_{e_2}e_1 - [e_1,e_2]_{\Eh}, e_3\rangle+\frac 12 \left(\langle  \nabla^{\Eh}_{e_3}e_1,e_2\rangle -\langle \nabla^{\Eh}_{e_3}e_2,e_1\rangle\right)\,.
\eea 
Note that the first term on the right hand side, sometimes called the naive torsion, is not linear in all arguments. The logic then is to add a suitable new piece to the naive definition such that it counterbalances its failure to be a tensor, indeed the second set of terms. 
It is useful to note that a similar definition exists for the Dorfman bracket: 
\bea \label{GTEd}
\mc T^{\Eh_{\circ}}(e_1,e_2,e_3)=\langle \nabla^{\Eh}_{e_1}e_2- \nabla^{\Eh}_{e_2}e_1 - e_1\circ e_2, e_3\rangle+ \langle \nabla^{\Eh}_{e_3}e_1,e_2\rangle\,,
\eea 
where the correction term was de-antisymmetrized.
It is useful to write down the local coordinate expressions for this tensor. In our conventions, the skew-based one reads
\bea 
{\cal T}_{abc}= \o_{[ab]c}+\frac 12 (\omega_{c[ab]}-C_{cab})\,,
\eea 
where indices are raised and lowered with the fiber metric on the Courant algebroid. 

A class of examples, which are the ones mostly appearing in physical applications, are exact Courant algebroids of the form $E\oplus E^{\ast}$. This includes the generalized tangent bundle $TM\oplus T^{\ast}M$, in which case we denote local sections as
$e=(X,\eta)$,
with vector field $X$ and 1-form $\eta$. The fiber metric is given as 
\be 
\langle e,e'\rangle=\frac 12 \left(X(\eta')+X'(\eta)\right)\,,
\ee 
whereas the corresponding antisymmetric combination will be denoted as 
\be 
\langle e,e'\rangle_{-}=\frac 12(X(\eta')-X'(\eta))\,.
\ee 
In this case, the Courant bracket can be written as 
\bea \label{Courantbracketfull}
[e,e']_{\Eh}= \left([X,X']+{\cal L}^{\ast}_{\eta}X'-{\cal L}^{\ast}_{\eta'} X+\dd_{\ast}\langle e,e'\rangle_{-},\, [\eta,\eta']_{\text{KS}}+{\cal L}_{X}\eta'-{\cal L}_{X'}\eta-\dd\langle e,e'\rangle_{-}\right)\,,\,\,\,\,
\eea 
where $\dd_{\ast}$ is the differential on the cotangent bundle and ${\cal L}^{\ast}$ the Lie derivative associated to it \cite{Liu:1995lsa}.
To make all this more specific, we are going to work out two simple examples. In both cases we consider the canonical $\widehat{E}$-connection $\mathbullet{\nabla}^{\Eh}$ induced by a connection $\nabla$ on $M$, possibly with (ordinary) torsion $T$. First the standard Courant algebroid, say $\widehat{E}_{A}$ where the anchor is the projection to the vector field, $\rho=\text{id}\oplus 0$ and the bracket reads 
\bea 
[e,e']_{\Eh_{A}}= \left([X,X'],\, {\cal L}_{X}\eta'-{\cal L}_{X'}\eta-\dd\langle e,e'\rangle_{-}\right)\,;
\eea 
 second the cotangent version of it, say $\widehat{E}_{B}$, where the anchor is $\Pi^{\sharp}:\widehat{E}\to TM$ given solely by a Poisson (or even twisted Poisson) structure $\Pi$ on $M$  
such that 
\be 
\Pi^{\sharp}(X)=0 \quad \text{and}\quad \Pi^{\sharp}(\eta)=\Pi^{\m\n}\eta_{\m}\partial_{\n}\,,
     \ee 
and the bracket being 
\be 
[e_1,e_2]_{\Eh_B}=\left({\cal L}_{\eta_1}X_2-{\cal L}_{\eta_2}X_1-\frac 12 \dd_{\Pi}\left(\eta_1(X_2)-\eta_2(X_1)\right),[\eta_1,\eta_2]_{\text{KS}}\right)\,.
\ee 
Here the 1-form part of the bracket is given by the binary Koszul-Schouten bracket (or a twisted extension thereof), see Appendix \ref{appa}, whereas the vector part contains the differential 
\be 
\dd_{\Pi}=[\Pi,\cdot\,]_{\text{SN}}\,,
\ee 
in terms of the Schouten-Nijenhuis bracket for multivector fields, and the associated Lie derivative defined via the Cartan equation 
\be 
{\cal L}_{\eta}=\dd_{\Pi}\circ \iota_{\eta}+\iota_{\eta}\circ\dd_{\Pi}\,.
\ee 
The above data indeed satisfy all axioms of a Courant algebroid. 
Let us now compute the Gualtieri torsion for these two examples. In the first case the result is 
\bea 
\mathbullet{\cal T}^{\Eh_A}(e_1,e_2,e_3)=\text{Cycl}_{123}\langle T(X_1,X_2),\eta_3\rangle\,,
\eea 
or in components 
\be 
\mathbullet{\cal T}^{\Eh_A}(e_1,e_2,e_3)=\frac 32\, T_{\m\n}^{\rho}X_{[1}^{\m}X_2^{\n}\eta_{3]\rho}\,.
\ee 
We observe that the Gualtieri torsion vanishes in this example when the torsion of $M$ vanishes. 
In the second case,  the result is 
\bea 
\mathbullet{\mc T}^{\Eh_B}(e_1,e_2,e_3)=-\frac 32 \bigg(\mathring{\nabla}_{\rho}\Pi^{\m\n}-\Pi^{\m\s}T_{\s\r}^{\n}+\Pi^{\n\s}T_{\s\r}^{\m}\bigg) \eta_{[1 \m}\eta_{2\n}X_{3]}^{\rho}
\,.
\eea 
Not surprisingly, when the torsion on the base manifold vanishes, the Gualtieri torsion in this case is not zero; it is given by the $E$-torsion of the cotangent Lie algebroid. 

Taking the above two examples as building blocks of a more general case and recalling that a class of Courant algebroids consists of Lie bialgebroids, we have the following remarks. The first example is the sum of the tangent Lie algebroid studied in the previous section and the trivial cotangent Lie algebroid with zero anchor map{\footnote{A totally intransitive Lie algebroid in the terminology of \cite{Mackenzie}. Recall that a transitive Lie algebroid is one where the anchor is a fiberwise surjective map and a regular Lie algebroid is one where the anchor is of constant rank.}} and vanishing bracket, which therefore would have vanishing $E$-torsion anyway. The second example is the sum of the cotangent Lie algebroid studied in the previous section with a trivial tangent Lie algebroid, one with zero anchor and bracket and zero $TM$-torsion. In the same spirit, one can now sum up the two examples and construct one with anchor 
\be 
\rho=\text{id}\oplus \Pi^{\sharp}\,,
\ee 
and bracket the full Courant bracket of Eq.\eqref{Courantbracketfull}. 
In this case, the Gualtieri torsion is essentially the sum of the previous ones, which appear as components of the ``double'', 
\bea  
\mathbullet{\cal T}^{\Eh}(e_1,e_2,e_3)=\frac 32\, T_{\m\n}^{\rho}X_{[1}^{\m}X_2^{\n}\eta_{3]\rho}-\frac 32 \bigg(\mathring{\nabla}_{\rho}\Pi^{\m\n}-\Pi^{\m\s}T_{\s\r}^{\n}+\Pi^{\n\s}T_{\s\r}^{\m}\bigg) \eta_{[1 \m}\eta_{2\n}X_{3]}^{\rho}\,.
\eea  
As a final comment on the generalized torsion, it is straightforward to show that 
\be 
(\mc T^{\Eh}-\mathbullet{\mc T}^{\Eh})(e_1,e_2,e_3)=\langle\phi(e_1,e_2)-\phi(e_2,e_1),e_3\rangle +\frac 12 \big(\langle \phi(e_3,e_1),e_2\rangle -\langle\phi(e_3,e_2),e_1\rangle \big)\,,
\ee 
where the endomorphism $\phi$ is defined in the same way as for Lie algebroids. 

\subsection{The basic curvature for Courant algebroids}\label{sec32}

Given our interest in the basic curvature tensor, one may now ask whether this notion generalizes to Courant algebroids too. As expected, directly considering the same map as before, namely \eqref{basiccurvature}, and replacing $E$ by $\widehat{E}$, does not result in a tensor. Indeed, linearity fails then and there are additional terms that should be compensated for. It turns out, however, that there exists a generalized notion of basic curvature that indeed is a tensor. It is obtained in a similar way as the Gualtieri torsion, adding suitable compensating terms. 
Specifically, given a Courant algebroid $\widehat{E}$ with anchor ${\rho}$ and a Courant algebroid connection $\nabla^{\Eh}_{e}e'$ with $e,e'\in\G(\widehat E)$, let the induced (basic) $\widehat E$-connection on $TM$ be  
\be \label{EhonTM}
\overline{\nabla}^{\Eh}_{e}X=\rho(\nabla_{X}e)+[\rho(e),X]\,.
\ee 
This definition is independent of whether the skew or nonskew binary operation is used. 
Then we define the Courant algebroid basic curvature as follows 
\begin{defn} \label{courantbasicdef} The basic curvature of a connection $\nabla$  on a Courant algebroid $\widehat{E}$ with anchor map $\rho$, skew-symmetric Courant bracket $[\cdot,\cdot]_{\Eh}$ and fiber metric $\langle\cdot,\cdot\rangle$ is a section $\mc S^{\Eh}\in\G(\w^2\widehat{E}\otimes \widehat{E}\otimes T^{\ast}M)$ given by the formula
 \begin{tcolorbox}[ams nodisplayskip, ams align]  
\mc S^{\Eh}(e_1,e_2,e_3)X&=\langle \nabla_{X}[e_1,e_2]_{\Eh}-[\nabla_Xe_1,e_2]_{\Eh}-[e_1,\nabla_Xe_2]_{\Eh}-\nabla_{\overline{\nabla}^{\Eh}_{e_2}X}e_1+\nabla_{\overline{\nabla}^{\Eh}_{e_1}X}e_2 , e_3\rangle \,\,\nn\\[4pt] 
&+\frac 12 \left(\langle\nabla_{\overline{\nabla}^{\Eh}_{e_3}X}e_1,e_2\rangle-\langle\nabla_{\overline{\nabla}^{\Eh}_{e_3}X}e_2,e_1\rangle\right)\,,
\label{basiccurvaturecourant}
\end{tcolorbox} 
where $\overline{\nabla}^{\Eh}$ is the induced $\widehat{E}$-connection on $TM$ given in Eq.\eqref{EhonTM}. 
\end{defn}
It is a simple and straightforward exercise to show that this basic curvature is linear in all arguments. Moreover, it is manifestly antisymmetric in its first two arguments. On the other hand, this last property ceases to be true in the corresponding definition in terms of the Dorfman bracket. This is given as 
\bea 
\mc S^{\Eh_{\circ}}(e_1,e_2,e_3)X&=&\langle \nabla_{X}(e_1\circ e_2)-\nabla_Xe_1\circ e_2-e_1\circ \nabla_Xe_2-\nabla_{\overline{\nabla}^{\Eh}_{e_2}X}e_1+\nabla_{\overline{\nabla}^{\Eh}_{e_1}X}e_2 , e_3\rangle \,\,\nn\\[4pt] 
&+& \langle\nabla_{\overline{\nabla}^{\Eh}_{e_3}X}e_1,e_2\rangle\,,
\label{basiccurvaturecourantdorfman}
\eea 
where once more the correction term in the second line has been de-antisymmetrized.

We can evaluate the basic curvature tensor in the two examples $\widehat{E}_{A}$ and $\widehat{E}_{B}$. Starting with the standard Courant algebroid where the anchor map is the projection to the tangent bundle, we obtain
\bea 
 \mc S^{\Eh_A}(e_1,e_2,e_3)X_4 &=&-\, \text{Cycl}_{123}\bigg(\langle R(X_1,X_2)X_4-\nabla_{X_1}T(X_2,X_4)+\nabla_{X_2}T(X_1,X_4),\eta_3\rangle\,\,\nn\\[4pt] 
&&\qquad \qquad\quad +\,\langle\text{Cycl}_{124}(T(T(X_1,X_2),X_4)), \eta_3\rangle\bigg)\,,
\eea 
observing that it results in the basic curvature of the tangent Lie algebroid, as found in \eqref{STM} and to the ordinary  curvature when the base manifold $M$ has no torsion.
In components, it reads 
\bea 
 \mc S^{\Eh_A}(e_1,e_2,e_3)X_4 =S^\sigma_{\rho\mu\nu}X^\mu_{[1}X^\nu_2\eta_{3]\sigma}X_4^\rho\,,\label{ShatEA}
 \eea 
 where the components $S^\sigma_{\rho\mu\nu}$ are defined in \eqref{STM}.
 In the second example, we obtain
 \bea 
 \mc S^{\Eh_B}(e_1,e_2,e_3)X_4 =\bigg(\frac 32\big(\nabla_\sigma\mathring{\nabla}_\rho\Pi^{\mu\nu}+2\Pi^{\k[\mu}R^{\nu]}_{\rho\s\kappa}\big) -2\,\nabla_{\s}\big(\Pi^{\k[\mu}T^{\nu]}_{\rho\kappa}\big)\bigg)\eta_{[1\mu}\eta_{2\nu}X^\rho_{3]}X^\sigma_4\,,\label{ShatEB}
 \eea 
 which is to be compared with Eq. \eqref{ST*M} to observe the relation to the basic curvature of the cotangent Lie algebroid in the torsionless case. As for the Gualtieri torsion, the Courant algebroid basic curvature tensor for the ``$A+B$'' example is the sum of the ones given in \eqref{ShatEA} and \eqref{ShatEB}, in other words it has as components the basic curvature tensors of the Lie algebroids over the tangent and cotangent bundles. This provides the answer to one of the questions we posed in the introduction; the basic curvature of a Courant algebroid connection can encode simultaneously the ordinary curvature of an affine connection on the base manifold and the basic curvature of the dual Lie algebroid on the cotangent bundle. 
 
Returning to the general discussion, it is possible to express the generalized basic curvature in terms of the Gualtieri torsion in a relation analogous to \eqref{screl} for Lie algebroids:
\bea  
\mc S^{\Eh}(e_1,e_2,e_3)X&=&-\nabla_{X}\mc T^{\Eh}(e_1,e_2,e_3)+\langle \mc C(e_2,X)e_1-\mc C(e_1,X)e_2,e_3\rangle \nn\\[4pt]
&& -\, \frac 12 \big(\langle\mc C(e_3,X)e_1,e_2\rangle-\langle \mc C(e_3,X)e_2,e_1\rangle\big)\,,
\eea  
in terms of the same $\mc C$ tensor \eqref{ctensor} as the one defined for Lie algebroids, now extended to be evaluated on Courant algebroid sections. For the nonskew case,
\be
\mc S^{\Eh_{\circ}}(e_1,e_2,e_3)X=-\nabla_{X}\mc T^{\Eh}(e_1,e_2,e_3)+\langle \mc C(e_2,X)e_1-\mc C(e_1,X)e_2,e_3\rangle  -\langle\mc C(e_3,X)e_1,e_2\rangle\,.
\ee
We are now in position to choose the vector field $X$ to be an anchored section of $\widehat{E}$, say $\rho(e_4)$. Then we find 
\bea 
\mc S^{\Eh}(e_1,e_2,e_3)\rho(e_4)&=&-\mathbullet{\nabla}^{\Eh}_{e_4}\mathbullet{\mc T}^{\Eh}(e_1,e_2,e_3)+\langle R(\rho(e_2),\rho(e_4))e_1-R(\rho(e_1),\rho(e_4))e_2,e_3\rangle \nn\\[4pt]
&& -\, \frac 12 \bigg(\langle R(\rho(e_3),\rho(e_4))e_1,e_2\rangle-\langle R(\rho(e_3),\rho(e_4))e_2,e_1\rangle\bigg) \nn\\[4pt] && + \, \langle\mathbullet{\nabla}^{\Eh}_{e_4}e_3,\phi(e_2,e_1)-\phi(e_1,e_2)\rangle \nn\\[4pt] && -\, \frac 12 \langle \mathbullet{\nabla}^{\Eh}_{e_4}e_2,\phi(e_3,e_1)\rangle+\frac 12 \langle\mathbullet{\nabla}^{\Eh}_{e_4}e_1,\phi(e_3,e_2)\rangle\,.  \label{SERCourant}
\eea 
The last two lines vanish when the canonical $\widehat{E}$-connection is considered. 

Even though what we have defined so far is enough for our purposes and for explaining the relation to field theory, there is a remaining challenge. This is to determine the relation of the Courant algebroid basic curvature to \emph{some} suitable notion of Courant algebroid  $\widehat{E}$-curvature tensor. However, as for the torsion tensor, the naive curvature defined through \eqref{rlie} is in general not a tensor. It is worth mentioning, however, that for the canonical $\widehat{E}$-connection it becomes a tensor; in other words, the naive  $\widehat{E}$-curvature $\mathbullet{R}^{\Eh}$ \emph{is} a tensor, unlike the naive torsion tensor $\mathbullet{T}^{\Eh}$. The question is whether there exists such a curvature tensor for an arbitrary $\widehat{E}$-connection.{\footnote{Other proposals have appeared before in \cite{Hohm:2012mf,Aschieri:2019qku,Jurco:2016emw}.}} Since this is not central for our purposes in this paper, we refer the reader to Appendix \ref{appb} for a proposal.

\section{Tensors in the dg manifold picture}
\label{sec4}

\subsection{Lie algebroids as dg manifolds and the Atiyah cocycle} \label{sec41}

Consider a smooth (nongraded) manifold $M$, a vector bundle $E$ over $M$ and shift its fiber degree by 1, denoted as $E[1]$. Locally, $E[1]$ has coordinates $x^{\m}$ and $a^{a}$ of degrees 0 and 1 respectively. Furthermore, considering that $E$ has the structure of a Lie algebroid is equivalent to equipping $E[1]$ with a degree-1 vector field 
\be 
Q=\rho^{\m}_{a}(x)a^{a} \frac{\partial}{\partial x^{\m}}-\frac 12 C^{a}_{bc}(x)a^ba^c\frac{\partial}{\partial a^a}\,,
\label{QLA}
\ee 
which is homological, meaning that 
\be  
\frac 12 [Q,Q] = Q^2 = 0\,.
\ee  
This correspondence was first discussed in \cite{Vaintrob} and it allows us to think of Lie algebroids in the context of dg manifolds. This has the advantage that one works with functions instead of sections at the cost of having to be cautious of signs. 

That $Q$ is a vector field of degree 1 on $\mc M=E[1]$ means that it is a degree 1 section of the graded tangent bundle $T\mc M$. Vector fields can be locally expanded in the basis $(\frac{\partial}{\partial x^{\m}},\frac{\partial}{\partial a^a})$ with degrees $0,-1$, as for example in Eq. \eqref{QLA}.  If we now consider the Lie derivative $\mc L_{Q}$,
 then $TE[1]$ turns out to be a dg vector bundle and a homological vector field on it is given by the tangent (or, more generally, the complete) lift $\mc Q$ of $Q$, see e.g. \cite{lift1,lift2}.{\footnote{One could as well work on $T[n]E[1]$ instead of $TE[1]$, without any changes in the logic. The definition of the complete lift and its relation to the Lie derivative along $Q$ depend on sign conventions though. In particular, we can define the Lie derivative along a vector field $X$ of degree $|X|$ as $\mc L_{X}=[\iota_{X},\dd]$ where the commutator is graded. Suppose we denote the form degree of any object as $(\cdot)$ and the super-degree (``ghost'') as $|\cdot|$. Then $\iota_{X}$ has bidegree $((\iota_X),|\iota_X|)=(-1,|X|)$ and total degree $|X|-1$. This means that ${\cal L}_{X}=\iota_X\dd - (-1)^{s}\dd \iota_X$, where $s$ depends on the sign convention as follows. For the Deligne convention $s=(\iota_X)(\dd)+|\iota_X||\dd|$ whereas for the Bernstein convention $s=((\iota_X)+|\iota_X|)((\dd)+|\dd|))$; see https://ncatlab.org/nlab/show/signs+in+supergeometry for a discussion on conventions. For the case of $X=Q$ with $|Q|=1$ and regardless of whether one works on $TE[1]$ or $T[n]E[1]$, this results in the complete lift $\mc Q=-(-1)^{s}\eta{\cal L}_{Q}$, where $s$ is 1 for the Deligne convention and 0 for the Bernstein convention, for the definition $\mc Q(\dd f)=\eta \,\dd (Qf)$, where $\eta=\pm 1$. In the text above we chose $\eta=-1$ and $s=0$.}} This is because 
\bea  
\mc Q(f)&=& Q(f)\,, \\[4pt] 
\mc Q(\dd f)&=& -\,\dd(Q(f))\,,
\eea  
for any function $f\in C^{\infty}(\mc M)$. The role of $\mc Q$ is played by $\mc L_{Q}$ here and hence with a little abuse of notation we will identify them, $\mc Q=\mc L_{Q}$.{\footnote{ $\mc Q$ and $\mc L_{Q}$ are different to start with. We thank T. Strobl for pointing this out.}}
As a first relation between the $E$-tensors defined in Section \ref{sec22} and the dg manifold data, notice that for a $TM$-connection $\nabla$ on $E$, which induces the canonical $E$-connection, the homological vector field can be rewritten as \cite{Chatzistavrakidis:2021nom}   
\be 
\label{QLAcovtriv} Q=a^{a} \rho_{a}{}^{\m}\, \DD^{(0)}_{\m}+\frac 12 \, \mathbullet{T}^{a}_{bc}\, a^ba^c\,  \DD^{(-1)}_a\,,
\ee 
with 
\be 
\DD^{(0)}_{\m} =\frac{\partial}{\partial x^{\m}} -\omega_{\m b}^{c}a^{b}\frac{\partial}{\partial a^c}\,, \qquad \DD^{(-1)}_a=\frac\partial{\partial a^a}\,.
\ee 
This shows that $Q$ is directly related to the $E$-torsion of the canonical $E$-connection on $E$. However, no sign of the basic curvature appears here; crucially this will drastically change when we study Courant algebroids in the dg manifold picture, where the Courant algebroid basic curvature will be part of the homological vector field.  
  
To make the connection between differential geometric notions on a Lie algebroid and functions and vector fields on the dg manifold  precise, let us introduce the contraction map 
\bea  \label{iotadef}
\iota: C^{\infty}(M,E) \to \mf X_{-1}(E[1])\,, \nn \\[5pt]
e\mapsto \iota(e)=e^a(x)\frac{\partial}{\partial a^{a}}\,,
    \eea  
    which takes a section of $E$ to a degree $-1$ vector field on $E[1]$---the space of degree $-1$ vector fields is denoted as $\mf X_{-1}$.  It satisfies 
    \bea  
\iota(e+e')=\iota(e)+\iota(e')\,,  \qquad
    \iota(fe)=f\iota(e)\,.
    \eea  
    This map can be used to establish the correspondence between Lie algebroids and dg manifolds. Specifically, the anchor and bracket of the Lie algebroid are obtained as follows, 
\bea  
\label{iotarho} \rho(e)f &=& [[Q,\iota(e)],f]\,, \\[4pt] 
\label{iotalie} \iota([e,e']_{\E})&=& [[Q,\iota(e)],\iota(e')]\,,
\eea  
where the bracket on the right-hand side is the obvious one for vector fields, taking into account the grading.  Recalling that in this case 
\be 
[Q,\iota(e)]=\mc L_{Q}\iota(e)=\mc Q\iota(e)\,, 
\ee 
these may be written in the evident alternative form
\bea  
\rho(e)f &=& [\mc Q\, \iota(e),f]=\mc Q\iota(e)f\,, \\[4pt] 
\iota([e,e']_{\E})&=& [\mc Q\, \iota(e),\iota(e')]\,;
\eea  
we shall stick to this kind of expressions below to avoid clutter with too many brackets, unless there is an additional point to be made. These equations reflect the philosophy of the derived bracket construction; the anchor and Lie bracket on the Lie algebroid are obtained from the graded Lie bracket of vector fields through the homological vector field.

On the tangent bundle $TE[1]$ ordinary yet graded connections can be defined in the usual way.{\footnote{One may define the notion of a graded Lie algebroid too and subsequently ``${E[1]}$-connections on ${V[1]}$'', however we shall not delve into details of such constructions that are not of particular use in the context of this paper. The interested reader may consult for example \cite{AtiyahDG}.}} Let us denote such a connection as $\grave\nabla$ and arbitrary vector fields on $E[1]$ as $\mathcal X, \mathcal Y$ etc.  Then we are looking at objects such as $\grave\nabla_{\mc X}{\mc Y}$. The connection $\grave{\nabla}$ itself does not have any grading, $|\grave{\nabla}|=0$, whereas the vector fields are of degree $|\mc X|$ and $|\mc Y|$. This means that its properties read  
\bea  
\grave\nabla_{f\mc X}\mc Y &=& f\grave\nabla_{\mc X}{\mc Y}\,, \\[4pt] 
\grave\nabla_{\mc X}(f{\mc Y}) &=& {\mc X}(f) \, \mc Y + (-1)^{|\mc X|\,|f|} f\grave\nabla_{\mc X}\mc Y\,,
\eea  
for any function $f\in C^{\infty}(E[1])$ of degree $|f|$. 
We can directly define the (ordinary yet graded) torsion of the connection $\grave\nabla$, 
\be 
\grave{T}(\mc X,\mc Y)=\grave\nabla_{\mc X}\mc Y-(-1)^{|\mc X||\mc Y|}\grave\nabla_{\mc Y}\mc X-[\mc X,\mc Y]\,,\ee
taking care of the correct grading.
Similarly we can also define its  curvature  
\be 
\grave{R}(\mc X,\mc Y)\mc Z=[\grave\nabla_{\mc X},\grave\nabla_{\mc Y}]\mc Z-\grave\nabla_{[\mc X,\mc Y]}\mc Z\,,
\ee 
where the commutator and Lie bracket are both graded. It is useful at this stage to denote the coordinates of $E[1]$ collectively as $x^{\a}=(x^{\m},a^{a})$. Then with respect to a basis $\mf e_{\a}$ of the tangent bundle $TE[1]$, the connection coefficients are 
\begin{equation}
    \grave\nabla_{\mf e_\a}\mf e_{\b}=\grave\omega_{\a\b}^{\g}\mf e_{\g}\,. \label{connectioncollective}
\end{equation}
Using the notation $|\cdot|$ for the degree, these coefficients satisfy 
\be |\grave\omega_{\a\b}^{\g}|=|\mf e_\a|+|\mf e_\b|-|\mf e_\g|\,. \ee
 One can then determine the degree of the various components, which turns out to be 
 \bea  
|\grave\omega_{\m\n}^{\rho}|=0\,, \quad |\grave\omega_{\m\n}^{a}|=1\,, \quad |\grave\omega_{\m a}^{\n}|=-1=|\grave\omega_{a\m}^{\n}|\,,\nn\\[4pt] |\grave\omega_{\m a}^b|=0=|\grave\omega_{a\m}^{b}|\,,   \quad |\grave\omega_{ab}^{\m}|=-2\,,\quad |\grave\omega_{ab}^{c}|=-1\,.
 \eea  
 We observe that since there are no negative degree coordinates, several of these coefficients must vanish. In this local picture, the components of the graded torsion and curvature tensors are given as 
 \bea   
\grave{T}^{\a}_{\b\g}&=&\grave\omega^{\a}_{\b\g}-(-1)^{|\partial_{\beta}|\,|\partial_{\g}|}\grave\omega^{\a}_{\g\b}\,, \\[4pt] 
\grave{R}^{\a}{}_{\b\g\d}&=& \partial_{\g}\grave\omega_{\d\b}^{\a}-(-1)^{|\partial_{\g}|\,|\partial_{\d}|}\partial_{\d}\grave\omega^{\a}_{\g\b} \nn\\[4pt] && +\, (-1)^{|\grave\omega^{\e}_{\d\b}|(|\partial_{\a}|+|\partial_{\e}|)}\grave\omega^{\a}_{\g\e}\grave{\omega}^{\e}_{\d\b}-(-1)^{|\partial_{\g}|\,|\partial_{\d}|}(-1)^{|\grave\omega^{\e}_{\g\b}|(|\partial_{\a}|+|\partial_{\e}|)}\grave\omega^{\a}_{\d\e}\grave{\omega}^{\e}_{\g\b}\,.
 \eea   
Equipped with this connection and the homological vector field $\mc Q$, one may now ask whether these two structures on the tangent bundle $T\mc M=TE[1]$ are compatible. This is measured by the Atiyah 1-cocycle that gives rise to the Atiyah class of the dg vector bundle $T\mc M\rightarrow \mc M$ relative to the dg Lie algebroid $T\mc M\rightarrow \mc M$, called the Atiyah class of the dg manifold $(\mc M,Q)$ in \cite{AtiyahDG98,AtiyahDG}. It is given as the map{\footnote{As for Lie algebroid connections, it is possible to define this map more generally, for a different vector bundle $V[1]$ such that it maps from $TE[1]\otimes TV[1]$ to $TV[1]$. Furthermore, one could think of it as a commutator by using the connection directly, that is by writing $\text{At}(\cdot,\mc Y)=[\mc Q,\grave\nabla]\mc Y$.}}
\bea   
\text{At}: T\mc M\otimes T\mc M &\to& T\mc M \nn\\[4pt]  (\mc X,\mc Y) &\mapsto & \text{At}(\mc X,\mc Y)=\mc Q\grave\nabla_{\mc X}\mc Y-\grave\nabla_{\mc Q\mc X}\mc Y-(-1)^{|\mc X|}\grave\nabla_{\mc X}\mc Q\mc Y\,.
\eea 
This is a 1-cocycle because using its definition we find
\bea  
\mc Q(\text{At})(\mc X,\mc Y)&=&\mc Q(\text{At}(\mc X,\mc Y))+\text{At}(\mc Q\mc X,\mc Y)+(-1)^{|\mc X|}\text{At}(\mc X,\mc Q\mc Y) \nn\\[4pt] 
&=&\mc Q\big(\mc Q\grave\nabla_{\mc X}\mc Y-\grave\nabla_{\mc Q\mc X}\mc Y-(-1)^{|\mc X|}\grave\nabla_{\mc X}\mc Q\mc Y\big) \,+ \nn\\[4pt] 
&& +\, \mc Q\grave\nabla_{\mc Q\mc X}\mc Y-\grave\nabla_{\mc Q^{2}\mc X}\mc Y-(-1)^{|\mc X|+1}\grave\nabla_{\mc Q\mc X}\mc Q\mc Y \, + \nn\\[4pt]    
&& +\, (-1)^{|\mc X|}\mc Q\grave\nabla_{\mc X}\mc Q\mc Y-(-1)^{|\mc X|}\grave\nabla_{\mc Q\mc X}\mc Q\mc Y-\grave\nabla_{\mc X}\mc Q^2\mc Y=0\,, 
\eea  
for every $\mc X, \mc Y$, using just $\mc Q^{2}=0$. Thus in general 
\be  \mc Q(\text{At})=0\,. \ee
The cohomology class of the Atiyah 1-cocycle does not depend on the choice of connection and it is called the Atiyah class.
The Atiyah class was also studied for Lie and Manin pairs in Refs. \cite{batak1,batak2}. Using the properties of the connection and of the graded Lie bracket, it is simple to show that for any $f\in C^{\infty}(\mc M)$  
\be  
\text{At}(f\mc X,\mc Y)=(-1)^{|f|}f\,\text{At}(\mc X,\mc Y)\,,\quad \text{At}(\mc X,f\mc Y)=(-1)^{|f|(|\mc X|+1)}f\text{At}(\mc X,\mc Y)\,.\label{atiyahlinear}
\ee 
This reflects the fact that by its definition the map ${\text{At}}$ has degree 1.  If the torsion of $\grave\nabla$ vanishes, then the Atiyah 1-cocycle is graded symmetric. Specifically, using the definitions above it is simple to show that 
\be 
\text{At}(\mc X,\mc Y)-(-1)^{|\mc X||\mc Y|}\text{At}(\mc Y,\mc X)=\mc Q\,\grave{T}(\mc X,\mc Y)\,.\label{atiyahantisymm}
\ee 
Specializing this expression to the image of the map $\iota$, we obtain 
\be \label{skw}\text{At}(\iota(e),\iota(e'))+\text{At}(\iota(e'),\iota(e))=\grave{T}(\iota(e),\mc Q\iota(e'))-\grave{T}(\mc Q\iota(e),\iota(e'))\,,\ee
for the specific degree $-1$ vector fields obtained  from sections of $E$. This holds in the Lie algebroid case because the quantity $\grave\nabla_{\iota(e)}\iota(e')$ vanishes. This is easily seen by the fact that it would require connection coefficients of degrees $-1$ or $-2$, however no negative degree coordinates are present. 

It is useful to express the Atiyah 1-cocycle in local coordinate form. By straightforward calculations we find 
 \begin{tcolorbox}[ams nodisplayskip, ams align]  
\text{At}(\mc X,\mc Y)= (-1)^{s_1}\mc X^{\a}\mc Y^{\b}\bigg(\grave\nabla_{\a}\grave\nabla_{\b}Q^{\g}+(-1)^{s_2}Q^{\d}\grave{R}^{\g}{}_{\b\d\a}+(-1)^{s_3}\grave\nabla_{\a}(Q^{\d}\grave{T}^{\g}_{\d\b})\bigg)\partial_{\g}\,.\,\,\,
\label{Atiyahcomponents}
\end{tcolorbox}
where the signs are given as 
\bea  
s_1=|\mc X|+|\mc Y|+|\mc Y^{\b}||\partial_{\a}|\,,\quad s_2=|\partial_\a|+|\partial_\b|\,,\quad s_3=|\partial_\b|\,.
\eea  
We observe that the Atiyah cocycle encodes the torsion, curvature and the second covariant derivatives of the homological vector field of the dg manifold $E[1]$. One can also see that in a local patch where the connection coefficients are set to zero, the Atiyah cocycle becomes \cite{AtiyahDG}
\be  
\text{At}(\partial_{\a},\partial_{\b})^{\g}= (-1)^{|\partial_{\a}|+|\partial_\b|}\partial_{\a}\partial_{\b}Q^{\g}\,.
\ee  
Thus locally it is identical to the second derivatives of the homological vector field. What about the first derivatives? Evidently, this has already been considered through the tangent lift. Indeed, locally we can write 
\be \label{local tangentlift}  
\mc Q(\partial_{\a})^{\b} =-(-1)^{|\partial_{\a}| } \partial_{\a}Q^{\b}\,,
\ee 
which is just the specification of ${\cal L}_{Q}=\mc Q$ in the local setting. Therefore, the tangent lift encodes first derivatives of $Q$ and the Atiyah cocycle its second derivatives. What happens with higher derivatives of Q? This question is very relevant in the field-theoretical context, since it is related to the closure of gauge transformations. To answer it, we calculate the graded Jacobiator of the map $\text{At}$. We start from applying the definition of the Atiyah cocycle to the first $\text{At}$ in $\text{At}(\mc Z,\text{At}(\mc X,\mc Y))$,   
\be
\text{At}(\mc Z,\text{At}(\mc X,\mc Y))=\mc Q\grave\nabla_{\mc Z}(\text{At}(\mc X,\mc Y))-\grave\nabla_{\mc Q\mc Z}(\text{At}(\mc X,\mc Y))-(-1)^{|\mc Z|}\grave\nabla_{\mc Z}\mc Q(\text{At}(\mc X,\mc Y))\,. 
\ee 
Next we expand the action of the covariant derivative on $\text{At}(\mc X,\mc Y)$ of the first two terms and also use the  property $\mc Q(\text{At})=0$ in the third term; this leads to 
\begin{align} 
\text{At}(\mc Z,\text{At}(\mc X,\mc Y))&=\mc Q\grave\nabla_{\mc Z}\text{At}(\mc X,\mc Y)+(-1)^{|\mc Z|}\mc Q(\text{At}(\grave\nabla_{\mc Z}\mc X,\mc Y))\nn\\[4pt]
& \quad +(-1)^{|\mc Z|(1+|\mc X|)}\mc Q(\text{At}(\mc X,\grave\nabla_{\mc Z}\mc Y))-\grave\nabla_{\mc Q\mc Z}\text{At}(\mc X,\mc Y) \nn\\[4pt] 
&\quad -(-1)^{|\mc Z|+1}\text{At}(\grave\nabla_{\mc Q\mc Z}\mc X,\mc Y) -(-1)^{(1+|\mc Z|)(1+|\mc X|)}\text{At}(\mc X,\grave\nabla_{\mc Q\mc Z}\mc Y)
\nn\\[4pt]
&\quad -(-1)^{|\mc Z|+1}\grave\nabla_{\mc Z}(\text{At}(\mc Q\mc X,\mc Y)+(-1)^{|\mc X|}\text{At}(\mc X,\mc Q\mc Y))\,.
\end{align}
Next we expand the action of $\mc Q$ (second and third terms above) and of the covariant derivative (terms in the fourth line above) and obtain the following collection of terms in order of appearance,
\begin{align}
\text{At}(\mc Z,\text{At}(\mc X,\mc Y))&=\mc Q\grave\nabla_{\mc Z}\text{At}(\mc X,\mc Y)+(-1)^{|\mc Z|+1}\text{At}(\mc Q\grave\nabla_{\mc Z}\mc X,\mc Y)+(-1)^{|\mc X|+1}\text{At}(\grave\nabla_{\mc Z}\mc X,\mc Q\mc Y)\nn\\[4pt]
&\quad +(-1)^{|\mc Z|(1+|\mc X|)+1}\text{At}(\mc Q\mc X,\grave\nabla_{\mc Z}\mc Y)+(-1)^{(1+|\mc Z|)(1+|\mc X|)}\text{At}(\mc X,\mc Q\grave\nabla_{\mc Z}\mc Y)\nn\\[4pt]
&\quad -\grave\nabla_{\mc Q\mc Z}\text{At}(\mc X,\mc Y)-(-1)^{|\mc Z|+1}\text{At}(\grave\nabla_{\mc Q\mc Z}\mc X,\mc Y)\nn\\[4pt]
&\quad -(-1)^{(1+|\mc Z|)(1+|\mc X|)}\text{At}(\mc X,\grave\nabla_{\mc Q\mc Z}\mc Y)-(-1)^{|\mc Z|+1}\grave\nabla_{\mc Z}\text{At}(\mc Q\mc X,\mc Y) \nn\\[4pt] &\quad +\text{At}(\grave\nabla_{\mc Z}\mc Q\mc X,\mc Y)-(-1)^{|\mc Z|(|\mc X|+1)+1}\text{At}(\mc Q\mc X,\grave\nabla_{\mc Z}\mc Y)\nn\\[4pt]
&\quad -(-1)^{|\mc X|+|\mc Z|+1}\grave\nabla_{\mc Z}\text{At}(\mc X,\mc Q\mc Y)-(-1)^{|\mc X|+1}\text{At}(\grave\nabla_{\mc Z}\mc X,\mc Q\mc Y) \nn\\[4pt] &\quad -(-1)^{(|\mc Z|+1)|\mc X|+1}\text{At}(\mc X,\grave\nabla_{\mc Z}\mc Q\mc Y)\,.
\end{align}
Out of the 14 terms on the right hand side, 4 cancel in pairs (the 3rd with the 13th and the 4th with the 11th), 6 are used in two pairs of 3 to form the other two terms in the Jacobiator of the Atiyah cocycle through its definition (the 2nd, 7th and 10th terms and the 5th, 8th and 14th terms) and 4 terms remain. The result is:
 \begin{tcolorbox}[ams nodisplayskip, ams align]
&\text{At}(\mc Z,\text{At}(\mc X,\mc Y))-(-1)^{|\mc Z|+1}\text{At}(\text{At}(\mc Z,\mc X),\mc Y)-(-1)^{(|\mc Z|+1)(|\mc X|+1)}\text{At}(\mc X,\text{At}(\mc Z,\mc Y))=\nn\\[4pt]
&=\mc Q\grave\nabla_{\mc Z}\text{At}(\mc X,\mc Y)-\grave\nabla_{\mc Q\mc Z}\text{At}(\mc X,\mc Y)\,- 
\nn\\[4pt] 
& \quad -(-1)^{|\mc Z|+1}\grave\nabla_{\mc Z}\text{At}(\mc Q\mc X,\mc Y) 
-(-1)^{|\mc Z|+|\mc X|+1}\grave\nabla_{\mc Z}\text{At}(\mc X,\mc Q\mc Y)\,.
\end{tcolorbox}
This was already reported in \cite{AtiyahDG} without the details of the calculation. 
Denoting the left hand side as $\text{Jac}_{\text{At}}$ and realizing that the right hand side is the Lie derivative along $Q$ of the covariant derivative on the Atiyah cocycle, we can write 
\be  
\text{Jac}_{\text{At}}= {\cal L}_{Q}(\grave\nabla\text{At})\,.
\ee 
It is observed that when $\grave\nabla\text{At}$ is $\mc Q$-closed then the Atiyah cocycle satisfies the graded Jacobi identity and therefore it gives rise to a graded Lie bracket. This, however, does not hold in general. The underlying structure is instead an L$_{\infty}[1]$ algebra \cite{AtiyahDG} similar to what happens in the nongraded case \cite{Kapranov}. 
Specifically, the authors of \cite{AtiyahDG, Seol:2021tol} show (Theorem 4.1. in the first reference) that for a dg manifold $(\mc M,Q)$ and a torsion-free affine connection on it, the space of vector fields on ${\cal M}$ admits an L$_{\infty}[1]$-algebra structure   with unary and binary brackets given by the tangent lift and the (opposite of the) Atiyah cocycle, 
\bea  
b_1&=&{\cal L}_{Q}\,, \\[4pt]  
b_2&=&-\,\text{At}\,.
\eea 
These are called Kapranov L$_{\infty}[1]$ algebras. Note that we work in the description of L$_{\infty}$ algebras in terms of brackets of equal degree (called b-picture in \cite{Hohm:2017pnh}). Indeed both of them have degree 1. A closed form for all higher brackets is reported in \cite{Seol:2021tol}. They depend on the Atiyah cocycle and the curvature $\grave{R}$ along with their higher derivatives. (We emphasize once again that the torsion is zero to ensure that the Atiyah cocycle is graded symmetric.) In case $\grave{R}=0$ too, then the third bracket is given by the (opposite of the) covariant derivative on the Atiyah cocycle, 
\be  b_3=-\grave\nabla\text{At}\,, \ee and recursively all higher brackets are the covariant derivative on the previous one. We shall explain the relevance of this Kapranov L$_{\infty}[1]$ algebra for field theory in Section \ref{sec6}. 

\subsection{E-tensors from E[1]-tensors}
\label{sec42}

Our next goal is to extend the logic of expressing the anchor and the Lie bracket of the Lie algebroid in terms of dg manifold data, Eqs. \eqref{iotarho} and \eqref{iotalie}, to $TM$- and $E$-connections on $E$. The idea is to start with a $T\mc M$-connection on $T\mc M$, say $\grave{\nabla}$, and examine how are $
\nabla$ and $\nabla^{\E}$ obtained. We emphasize that we do not suggest this as a minimal approach; an $E$-connection on $E$ could minimally be described on the dg manifold $E\oplus E[1]$. Considering $TE[1]$ is not a minimal choice but on the other hand it will allow us to encode a $TM$-connection on $E$ as well. 
We recall that $\grave\nabla$ does not carry any degree by itself; in other words, the degree of $\grave\nabla_{\mc X}{\mc Y}$ is equal to the sum of degrees of $\mc X$ and $\mc Y$. 

We begin with the connection $\grave{\nabla}$ on $\mc M$, which acts on vector fields of degree 0 and $-1$. As already mentioned, for degree reasons 
\be 
\grave{\nabla}_{\iota(e)}\iota(e')=0\,. \label{iota on iota lie}
\ee 
Second, for an arbitrary degree 0 vector field on $\mc M$, say $\mc X_0$, $\grave{\nabla}_{\mc X_0}\iota(e)$ is of degree $-1$ and it lies in the image of the (surjective) map $\iota$---note that the degree 0 vector fields on $\mc M$ are not the same thing as the ones on $M$. Then we have 
\be \label{ordinaryconfromdg}
\grave{\nabla}_{\mc X_0}\iota(e)=\iota(\nabla_{X}e)\,,
\ee 
where $\nabla$ is a $TM$-connection on $E$ and $X$ is the vector field on $M$ that is, roughly speaking, the projection of the vector field $\mc X_0$ to to its pure degree 0 part $\partial/\partial x^{\m}$.{\footnote{More precisely, given the Lie algebroid Atiyah sequence $$ 0\to \text{End}(E^\ast)\to \text{At}(E^\ast)\xrightarrow{\s} TM\to 0$$ we have $\s(\mc X_0)=X$, for $\mc X_0\in \mathfrak{X}_0(\mc M)\simeq \Gamma(\text{At}(E^\ast))$ and $X\in \mathfrak{X}(M)$, and $\s$ is the anchor of the Atiyah Lie algebroid. We thank D. Roytenberg for clarifying this point to us.}} Then if $\omega_{\m a}^{b}$ are the connection coefficients of $\nabla$, we have the relation 
\be \label{omega is omega}
\grave\omega_{\m a}^{b}=\omega_{\m a}^{b}\,.
\ee 
To fully characterize $\grave\nabla$, one should also find the action on degree 0 vector fields, but this is beyond the scope of the present work, since we already obtained the $TM$-connection. 

Next, we turn to $E$-connections. First we compute the following quantities,
\begin{eqnarray}
\grave\nabla_{\iota(e)}\mc Q\iota(e')&=& e^{a}\big(\rho_{a}{}^{\m}\partial_{\m}e'^{c}+(\rho_{b}{}^{\m}\grave{\omega}^c_{a\m}+C^c_{ab})e'^b\big)\frac{\partial}{\partial a^{c}}\,, \label{gravenablaonQiota}\\[4pt]
\grave\nabla_{\mc Q\iota(e)}\iota(e')&=&e^{a}\big(\rho_{a}{}^{\m}\partial_{\m}e'^{c}+\rho_{a}{}^{\m}\grave\omega^c_{\m b} e'^b\big)\frac{\partial}{\partial a^{c}}\,. \label{gravenablaoniota}
\end{eqnarray}
 Note that the coefficients appearing in \eqref{gravenablaonQiota} and \eqref{gravenablaoniota} are both of zero degree, therefore ordinary functions of $x^{\m}$. This is of course consistent with the degree counting of these equations, which give a degree $-1$ result that lies in the image of $\iota$.  
This way an $E$-connection is obtained and using definition \eqref{iotadef} we find
\be
\iota(\nabla^\E_e e')=e^{a}\big(\rho_{a}{}^{\m}\partial_{\m}e'^{c}+\omega^c_{a b} e'^b\big)\frac{\partial}{\partial a^{c}}\,.
\ee
To facilitate comparison to the two equations \eqref{gravenablaonQiota} and \eqref{gravenablaoniota} we combine them into
\be
\k \grave\nabla_{\mc Q\iota(e)}\iota(e')+ \l \grave\nabla_{\iota(e)}\mc Q\iota(e')=e^{a}\bigg((\k+\l)\rho_{a}{}^{\m}\partial_{\m}e'^{c}+(\l\rho_{b}{}^{\m}\grave{\omega}^c_{a\m}+\k\rho_{a}{}^{\m}\grave\omega^c_{\m b}+\l C^c_{ab})e'^b\bigg)\frac{\partial}{\partial a^{c}}\,. 
\ee
Evidently, at face value there is no unique way to determine an $E$-connection from the connection on $\mc M$; we find the one-parameter family of connections
 \begin{tcolorbox}[ams nodisplayskip, ams align]
\iota(\nabla^{\E}_e e')^{(\k)}=\k \grave\nabla_{\mc Q\iota(e)}\iota(e')+ (1-\k) \grave\nabla_{\iota(e)}\mc Q\iota(e')\,. \label{iotanablaEk}
\end{tcolorbox}
The connection coefficients of $\nabla^{\E}$ and $\grave\nabla$ are related according to 
\be {\o^c_{ab}=\k\rho_{a}{}^{\m}\grave\o_{\m b}^{c}+(1-\k)(\rho_{b}{}^{\m}\grave\o_{a\m}^{c}+C_{ab}^c)}\,. \label{omegagraveomegarelation}\ee
For every value of $\k$ the connection $\grave{\nabla}$ determines one $E$-connection on $E$ via $\mc Q$. Using \eqref{omega is omega}, we see that 
\be 
{\o^c_{ab}-\k\rho_{a}{}^{\m}\o_{\m b}^{c}=(1-\k)(\rho_{b}{}^{\m}\grave\o_{a\m}^{c}+C_{ab}^c)}\,. \label{omegagraveomegarelation2}
\ee
It becomes obvious that for $\k=1$, this results in the canonical  $E$-connection $\mathbullet{\nabla}^{\E}$, due to \eqref{omega is rho omega}. For most of this section, we shall not explore further the  cases of $\k\ne 1$---some aspects, not important for our purposes in this paper, are delegated to Appendix \ref{appd}---and we shall hence take $\k=1$, denoting $\iota(\cdot)^{(1)}$ simply as $\iota(\cdot)$. 
In that case, the connection coefficients are related as 
$
\omega_{ab}^{c}=\rho_{a}{}^{\m}\grave\omega_{\m b}^c\,.
$  

The next step is to express the various tensors on the Lie algebroid $E$ we discussed previously in terms of suitable quantities on the corresponding dg manifold. Starting with the $E$-torsion, we find 
    \bea 
    \iota(\mathbullet{T}^{\E}(e,e'))&=& \grave\nabla_{\mc Q\iota(e)}\iota(e')-\grave\nabla_{\mc Q\iota(e')}\iota(e) - [\mc Q\iota(e),\iota(e')]\,.
    \label{iotaTE}
    \eea 
There exists, however, another interesting way to express this:
\bea 
\iota(\mathbullet{T}^{\E}(e,e'))&=&\grave T(\mc Q\iota(e),\iota(e'))+\text{At}(\iota(e'),\iota(e))\,,
\eea
using the definitions of $\grave{T}$ and $\text{At}$, which leads to the following proposition: 
\begin{prop} \label{derivedtorsionLie}
The $E$-torsion of the $E$-connection $\mathbullet{\nabla}^{\E}$ canonically induced from a fixed $TM$-connection $\nabla$ on a Lie algebroid $E$ through its anchor, is obtained as
\bea 
\iota(\mathbullet T^{\E}(e,e'))=-{\rm At}(\iota(e),\iota(e'))\,,
\eea
where ${\rm At}$ is the Atiyah 1-cocycle of a torsion-free connection $\grave{\nabla}$ on the dg manifold $E[1]$.{\footnote{Such a connection exists for every dg manifold \cite{AtiyahDG}; it can be constructed as the averaged connection $\frac 12(\grave\nabla+\grave\nabla^{\text{op}}) $, where the opposite connection is given as $\grave\nabla^{\text{op}}_{\mc X}\mc Y=\grave\nabla_{\mc X}\mc Y-\grave{T}(\mc X,\mc Y)$.}}
\end{prop} 
\begin{rmk}
    One may extent this result to other $E$-connections and for all values of the parameter $\k$, see Appendix \ref{appd}.
\end{rmk}
This allows us to also express the Atiyah cocycle for vector fields in the image of $\iota$ differently, 
\be  
 \text{At}(\iota(e),\iota(e'))=\grave\nabla_{\iota(e)}\mc Q\iota(e') - \iota(\nabla^{\E}_{e}e')\,. \label{Atiyah2}
\ee  
 One could then think of the Atiyah 1-cocycle evaluated on vector fields in the image of the map $\iota$ as an obstruction to a strict correspondence between the connections $\nabla^{\E}$ on $E$ and $\grave\nabla$ on $E[1]$ in the sense of Eq. \eqref{Atiyah2}. We also see that vanishing $\grave{T}$ torsion does not necessarily imply that the $E$-torsion vanishes too. 

Let us proceed to discuss the basic curvature of $\nabla$ and its relation to the dg manifold data. 
Based on \eqref{ordinaryconfromdg}, we can write all the ingredients in its definition as:
\begin{align}
\iota(\nabla_{X}[e,e']_{\E})&=\grave\nabla_{\mc X_0}[\mc Q\iota(e),\iota(e')]\,, \\[4pt]
\iota([\nabla_{X}e,e']_{\E})&=[\mc Q\grave\nabla_{\mc X_0}\iota(e),\iota(e')]\,, \\[4pt]
\iota(\nabla_{\overline{\nabla}^{\E}_{e}X}e')&=\grave\nabla_{\mc Q\grave\nabla_{\mc X_0}\iota(e)}\iota(e')+\grave\nabla_{[\mc Q\iota(e),\mc X_0]}\iota(e')\,.
\end{align}
One may then directly write an expression for the basic curvature: 
\begin{align}\label{iS}
\iota(S^{\E}(e,e')X)&=\grave\nabla_{\mc X_0}[\mc Q\iota(e),\iota(e')]-[\mc Q\grave\nabla_{\mc X_0}\iota(e),\iota(e')]-[\iota(e),\mc Q\grave\nabla_{\mc X_0}\iota(e')] - \nn\\[4pt] 
& - \, \grave\nabla_{\mc Q\grave\nabla_{\mc X_0}\iota(e')}\iota(e)+\grave\nabla_{\mc Q\grave\nabla_{\mc X_0}\iota(e)}\iota(e')-\grave\nabla_{[\mc Q\iota(e'),\mc X_0]}\iota(e)+\grave\nabla_{[\mc Q\iota(e),\mc X_0]}\iota(e')\,.
\end{align}
Note that this expression is independent of the value of $\k$, which we chose to be 1 earlier, since it refers to the $TM$-connection.
The question now becomes what is the meaning of the right-hand side and what does it amount to on the dg manifold. To see this, recall the definition of the torsion $\grave{T}$ and of the curvature $\grave{R}$ and note also that the curvature $\grave{R}$ and the Atiyah cocycle satisfy,
\begin{align}
&\grave R(\iota(e),\mc X_0)\iota(e')=0\,,\nn\\[4pt]
&\grave R(\iota(e),\mc X_0)\mc Q\iota(e')+\grave R(\iota(e),\mc Q\mc X_0)\iota(e')-\grave R(\mc Q\iota(e),\mc X_0)\iota(e')=\nn\\[4pt]
&=\text{At}(\grave T(\iota(e),\mc X_0),\iota(e'))-\grave\nabla_{\iota(e)}\text{At}(\mc X_0,\iota(e'))-\grave\nabla_{\mc X_0}\text{At}(\iota(e),\iota(e'))\,.
\end{align}
The latter should not be confused with the additional graded algebraic Bianchi identity that $\grave{R}$ also satisfies. Combining the above definitions, we obtain after direct algebraic manipulations that 
 \begin{tcolorbox}[ams nodisplayskip, ams align]
\iota(S^{\E}(e,e')X)&=\grave R(\mc X_0,\mc Q\iota(e))\iota(e')-\grave R(\mc X_0,\mc Q\iota(e'))\iota(e)-\grave\nabla_{\mc X_0}\text{At}(\iota(e'),\iota(e)) \nn\\[4pt]
&-\grave\nabla_{\mc X_0}\grave T(\mc Q\iota(e),\iota(e'))
-\grave T(\grave\nabla_{\mc X_0}\mc Q\iota(e),\iota(e'))+\grave T(\grave\nabla_{\mc X_0} \iota(e),\mc Q\iota(e'))\,. \label{iotaSE}
\end{tcolorbox}
We observe now that when we choose an affine connection without torsion on the dg manifold, we find that 
\begin{prop} \label{derivedbasicLie}
The basic curvature of a connection $\nabla$ on a Lie algebroid $E$ is obtained from the ordinary curvature and the Atiyah cocycle of a torsion-free connection $\grave{\nabla}$ on the dg manifold $E[1]$ as
\be  
\iota(S^{\E}(e,e')X)=\grave R({\mc X_0},\mc Q\iota(e))\iota(e')-\grave R({\mc X_0},\mc Q\iota(e'))\iota(e)-\grave\nabla_{\mc X_0}\text{At}(\iota(e'),\iota(e)) \,.
\ee  
\end{prop} 
Finally, we examine the image of the $E$-curvature of a Lie algebroid $E$-connection under the map $\iota$. The result is $\k$-dependent and reads
\begin{align}
\iota(R^{\E}(e,e')e'')^{(\k)}&=\k\grave R(\mc Q\iota(e),\mc Q\iota(e'))\iota(e'')+(1-\k)\grave R(\mc Q\iota(e),\iota(e'))\mc Q\iota(e'')-\nn\\[4pt]
&-(1-\k)\grave\nabla_{\iota(e')}\text{At}(\iota(e),\mc Q\iota(e''))+\nn\\[4pt]
&+\k(1-\k)\Big(\text{At}(\iota(e'),\text{At}(\iota(e),\iota(e'')))-\text{At}(\iota(e),\text{At}(\iota(e'),\iota(e'')))\Big)\,. \label{iotaRE}
\end{align}
We observe that for our working value $\k=1$ the $E$-curvature tensor is obtained from the curvature tensor $\grave{R}$ according to 
    \begin{tcolorbox}[ams nodisplayskip, ams align]
\iota(\mathbullet{R}^{\E}(e,e')e'')=\grave R(\mc Q\iota(e),\mc Q\iota(e'))\iota(e'')\,. \label{iotaRE1}
\end{tcolorbox}  
Together with the formulas for the torsion, Eqs. \eqref{iotaSE} and \eqref{iotaRE} imply that the $E$-torsion,  $E$-curvature and basic curvature for $E$- and $TM$-connections on a Lie algebroid $E$ map on its description as a dg manifold $E[1]$ to the ordinary yet graded torsion and curvature together with the Atiyah cocycle.  Moreover we observe that in the torsion-free case $\grave T=0$,  the  $E$-curvature and the basic curvature are both associated to the curvature tensor $\grave{R}$ via the action of $\mc Q$, once with two appearances of $\mc Q$ and once with a single appearance.

\subsection{Courant algebroids as dg manifolds \& basic curvature} \label{sec43}

Moving on to Courant algebroids $\widehat{E}$, their description as dg manifolds was given in \cite{dimaphd}. On the graded side the relevant manifold is a symplectic submanifold of $T^{\ast}[2]\widehat{E}[1]$. For exact Courant algebroids $\widehat{E}=E\oplus E^{\ast}$, on which we focus from now on, the dg manifold is $\mc M=T^{\ast}[2]E[1]$. Being a cotangent bundle, it comes naturally with a symplectic structure, which was generically absent in the dg manifold description of Lie algebroids. This turns the manifold into a QP2 one, a symplectic dg manifold with a degree 2 symplectic structure $\Omega$  invariant under the flow generated by the homological vector field $Q$ in this case, namely ${\mc L}_{Q}\Omega=0$. 

For coordinates $(x^{\m},a^{a}, b_{\m})$  on $\mc M$ of degrees $(0,1,2)$ respectively, we have the general homological vector field 
\bea \label{QCA}
Q=\rho_{a}{}^{\m}a^{a}\frac{\partial}{\partial x^{\m}} - \big(\eta^{ab}\rho_b{}^{\m}b_{\m}+\frac 12 C^{a}_{bc}a^{b}a^{c}\big)\frac{\partial}{\partial a^{a}}-\big(\partial_{\m}\rho_{a}{}^{\n}b_{\n}a^{a}+\frac 1{3!}\partial_{\m}C_{abc} a^{a}a^{b}a^{c}\big)\frac{\partial}{\partial b_{\m}}\,.
\eea 
It is a section of the tangent bundle $T{\cal M}$  spanned in addition  by the  derivations $(\partial_\m:=\frac{\partial}{\partial x^{\m}},\partial_a:=\frac{\partial}{\partial a^{a}}, \partial^\m:=\frac{\partial}{\partial b_{\m}})$ of   degrees $(0,-1,-2)$ respectively. That the pair $(\mc M,Q)$ is a dg manifold is equivalent to the functions $\rho_{a}{}^{\m}$ and $C^{a}_{bc}$ being the components of the anchor map and bracket of $\widehat{E}$ and to $\eta^{ab}$ being the inverse of the nondegenerate fiber metric on it \cite{dimaphd}. 
It is possible to give a target space covariant formulation of $Q$ if one considers in addition a (noncanonical) $TM$-connection $\nabla$ and its canonical generalized connection $\mathbullet{\nabla}^{\Eh}$:
 \begin{tcolorbox}[ams nodisplayskip, ams align]   
&Q= a^{a} \rho_{a}{}^{\m} \DD^{(0)}_{\m} 
-\big(\eta^{ac}\rho_{a}{}^{\m}b_\m^\nabla-\mathbullet{\mc T}_{ab}{}^ca^aa^b\big)\DD^{(-1)}_{c}- \big( \nabla_{\n}\rho_{a}{}^{\m}b^\nabla_{\m}+\frac 13 {\mc S}_{\n abc}a^ba^c\big)a^{a}\DD^{\n}_{(-2)}\,. \,\,\,\,  \label{Qcourcov}
\end{tcolorbox}
Here the three derivations are defined as 
\bea  
\DD^{(0)}_{\m}&=& \frac{\partial}{\partial x^{\m}}-\frac 12\frac{\partial \omega_{\n bc}}{\partial x^{\m}} a^ba^c\frac{\partial}{\partial b_{\n}}+\Gamma_{\n\m}^\s b_\s^\nabla \frac{\partial}{\partial b_{\n}}-\o_{\m b}{}^c a^b\DD_c^{(-1)} \,, \\[4pt] 
\DD_a^{(-1)}&=&\frac{\partial}{\partial a^{a}}+\o_{\m ba}a^b\frac{\partial}{\partial b_{\m}}\,, \\[4pt] 
\DD^{\m}_{(-2)}&=&\frac{\partial}{\partial b^{\nabla}_{\m}}\,,
\eea 
and the degree 2 covariant coordinate $b^{\nabla}$ is given by the nonlinear redefinition 
\be 
b_{\m}^{\nabla} := b_{\m} +\frac 12\o_{\m ab}a^{a}a^b\,,
\ee
mixing the degree $2$ and degree $1$ coordinates. Thus we observe that when written in a covariant form  the homological vector field for a Courant algebroid is given in terms of the Gualtieri torsion and the basic curvature given in Definition \eqref{courantbasicdef}. This can be utilized to determine the covariant form of the Courant sigma model and its relation to the basic curvature we introduced here; we report on this separately in \cite{CIJ}, together with Noriaki Ikeda, where a more detailed derivation of Eq.\eqref{Qcourcov} is given. 

Mapping sections of $\widehat{E}$ to vector fields on ${\cal M}$ is somewhat different than for Lie algebroids. This is mainly due to the additional symplectic structure on one hand and due to the degree 2 coordinate and degree $-2$ derivation $\partial^{\m}$ on the other hand. We introduce the following two maps. First an analog of the $\iota$ map, which we do not fix completely yet: 
\bea  
&& \widehat\iota: C^{\infty}(M,\widehat E) \to \mf{X}_{-1}({\cal M})\,, \nn \\[4pt]
&& e\mapsto \widehat\iota(e)=e^a(x)(\frac{\partial}{\partial a^{a}}-f_{ab\m}a^b\frac\partial{\partial b_{\m}})-g\eta_{ab}\partial_{\m}e^{a}a^{b}\frac{\partial}{\partial b_{\m}}\,,\label{iotadef2}
    \eea  
    where $f_{ab\m}$ is a so far unspecified degree 0 function and $g$ is a parameter. One canonical choice is $f_{ab\m}=0$ and $g=1$, which corresponds to the operation
    \be 
\widehat{\iota}(e)|_{f_{ab\m}=0,g=1}=\{e,\cdot\}\,,
    \ee 
    with respect to the natural Poisson bracket on the QP manifold $\mc M$. This choice leads to Hamiltonian vector fields. In our analysis, which includes a noncanonical choice of connection, we will eventually require a different choice of map, which will have $g=0$ so that it is linear in the sense that $\widehat{\iota}(fe)=f\,\widehat{\iota}(e)$, like its Lie algebroid counterpart. 
    Second we define the auxiliary map 
    \bea  
 \label{iota0def}
&&\widehat\iota_0: C^{\infty}(M,\widehat E) \to C_1({\cal M})\,, \nn \\[4pt]
&& e\mapsto \widehat\iota_0(e)=e^a(x)a_{a}
    \eea  
    from sections of $\widehat{E}$ to degree 1 functions on $\mc M$. It is instructive to note that the two maps may be related as follows:
\be  
\widehat\iota_0(e)= (\,\widehat\iota(e), \mf a\,)\,, \quad \mf a:=a_b\,{d}a^b\,,
\ee
where $(\cdot,\cdot)$ denotes the natural isomorphism between the tangent and cotangent bundles of $\mc M$. This provides a way to turn functions into vector fields through the degree 2 map 
\be \label{jmap}
\widehat{j}\equiv (\cdot,\mf a): \G(T\mc M)\to C_1(\mc M)\,.
\ee 
To streamline the notation, as is customary we identify $\G(\widehat E)$ and $C_1(\mc M)$ thus dispensing with the redundant map $\widehat{\iota}_{0}$ which is now taken as the identity map. In other words $\widehat{j}\circ\widehat{\iota}=\text{id}$ and as such $\widehat j$ is the inverse of the map $\widehat{\iota}$. 

The anchor map of the Courant algebroid can be expressed as before, namely 
\be  
\rho(e)f=\big(\mc Q\,\widehat{\iota}(e)\big)f\,, 
\ee  
where once more $\mc Q={\cal L}_{Q}$. For the Courant bracket (or the non-skew Dorfman one), we introduce the degree $3$ Hamiltonian function 
\be 
\Theta=-\rho_{a}{}^{\m}a^{a}b_{\m}-\frac 1{3!}C_{abc}a^{a}a^{b}a^{c}\,,
\ee  
which satisfies $\{\Theta,e\}= Q\,e$. Together with the Poisson brackets on ${\cal M}$ induced by the natural symplectic structure 
\bea  \label{poissonbracketscourant}
\{x^{\m},b_{\n}\}=\d^{\m}_{\n}\,, \quad \{a^a,a^b \}=\eta^{ab}\,,
\eea  
the skew-symmetric Courant bracket is obtained as a derived bracket according to
\bea  
[e,e']_{\Eh}&=&\frac 12 \bigg(\{\{\Theta,e\},e'\}-\{\{\Theta,e'\},e\}\bigg)\, \nn\\[4pt] 
&=& \frac 12 \bigg(\{ Q\,e,e'\}-\{Q\,e',e\}\bigg)\,.
\eea  
The Dorfman bracket, which is a Leibniz bracket, is just the first term, without the skew-symmetrization.
The last main ingredient of the Courant algebroid is the nondegenerate fiber metric $\langle\cdot,\cdot\rangle$ with components $\eta_{ab}$. This is associated to the Poisson bracket structure on the graded cotangent bundle as follows 
\be  \langle e,e'\rangle=\frac 12 \{e,e'\}\,. \ee
This is simple to see with the aid of the corresponding local expressions since 
\be  \langle e,e'\rangle=\frac 12 \eta_{ab}e^{a}e'^{b}=\frac 12 \{a_a,a_b\}e^{a}e'^{b}=\frac 12 \{e^{a}a_a,e'^{b}a_b\}=\frac 12\{e,e'\}\,, \ee
for $e^{a}=e^{a}(x)$ and similarly for $e'^{a}$, where we used both graded Poisson brackets in \eqref{poissonbracketscourant}. 
These are well known due to \cite{dimaphd}; our main goal here is to express the Courant algebroid basic curvature that we defined in Section \ref{sec32} in terms of data on the QP2 manifold. To do so, and bearing in mind again that we do not necessarily work with the minimal data, let us choose a $T\mc M$-connection on $T\mc M$. This is defined in the same way and satisfies the same basic properties regardless if $\mc M=E[1]$ or $\mc M=T^{\ast}[2]E[1]$. Then it is very welcome that in the collective notation $x^{\a}$ for coordinates on $\mc M$, the definition of connection coefficients \eqref{connectioncollective} formally remains unchanged. 

Nevertheless, one must be careful with the degrees of the various components, which now change completely. Once more, there are no negative degree coordinates, which means that various coefficients vanish identically. 
The nonvanishing ones are 
 \bea 
 && \text{degree 0:} \quad \grave\omega_{\m\n}{}^{\rho}, \, \grave\omega_{\m a}{}^{b},\,\grave{\omega}_{a\m}{}^{b},\,\grave\omega_{ab\m},\,\grave\omega_{\m}{}^{\rho}{}_{\n},\,\grave\omega^{\m}{}_{\n\rho}\,,  \nn\\[4pt]
 && \text{degree 1:} \quad \grave\omega_{\m\n}{}^{a},\,\grave\omega_{\m a\n}\,, \grave\omega_{a \m \n}\,,  \\[4pt] 
 && \text{degree 2:} \quad \grave\omega_{\m\n\rho}\,.\nn
 \eea  
 Every other connection coefficient not listed here vanishes. Note that we have paid attention to the position of the indices because in the present case there are degree 0 and degree 2 coordinates carrying the same type of index, $x^{\m}$ and $b_{\m}$ respectively. Later we shall not insist on this notation when no confusion can arise. It is with these coefficients that all local coordinate expressions for the torsion, curvature and Atiyah cocycle written in Section \ref{sec41} continue to be valid here formally. 

\subsection{\texorpdfstring{$\widehat {\rm E}$}{e}-tensors  from T\texorpdfstring{$^\ast$}{*}[2]E[1]-tensors}
\label{sec44} 

Our goal now is to extent the logic of Section \ref{sec42} to Courant algebroids and QP2 manifolds. We start by choosing a $T\mc M$-connection on $T\mc M$ where $\mc M$ is the QP2 manifold. To proceed we must specify the map $\widehat\iota$. In general, the covariant derivative of degree $-1$ vector fields in its image along vector fields of the same type will not vanish, contrary to Lie algebroids, since now there exist derivations of degree $-2$. More specifically, working with Hamiltonian vector fields would lead to a nonzero result. We take a different route here and work with the set of vector fields of degree $-1$ that satisfy
\be  \label{nablaiotaiotaCourant}
\grave{\nabla}_{\widehat{\iota}(e)}\widehat\iota(e')\overset{!}=0\,.
\ee  
This fixes the map $\widehat\iota$ to be 
\be 
\widehat\iota(e)=e^a(x)(\frac{\partial}{\partial a^{a}}-\grave\o_{ba\m}a^b\frac\partial{\partial b_{\m}})\,,
\ee 
which contains the specific degree 0 coefficient $\grave{\omega}_{ab\m}$ and corresponds to $g=0$.
Obviously then $\widehat\iota(fe)=f\,\widehat\iota(e)$ for any degree 0 function $f$, which simplifies the analysis. Next we consider an arbitrary degree 0 vector field, denoted again as $\mc X_0$ and we ask for which connection coefficients the result of $\grave{\nabla}_{\mc X_0}\widehat\iota(e)$ is in the image of $\widehat\iota$. We find that this requires   
\be \label{omegacour2}
\grave\omega_{a\m b\n}=\partial_\m\grave\omega_{ab\n}-\grave\omega_{ad\n}\grave\omega_{\m b}{}^d+\grave\omega_{ab\rho}\grave\omega_\m{}^{\rho}{}_{\n}\,,
\ee 
where $\grave\omega_{a\m b\n}$ are the degree $0$ functions obtained from expanding the degree $1$ coefficient $\grave\omega_{\m b\n}$ as 
\be  
\grave\omega_{\m b\n}(x^{\a})=\grave\omega_{a\m b\n}(x^{\m})a^a\,.
\ee  
We note that although this condition looks basis dependent, we have checked that performing a coordinate dependent change of basis it transforms covariantly and therefore it is an acceptable condition. This allows us to write for this choice of $T\mc M$-connection
\be 
\grave\nabla_{\mc X_0}\widehat\iota(e)=\widehat\iota(\nabla_{X}e)\,,
\ee
for a $TM$-connection $\nabla$ on the Courant algebroid $\widehat{E}$ over $M$, together with the relation \eqref{omega is omega} among their coefficients. As for Lie algebroids, this does not characterize fully the connection but it is sufficient to discuss below how the Courant algebroid basic curvature is obtained from the dg manifold data, which is our original goal. 

We now turn to $\widehat{E}$-connections. First we compute 
\bea
\mc Q\, \widehat\iota(e)&=&e^a\r_a{}^\m\partial_\m + \nn\\[4pt] 
 && +\, \bigg(\r_a{}^\m \partial_\m e^b-e^c(C^b_{ca}-\eta^{bd}\rho_{d}{}^{\m}\grave\omega_{ac\m})\bigg)a^{a}\partial_b-  \rho_{a}{}^{\m}\partial_{\m}e^b\grave\omega_{cb\n}a^aa^c \partial^{\n} - 
 \nn\\[4pt] && - \, e^a\bigg((\partial_\m \r_a{}^\n-\eta^{cd}\rho_{d}{}^{\n}\grave{\omega}_{ca\m}) b_\n+ \nn\\[4pt] 
 && \quad \quad \quad  +\, \frac 12 (\partial_\m C_{abc}- \grave{\omega}_{da\m}C_{bc}^{d}-2\rho_c{}^{\n}\partial_{\n}\grave\omega_{ba\m}-2\grave\omega_{ba\n}\partial_{\m}\rho_c{}^{\n})a^ba^c\bigg)\partial^\m\,,
\eea
and with the help of it, 
\bea\label{nab}
\grave\nabla_{\mc Q\,\widehat\iota(e)}\,\widehat\iota(e')=e^a(\r_a{}^\m\partial_\m e'^d+\r_a{}^\m e'^b\grave\omega_{\m b}{}^d)(\partial_d-\grave\o_{cd\n}a^c\partial^\n)~,
\eea
using the previously found relations among the connection coefficients. This results in the $\widehat\iota$-image of the canonical $\widehat{E}$-connection, 
 \begin{tcolorbox}[ams nodisplayskip, ams align]  
\widehat\iota(\mathbullet{\nabla}^{\Eh}_ee')= \grave\nabla_{\mc Q\, \widehat\iota(e) }\, \widehat\iota(e')\,. \label{connectioncourantderived}
\end{tcolorbox}
This is essentially the analog of the  $\kappa=1$ case for Lie algebroids. Whether there exist further connections in the present case is beyond the scope of this work.
\begin{rmk} 
It is instructive to follow the linearity in the first entry of both sides in \eqref{connectioncourantderived}. In the left hand side, obviously $\widehat\iota(\nabla^{\Eh}_{fe}e')=f\,\widehat\iota(\nabla^{\Eh}_{e}e')$ for any function $f\in C^{\infty}(M)$. How is this consistent with the right hand side? The question is not trivial because $\mc Q\,\widehat\iota(fe)\ne f\mc Q\,\widehat\iota(e)$. Nevertheless,  
\be \grave\nabla_{\mc Q\,\widehat\iota(fe)}\widehat\iota(e')=f\grave\nabla_{\mc Q\,\widehat\iota(e)}\widehat\iota(e')\,,  
\ee 
due to the way the map $\widehat\iota$ was defined. This guarantees the consistency of \eqref{connectioncourantderived}. The proof follows from the following simple computation 
\bea    
\grave\nabla_{\mc Q\,\widehat\iota(fe)}\widehat\iota(e') = \grave\nabla_{(\mc Q f) \,\widehat\iota(e)}\widehat\iota(e')+\grave\nabla_{f\,\mc Q\,\widehat\iota(e)}\widehat\iota(e')=\mc Q f\,\grave\nabla_{ \,\widehat\iota(e)}\widehat\iota(e')+f\,\grave\nabla_{\mc Q\,\widehat\iota(e)}\widehat\iota(e')\,,
\eea 
which gives the desired result using \eqref{nablaiotaiotaCourant}.
\end{rmk} 
The next step is to express the Gualtieri torsion and the basic curvature  
in terms of QP2 manifold data. For the Gualtieri $\widehat E$-torsion defined in \eqref{GTE} we obtain at face value
\begin{align}
\label{GTEdg}
\mc T^{\Eh}(e,e',e'')&=\frac{1}{2}\bigg\{\widehat{j}\big(\grave\nabla_{\mc Q\,\widehat\iota(e)}\widehat\iota(e')-\grave\nabla_{\mc Q\,\widehat\iota(e')}\widehat\iota(e)\big),e''\bigg\}-\frac 14 \bigg\{\{Q\,e,e'\}-\{Q\,e',e\},e''\bigg\}+\nn\\[4pt] 
&+\frac{1}{4}\bigg\{\widehat{j}\big( \grave\nabla_{\mc Q\,\widehat\iota(e'')}\widehat\iota(e)\big),e'\bigg\}-\frac{1}{4}\bigg\{\widehat{j}\big( \grave\nabla_{\mc Q\,\widehat\iota(e'')}\widehat\iota(e')\big),e\bigg\}\,,
\end{align}
recalling that the map $\widehat{j}$ is defined in \eqref{jmap}.
 For the one defined with Dorfman bracket in \eqref{GTEd} we have
\begin{align}
\mc T^{\Eh_{\circ}}(e,e',e'')&=\frac{1}{2}\bigg\{\widehat{j}\big( \grave\nabla_{\mc Q\,\widehat\iota(e)}\widehat\iota(e')-\grave\nabla_{\mc Q\,\widehat\iota(e')}\widehat\iota(e)\big)-\{ Q\,e,e'\},e''\bigg\} +\frac{1}{2}\bigg\{\widehat{j}\big( \grave\nabla_{\mc Q\,\widehat\iota(e'')}\widehat\iota(e)\big),e'\bigg\}\,. \label{gualtieritorsiondorfmanderivedbad}
\end{align}
Similar expressions in a slightly different spirit (working directly with functions) were obtained in \cite{Deser:2016qkw}. 
It would be desirable though to have an expression which on the right hand side contains the torsion tensor of the connection on $\mc M$, unlike the rather unimaginative Eq. \eqref{gualtieritorsiondorfmanderivedbad}. 
Direct algebraic manipulations lead to the formula
 \begin{tcolorbox}[ams nodisplayskip, ams align] 
\mc T^{\Eh_{\circ}}(e,e',e'')&=\frac 12\bigg\{\widehat{j}\big(\grave T(\mc Q\,\widehat\iota(e),\widehat\iota(e'))+\text{At}(\widehat\iota(e'),\widehat\iota(e))\big),e''\bigg\}+\frac 12 \Lambda(\widehat\iota(e),\widehat\iota(e'),\widehat\iota(e''))\,,
\end{tcolorbox}
where for vector fields $\mc X,\mc Y,\mc Z\in \G(T\mc M)$ such that they are all in the image of the map $\widehat{\iota}$, namely they are of degree $-1$ and have the form appearing on the right hand side of Eq. \eqref{iotadef2} we defined the $C^{\infty}(M)$-tensor  
    \be  
 \Lambda(\mc X,\mc Y,\mc Z)=\big\{\,\widehat{j}\big([\mc Q\,\mc X, \mc Y]\big)-\{ Q\,\widehat{j}(\mc X),\widehat{j}(\mc Y)\},\widehat{j}(\mc Z)\big\}+\big\{\,\widehat{j}\big( \grave\nabla_{\mc Q\,\mc Z}\mc X\big),\widehat{j}(\mc Y)\big\}\,.
    \ee  
This indeed satisfies $C^{\infty}(M)$-linearity in all arguments, $\Lambda(f\mc X,g\mc Y,h\mc Z)=fgh\Lambda(\mc X,\mc Y,\mc Z)$, as can be shown by degree counting and by recalling the form of the symplectic structure and Poisson brackets in the present case. This expresses the Gualtieri torsion as a $\widehat{E}$-tensor obtained from the ordinary torsion, the Atiyah cocycle and the $\Lambda$-tensor of $(\mc M,Q,\grave\nabla)$.
The $\Lambda$-tensor arises in the Courant algebroid case because apart from the Leibniz algebra $(C_1(\mc M), \{Qe,e'\}:=e\circ e')$ which corresponds to the Dorfman bracket on sections of $\widehat{E}$, 
there is a second Leibniz algebra on the space of vector fields on $\mc M$, specifically $(\mathfrak{X}(\mc M),[\mc Q\mc X,\mc Y]:=\mc X \bullet \mc Y)$. One may then ask whether the map $\widehat{j}$ is a Leibniz homomorphism.{\footnote{This explanation of the $\Lambda$-tensor (and of the $V$-tensor that appears below) was suggested to us by S. Lavau and P. Batakidis.}} To this end we first introduce 
\be 
\l(\mc X,\mc Y)=\widehat{j}\big( \mc X \bullet \mc Y\big)-\widehat{j}(\mc X) \circ \widehat{j}( \mc Y)\,.
\ee 
If $\lambda$ was zero then $\widehat{j}$ would have been a Leibniz homomorphism but in general this is nonvanishing and  it is not a tensorial quantity. The corresponding tensorial quantity is precisely the $\Lambda$-tensor which is given as  
\be 
\Lambda(\mc X,\mc Y,\mc Z)=\{\lambda(\mc X,\mc Y),\widehat{j}(\mc Z)\}+\{\widehat{j}\big(\grave\nabla_{\mc Q\mc Z}\mc X\big),\widehat{j}(\mc Y)\}\,.
\ee 
We now turn to the basic curvature of a connection $\nabla$ on a Courant algebroid.
The necessary ingredients in its non-skew-symmetric version $\mc S^{\Eh_\circ}$ are written in terms of dg manifold data as
\begin{align}
\langle\nabla_{X}(e\circ e'),e''\rangle_{\Eh}&=\nabla_X\langle e\circ e',e''\rangle_{\Eh}-\langle e\circ e', \nabla_X e''\rangle_{\Eh} \nn\\[4pt] 
&= \frac 12 \bigg(X\{e\circ e',e''\}-\{e\circ e',\nabla_X e''\}\bigg) \nn\\[4pt] 
&=\frac 12 \bigg( {\mc X_0}\{\{Q\,e,e'\},e''\}-\{\{Q\,e,e'\},\widehat{j}\big(\grave\nabla_{\mc X_0}\widehat\iota(e'')\big)\}\bigg)\,,
\\[4pt]
\langle\nabla_X e\circ e',e''\rangle_{\Eh}&= \frac 12 \{\{Q\,\widehat{j}\big(\grave\nabla_{\mc X_0}\,\widehat\iota(e)\big),e'\},e''\}\,, \\[4pt]
\langle\nabla_{\overline{\nabla}^{\Eh}_eX}e',e''\rangle_{\Eh}&= \frac 12 \{\widehat{j}\big(\grave\nabla_{\mc Q\,\grave\nabla_{\mc X_0}\,\widehat\iota(e)}\widehat\iota(e')+\grave\nabla_{[\mc Q\,\widehat\iota(e),{\mc X_0}]}\widehat\iota(e')\big),e''\}\,.
\end{align}
A similar set of formulas holds for the skew-symmetric ingredients, but we refrain from presenting them here too, since they do not add anything substantial to the discussion. 
Applying these term by term in the definition \eqref{basiccurvaturecourantdorfman} we can directly obtain an expression for the Dorfman-version of the basic curvature tensor. This expression would not be very suggestive though. 
To make things more transparent,  
we ask the question whether the basic curvature can be written in a form where the curvature of $\grave\nabla$ appears on the right hand side, in general together with the torsion $\grave{T}$ and the Atiyah cocycle. To this end, we first define the following $C^{\infty}(M)$-tensor for ${\mc X_0},\mc X,\mc Y,\mc Z\in \G(T\mc M)$ such that ${\mc X_0}\in \mf{X}_{0}({\cal M})$, $\mc X, \mc Y, \mc Z \in \text{Im}(\widehat\iota)$ and $X\in \G(TM)$,
\begin{align}  
V(\mc X,\mc Y,\mc Z)X&:= - \bigg\{ \widehat{j}\big(\grave\nabla_{\mc X_0}[\mc Q\mc X,\mc Y]-[\mc Q\mc X,\grave\nabla_{\mc X_0}\mc Y]-[\mc Q\grave \nabla_{\mc X_0}\mc X,\mc Y]\big) +\{Q\widehat{j}(\mc X), \widehat{j}\big(\grave\nabla_{\mc X_0}\mc Y\big)\} \nn\\[4pt] &\qquad +\{Q\,\widehat{j}\big(\grave\nabla_{\mc X_0}\mc X\big),\widehat{j}(\mc Y)\},\widehat{j}(\mc Z)\bigg\}+ {\mc X_0}\big\{\{ Q\, \widehat{j} (\mc X),\widehat{j}(\mc Y)\},\widehat{j}(\mc Z)\big\} \nn\\[4pt] &\,\,\quad - \big\{\{ Q \,\widehat{j}(\mc X),\widehat{j}(\mc Y)\},\widehat{j}\big(\grave\nabla_{\mc X_0}\mc Z\big)\big\}+\big\{\,\widehat{j}\big(\grave\nabla_{\mc Q\grave\nabla_{\mc X_0}\mc Z}\mc X+\grave\nabla_{[\mc Q\mc Z,{\mc X_0}]}\mc X\big),\widehat{j} (\mc Y)\big\}\,.\nn
\end{align}  
Then it is a matter of straightforward algebra to show that the basic curvature tensor for a Courant algebroid in terms of tensorial quantities in its dg manifold description is 
\begin{tcolorbox}[ams nodisplayskip, ams align]
\mc S^{\Eh_{\circ}}(e,e',e'')X&=\frac 12\bigg\{\widehat{j}\bigg( \grave R(\mc Q\,\widehat\iota(e'),{\mc X_0})\widehat\iota(e)-\grave R(\mc Q\,\widehat\iota(e),{\mc X_0})\widehat\iota(e')-\grave\nabla_{\mc X_0}\text{At}(\widehat\iota(e'),\widehat\iota(e))\bigg)-\nn\\[4pt]
& - \widehat{j}\bigg( \grave\nabla_{\mc X_0}\grave T(\mc Q\,\widehat\iota(e),\widehat\iota(e'))+\grave T(\widehat\iota(e'),\mc Q\grave \nabla_{\mc X_0}\widehat\iota(e))+\grave T(\grave\nabla_{\mc X_0}\mc Q\,\widehat\iota(e),\widehat\iota(e'))\bigg), e''\!\bigg\} \nn\\[4pt] & +\frac 12 \, V(\widehat\iota(e),\widehat\iota(e'),\widehat\iota(e''))X\,.
\end{tcolorbox}
Thus we obtained the basic curvature for a connection on a Courant algebroid from the ordinary curvature and torsion tensors of a connection on $\mc M$, the Atiyah cocycle and the additional tensor $V$ defined above.
The $V$-tensor can be written in terms of the two Leibniz products $\circ$ and $\bullet$. This goes through the question of whether $\grave\nabla$ is a derivation for the two Leibniz products; the $V$-tensor is then the difference of the failure of $\grave\nabla$ to be a derivation for each of the two products, suitably adjusted so as to be a tensorial quantity: 
\begin{align}
&V(\mc X,\mc Y,\mc Z)X= \nn\\[4pt] & \qquad -\bigg\{\widehat{j}\bigg(\grave\nabla_{\mc X_0}(\mc X\bullet\mc Y)-\grave\nabla_{\mc X_0}\mc X\bullet \mc Y-\mc X\bullet\grave\nabla_{\mc X_0}\mc Y\bigg),\widehat{j}(\mc Z)\bigg\} \nn\\[4pt] 
& \qquad +\, \bigg\{\grave\nabla_{\mc X_0}\big(\widehat{j}(\mc X) \circ \widehat{j}(\mc Y)\big)-\grave\nabla_{\mc X_0}\big(\widehat{j}(\mc X)\big)\circ \widehat{j}(\mc Y)-\widehat{j}(\mc X)\circ \grave\nabla_{\mc X_0}\big(\widehat{j}(\mc Y)\big),\widehat{j}(\mc Z)\bigg\} \nn\\[4pt] 
& \qquad + \, \bigg\{\grave\nabla_{\mc X_0}\widehat{j} (\mc X)\circ\widehat{j}(\mc Y),\widehat{j}(\mc Z)\bigg\}+\bigg\{\widehat{j}(\mc X)\circ\grave\nabla_{\mc X_0}\widehat{j}(\mc Y),\widehat{j}(\mc Z)\bigg\}+\bigg\{\widehat{j}(\mc X)\circ\widehat{j}(\mc Y),\grave\nabla_{\mc X_0}\widehat{j}(\mc Z)\bigg\} \nn\\[4pt] & \qquad +\, \bigg\{\widehat{j}\big(\grave\nabla_{\mc Q\grave\nabla_{\mc X_0}\mc Z}\mc X+\grave\nabla_{\mc Z\bullet {\mc X_0}}\mc X\big),\widehat{j}(\mc Y)\bigg\}\,.
\end{align}

\section{2D topological sigma models} 
\label{sec5}
\subsection{Poisson sigma model and basic curvature}\label{sec51}

We would now like to explain how the basic curvature tensor appears in field theory models. As a first example, we consider the well studied Poisson sigma model. This is a topological field theory in two dimensions induced by a Poisson structure on the target space. It is also the AKSZ sigma model in 2D due to the correspondence between QP1 manifolds and Poisson manifolds \cite{Roytenberg:2006qz}. The AKSZ formulation of Poisson sigma models appears in  \cite{Cattaneo:2001ys}. We also note that a manifestly target space covariant formulation in terms of an auxiliary connection was given in \cite{Baulieu:2001fi}, where the basic curvature tensor already appears in the gauge fixed action, although its geometric significance was not studied. In the more recent approach of \cite{Ikeda:2019czt}---which, importantly, is also valid in the non-AKSZ twisted Poisson case  \cite{Severa:2001qm},---the relation of the 4-fermion terms in the BV action to the basic curvature of the Poisson Lie algebroid was highlighted.   
Here we present a slightly different perspective. Instead of working with the BV action, we work with BRST transformations.{\footnote{Although this is not relevant in the present case, in general this approach can be followed even in the absence of a given action functional.}} We shall show that since they do not square to zero off-shell, when properly extended to off-shell nilpotent ones, the basic curvature directly appears in it and controls this extension. This is of course not surprising, since the BV-BRST differential is obtained from the BV action with the use of the BV antibracket. 

The description of the Lie algebroid on $T^{\ast}M$ of a Poisson manifold $M$ as a dg manifold $T^{\ast}[1]M$ features a homological vector field of the form
\be \label{Qhp}
Q^{\ast}=\Pi^{\m\n}(x)a_\m\frac\partial{\partial{x^{\n}}}-\frac 12 \partial_\rho\Pi^{\m\n}(x)a_{\m}a_{\n}\frac{\partial}{\partial{a_{\rho}}}\,,
\ee     
where the asterisk is a reminder that we work on the cotangent bundle.
If we consider in addition an ordinary connection $\nabla$ on it, this is rewritten in a  covariant fashion as \cite{Chatzistavrakidis:2021nom}
\be \label{Qhpb}
Q^{\ast}=\Pi^{\m\n}a_\m \DD_{\n}^{(0)}-\frac 12 \mathring{\nabla}_\rho\Pi^{\m\n}a_\m a_\n\frac\partial{\partial{a_\rho}}\,,
\ee 
where the degree 0 covariant derivative is
\be  \DD_{\n}^{(0)}=\frac\partial{\partial{x^{\n}}}+\Gamma^{\rho}_{\s\n}a_\rho\frac\partial{\partial{a_\sigma}}\,. \ee
Recall now that $-\mathring\nabla\Pi$ that appears in the second term of $Q^{\ast}$ is the Lie algebroid $T^{\ast}M$-torsion of the canonical $T^{\ast}M$-connection, as discussed in Section \ref{sec22}. 

Let us now give a quick reminder of the Poisson sigma model and its basic gauge structure. 
The classical action functional is
\be \label{SHPSM}
S_{\text{PSM}}[X,A]=\int_{\S_{2}}\left(A_\m\w\dd X^{\m}+\frac 12 \, (X^{\ast}\Pi^{\m\n})A_\m\w A_\n\right)\,,
\ee  
where the fields are scalars $X^{\m}=X^{\m}(\s^{m})$ and 1-forms $A_{\m}=A_{\m m}(X)\dd \s^{m}$ with $(\s^{m})$ local coordinates on the surface $\S_{2}$. 
The theory has gauge invariances under which the fields transform as 
\bea 
\label{PSMtrafo1}
\d X^{\m}&=&\Pi^{\n\m}\epsilon_\n\,,\\ \label{PSMtrafo2}
\d A_\m&=& \dd\epsilon_\m+\partial_\m\Pi^{\n\rho}A_\n\epsilon_\rho\,,
\eea   
with respect to the scalar gauge parameter $\epsilon_\m=\epsilon_\m(\s^{m})$. Having only a scalar gauge parameter is a feature that simplifies matters, as opposed to higher dimensional analogs of the model which have a richer structure but also the additional complication of higher form parameters that lead to reducible gauge symmetries \cite{Chatzistavrakidis:2021nom,Ikeda:2021rir}. 
The classical field equations of the theory are 
\bea \label{HPSMeom1}
F^{\m}&:=&\dd X^{\m}+\Pi^{\m\n}A_\n=0\,,
\\ \label{HPSMeom2}
G_\m&:=&\dd A_\m+\frac 12 \, \partial_\m\Pi^{\n\rho}A_\n\w A_\rho=0\,.
\eea  
Hence $F^{\m}$ and $G_{\m}$ are the field strengths of $X^{\m}$ and $A_\m$ respectively. 
These field strengths 
 transform covariantly under gauge transformations, specifically 
 \bea  
 \label{deltaFmu}
\d F^{\m}&=&\partial_\r\Pi^{\nu\mu} F^\r\epsilon_\nu\,,\\[4pt] 
\label{deltaGmu}
\d G_{\m}&=&\partial_\m\Pi^{\nu\r}G_\nu\epsilon_\r+\partial_\m\partial_\s\Pi^{\nu\r}F^\s\w A_\n\epsilon_\r\,.
 \eea  
 Furthermore, they satisfy the Bianchi identities 
 \bea   
\label{bipsm1} \dd F^{\m}&=&\partial_{\rho}\Pi^{\m\n}F^{\rho}\w A_{\n}+\Pi^{\m\n}G_{\n}\,, \\[4pt] 
\label{bipsm2} \dd G_{\m}&=& \frac 12 \partial_{\m}\partial_{\n}\Pi^{\k\l}F^{\n}\w A_\k\w A_\l+\partial_{\m}\Pi^{\k\l}G_{\k}\w A_{\l}\,.
 \eea  
 More details on the model may be found in other works, see for example \cite{Bojowald:2004wu}. Approaches to the covariantization of the model, including its BV action and its Wess-Zumino-Witten extension, were discussed in \cite{Ikeda:2019czt}. 
 It is useful to note that the field strengths defined in \eqref{HPSMeom1} and \eqref{HPSMeom2} satisfy two simple conditions. Denoting them collectively as $F^{\a}=(F^{\m},G_{\m})$, these conditions are
 \bea
\d F^{\a}|_{F^{\b}=0} &=& 0\,, \\[4pt] 
\dd F^{\a}|_{F^{\b}=0} &=&0\,.
 \eea 
In more generality, these are written in \cite{Grutzmann:2014hkn} in terms of the ideal $\mc I$ generated by the expressions giving the field strengths in the form 
 \bea 
\label{ideal1} && \d\, \mc I \subset \mc I\,, \\[4pt] 
 \label{ideal2} && \dd\, \mc I\subset \mc I\,,
 \eea 
 and they are the main two assumptions in defining a higher gauge theory. 

 The BRST transformations  $Q_{\text{\tiny{BRST}}}$ of the model are obtained by introducing a ghost of degree 1 for the gauge parameter; not to introduce too much additional notation, we shall  call the ghost $\e_{\m}$ too. Then $Q_{\text{\tiny{BRST}}}$ is of degree 1 and it replaces $\delta$ in Eqs. \eqref{PSMtrafo1} and \eqref{PSMtrafo2}, which we do not repeat, now with $\e_{\m}$ being the ghost. Then we need to specify its action on the ghost; this is known to be 
 \be  
Q_{\text{\tiny{BRST}}}\,\e_{\m}=-\frac 12 \partial_{\m}\Pi^{\n\rho}\e_{\n}\e_{\rho}\,.
 \ee  
 Using the defining relation of a Poisson structure $([\Pi,\Pi]_{\text{SN}}=0$ with the Schouten-Nijenhuis bracket), it is concluded that 
 \bea  
&& Q_{\text{\tiny{BRST}}}^2X^{\m}=0\,,\quad Q_{\text{\tiny{BRST}}}^{2}\e_{\m}=0\,, \\[4pt] 
&& Q_{\text{\tiny{BRST}}}^2A_{\m}=-\frac 12 \partial_{\m}\partial_\n\Pi^{\rho\s} F^{\n}\e_{\rho}\e_{\s}\,.
 \eea  
That the BRST operator does not square to zero on all fields unless the field equations are taken into account is  a signal of the openness of the gauge algebra. Indeed one can see this directly by computing the commutator of two gauge transformations on the 1-form $A_{\m}$; the result is again the same, namely the second derivative on the Poisson structure and a field strength of $X^{\m}$. 

Let us now ask for an extension of the operator $Q_{\text{\tiny{BRST}}}$, say $Q_{\text{\tiny{BV}}}$, such that it is nilpotent off-shell. The answer is obvious within the BV-BRST formalism and homological perturbation theory. Introduce a new field, denoted $A^{\dagger \m}$ and called the antifield of $A_{\m}$, which is a 1-form of ghost degree $-1$. Now ask from this field to transform as 
\be  \label{QBVAdaggerPSM}
Q_{\text{\tiny{BV}}} A^{\dagger \m}= -F^\m+\partial_\r\Pi^{\m\n}A^{\dagger \r}\epsilon_\n
\ee  
and introduce a correction to the BRST transformation of $A_{\m}$ as 
\be  
\D Q_{\text{\tiny{BRST}}} A_{\m} = -\frac{1}{2}\partial_\m\partial_\r\Pi^{\n\s} A^{\dagger \r}\epsilon_\n\epsilon_\s\,,
\ee 
with $\D Q_{\text{\tiny{BRST}}}X^{\m}=0=\D Q_{\text{\tiny{BRST}}}\e_{\m}$.
We observe that if we identify $Q_{\text{\tiny{BV}}}=Q_{\text{\tiny{BRST}}}+\D Q_{\text{\tiny{BRST}}}$, in which case 
\be  
Q_{\text{\tiny{BV}}}A_{\m}=\dd\epsilon_\m+\partial_\m\Pi^{\n\rho}A_\n\epsilon_\rho-\frac{1}{2}\partial_\m\partial_\r\Pi^{\n\s} A^{\dagger \r}\epsilon_\n\epsilon_\s\,,
\ee  
then 
\be  Q_{\text{\tiny{BV}}}^2=0
\ee 
on all fields and antifields considered so far, namely on $X^{\m}, A_{\m}, \e_{\m}, A^{\dagger\m}$. To have a complete description of the BRST structure we must assign the correct transformation rule to the remaining fields, namely the antifields $X^{\dagger}_{\m}$ and $\e^{\dagger\m}$ which must be introduced for consistency of the BV-BRST procedure.
We shall not delve in the non-illuminating details of this consistency check, since the answer is known, see for example \cite{Ikeda:2019czt}. 
This completes the derivation of the BV-BRST differential for the Poisson sigma model. We are now ready to reexamine it from a target space covariant perspective, following and extending the analysis of \cite{Ikeda:2019czt}.

As already mentioned earlier, we introduce a vector bundle connection on $T^{\ast}M$ and the canonical $T^{\ast}M$-connection on it given by 
\be  \mathbullet{\nabla}^{\E}_{\eta}\eta'=\mathring\nabla_{\Pi^{\sharp}(\eta)}\eta'\,, 
\ee  
for $\eta, \eta'\in \G(T^{\ast}M)$ and $\Pi^{\sharp}:T^{\ast}M\to TM$ the anchor map induced by the Poisson structure on $M$, see Appendix \ref{appa} for more details. As shown in \cite{Ikeda:2019czt}, the BV action is independent of the choice of this auxiliary connection, as expected since the answer found in \cite{Cattaneo:2001ys} does not require a connection. An important point though is that the geometric form of the BV-BRST differential remains the same for the 3-form-twisted model, which does not admit a canonical 2D AKSZ construction. The $E$-torsion with $E=T^{\ast}M$ in this case is given by $\mathbullet{T}^{\E=T^{\ast}M}:=T^{\ast}=-\mathring\nabla\Pi$ and the basic curvature $S^{\ast}:=S^{\E=T^{\ast}M}$ is as in Eq. \eqref{Sstar}. 
Then the action of the BV-BRST differential on the fields of the theory and on the antifield $A^{\dagger}$ in a fully basis independent form is given as 
 \begin{tcolorbox}[ams nodisplayskip, ams align] 
\label{qbvpsm1}&Q_{\text{\tiny{BV}}}X=\Pi(\e)\,, \\[4pt] 
&Q_{\text{\tiny{BV}}}\e= T^{\ast}(\e\w\e)\,,\\[4pt] 
\label{qbvpsm3} &Q_{\text{\tiny{BV}}}A^{\mathring\nabla}=\mathring{\DD}\e-T^{\ast}(A^{\mathring\nabla},\e)-S^{\ast}(\e\w\e)A^{\dagger}\,, \\[4pt] 
\label{qbvpsm4}&Q_{\text{\tiny{BV}}}A^{\dagger}=-F^{\mathring\nabla}-T^{\ast}(A^{\dagger},\e)
\end{tcolorbox}
where $\mathring{\DD}$ is the covariant exterior derivative induced by $\mathring\nabla$ and we have made the field redefinition 
\be   A^{\mathring{\nabla}}_{\m}:=A_{\m}+\mathring{\G}_{\m\n}^{\rho}A^{\dagger \n}\e_{\rho}\,,\ee 
with $F^{\mathring\nabla}$ being the same as $F$ but with $A$ replaced by $A^{\mathring\nabla}$. 
These equations contain all the necessary information for our purposes, therefore we do not present the additional complicated actions on the two remaining antifields. 

The main conclusion of this analysis is that the nonlinear terms of the  BV-BRST differential of the Poisson sigma model are given by the $E$-torsion and basic curvature tensors. Specifically, it is on the 1-form of the theory that the BV-BRST differential contains a basic curvature dependence. This is in agreement with the solution of the classical master equation, even though the elegance of the AKSZ formalism hides the relation to the basic curvature. We emphasize once more that although we have only discussed the Poisson sigma model here, the most interesting aspect of this approach is that the form of the BV-BRST differential we presented remains the same in presence of a Wess-Zumino-Witten 3-form with the difference that the auxiliary connection on the target is one with torsion controlled by the 3-form. In the vanilla Poisson sigma model, the expressions presented here can be related to the standard result of \cite{Cattaneo:2001ys} by nonlinear field redefinitions that involve the connection coefficients.

We close this section with a technical yet important point. When one studies the covariantization of the gauge transformations and the field equations of the theory at the BRST level, i.e. without antifields, one is naturally led to make a distinction between the transformations given by $\d$ and improved, covariant transformations denoted by $\d^{\mathring{\nabla}}$ in \cite{Ikeda:2019czt}. These are \emph{different} rules, not a rewriting, and they are related by terms that vanish on-shell. 
For instance, in the present case of the Poisson sigma model we would have 
\be  
\d^{\mathring{\nabla}}A_{\m}=\d A_\m -\mathring{\G}_{\m\n}^{\rho}F^{\n}\e_{\rho}\,.
\ee  
Promoting this to a BRST transformation would immediately result to 
\be  
(Q^{\mathring\nabla}_{\text{\tiny{BRST}}})^2 A_{\m}=-\frac 12 S^{ \k\l}_{\m\n}F^\n\e_\k\e_\l\,,
\ee 
in terms of the basic curvature tensor directly without need of further covariantization. Similarly to what we presented above, this now results in the basis independent form of the BV-BRST differential 
\be  \label{QBVAPSM2}
Q_{\text{\tiny{BV}}}A=\mathring\DD\e-T^{\ast}(A,\e)-S^{\ast}(\e\w\e)A^{\dagger}\,,
\ee  
without the need for a field redefinition any longer.
 This is consistent with the BV-BRST differential being unique up to field redefinitions. In fact, in more general cases such as Dirac sigma models this second approach is more convenient and we shall follow it in the next section. 

\subsection{Poisson sigma model and Atiyah cocycle} \label{sec52}

To see how structures such as the ones that we described in Section \ref{sec42} appear in the covariant form of the Poisson sigma model, we start with the local coordinate expressions. Recall that the homological vector field in the present case is given in \eqref{Qhp}. Then the components of its tangent lift, see \eqref{local tangentlift}, are 
\bea   
&& \mc Q(\partial_{\m})^{\n}=\partial_{\m}\Pi^{\n\rho}(x)a_{\rho} = -\mc Q(\partial^{\n})_{\m}\,, \\[4pt] 
&& \mc Q(\partial_{\m})_{\n}=-\partial_\m\partial_\n\Pi^{\k\l}(x)a_\k a_\l\,, \\[4pt] 
&&\mc Q(\partial^\m)^\n=\Pi^{\m\n}(x)\,,
\eea  
where for simplicity we have used the notation $\partial_{\m}=\partial/\partial x^{\m}$ and $\partial^{\m}=\partial/\partial a_{\m}$ and we have explicitly denoted the $x^{\m}$-dependence for clarity. It is then straightforward to observe via Eqs. \eqref{PSMtrafo1} and \eqref{PSMtrafo2} that 
\bea   
\d X^{\m}&=&\mc Q(\partial^{\n})^{\m}\e_{\n}\,,
\\[4pt]
\d A_{\m}&=&\dd \e_{\m}+\mc Q(\partial^{\n})_{\m}\e_{\n}\,.
\eea   
To understand these  equations better we should now rethink about the model through the lens of the shift functor $[1]$. For the scalar fields nothing changes, we still think of them as components of the map $X: \S\to M$. What changes is that we now consider as source space the dg manifold $T[1]\S$ and as target space the symplectic dg manifold $\mc M=T^{\ast}[1]M$, as in the AKSZ construction. Then we introduce a degree-preserving map $A:T[1]\S\to T^{\ast}[1]M$, whose pull-back is the 1-form  $A_{\m}$, now thought of as function on $T[1]\S$. 
To be more precise, we introduce local coordinates $(\s^{m},\theta^{m})$ on $T[1]\S$ of degrees 0 and 1 respectively and coordinates $x^{\m}$ and $a_{\m}$ on $T^{\ast}[1]M$. The homological vector field on $T[1]\S$ is given by the de Rham differential, 
\be  \dd=\theta^{m}\frac \partial{\partial \s^{m}}\,, \ee 
and we keep the same notation $\dd$ for it. Then according to the above, 
\be  X^{\m}=X^{\ast}(x^{\m}) \quad \text{and} \quad A_{\m}=A^{\ast}(a_{\m})\,.\ee 
For notational convenience we collect the two maps above in a single degree-preserving one 
\be  
\phi: T[1]\S\to \mc M\,,
\ee  
with base map $X$. It is also useful to extend this map to the product dg manifold
\be  \label{phitimes}
\phi_{\times}: T[1]\S \to T[1]\S\times \mc M\,,
\ee 
such that it is the identity in the first factor (this was first suggested in \cite{Grutzmann:2014hkn}.) This extension allows us to think of the gauge parameter $\e_{\m}$, which is an explicit function of the coordinates on $\S$, also as a pull back from the extended target through the map $\phi_{\times}$. Considering the tangent bundle $T(T[1]\S\times \mc M)$ we can construct the following vector field as a section of it, 
\be   
\e=\e_{\m}(\s)\frac{\partial}{\partial a_{\m}}\,.
\ee  
This is a very special vector field of course, but it is now well-defined. 
We are now ready to write down the gauge  transformation rules in a better form: 
\begin{tcolorbox}[ams nodisplayskip, ams align]
\d X^{\m}&= \phi^{\ast}_{\times}\big(\mc Q(\e)^{\m}\big)\,, \\[4pt] 
\d A_{\m}&=\dd \big(\phi^{\ast}_{\times}(\epsilon_{\m})\big) +\phi^{\ast}_{\times}\big(\mc Q(\epsilon)_{\m})\,,\label{deltaphiPSM}
\end{tcolorbox}
where we have explicitly indicated the pull-backs. 
These equations clarify the appearance of the tangent lift in the gauge transformations. 

In a similar spirit, considering a $T\mc M$-connection $\grave\nabla$ and working for the moment on a local patch where the connection coefficients are set to zero, we can compute all local coordinate components of the Atiyah 1-cocycle of $(T^{\ast}[1]M,Q^{\ast},\grave\nabla)$. They are 
\bea   
&& \text{At}(\partial_\m,\partial_\n)^{\rho} =-\partial_\m\partial_\n\Pi^{\rho\s}(x)a_\s=\text{At}(\partial_\m,\partial^{\rho})_{\n}\,,
\\[4pt]  
&& \text{At}(\partial_\m,\partial_\n)_{\rho} =-\frac 12 \partial_\m\partial_\n\partial_\rho\Pi^{\k\l}(x)a_\k a_\l\,,
\\[4pt]&& \text{At}(\partial_{\m},\partial^{\n})^{\rho}=\partial_{\m}\Pi^{\n\rho}(x)=\text{At}(\partial^{\n},\partial^{\rho})_{\m}\,,
\\[4pt]&& \text{At}(\partial^\m,\partial^\n)^\rho =0\,.
\eea  
We can then immediately write down the gauge transformation $\d$ of the field strengths given in Eqs. \eqref{deltaFmu} and \eqref{deltaGmu} in the following way, 
\bea  
\d F^{\m}&=&  F^{\rho}\e_{\n}\text{At}(\partial_{\rho},\partial^{\n})^{\m}\,, \\[4pt] 
\d G_{\m}&=&  G_{\n}\e_{\rho}\text{At}(\partial^{\n},\partial^{\rho})_{\m}+F^{\n}\e_{\rho}\text{At}(\partial_\n,\partial^{\rho})_{\m}\,,
\eea 
where pull-backs to the space of fields are understood. We can already observe an interesting outcome of this analysis; the second term in $\d G_{\m}$ encodes the gauge field $A_{\m}$ of the second term in \eqref{deltaGmu} in the Atiyah cocycle. This is an economic feature of this approach, but more importantly it allows us to find  the corresponding geometric expressions directly.  
Due to the linearity properties of the map $\text{At}$ laid out in \eqref{atiyahlinear}, these expressions read 
\begin{tcolorbox}[ams nodisplayskip, ams align]
\d F^{\m}&=\phi^{\ast}_{\times}\big(\text{At}(F,\e)^{\m}\big)\,, \\[4pt] 
\d G_{\m} &= \phi^{\ast}_{\times}\big(\text{At}(F+G,\e)_{\m}\big)\,,
\end{tcolorbox}
where $F=F^{\m}(x)\partial/{\partial x^{\m}}$ and $G=G_{\m}(x)\partial/{\partial a_{\m}}$ are vector fields on $\mc M$. 
This gives the relation of the Atiyah cocycle of the dg manifold with connection to the transformation of the field strengths in the Poisson sigma model. The second equation is rather remarkable, containing the sum of $F$ and $G$ that have different form degree. Note though that the Atiyah cocycle has different components when it is evaluated in each of the two; as obvious from the local coordinate expressions, it comes with an additional $a_{\m}$ on $F$ with respect to $G$, which upon pulling it back to the space of fields yields the ``missing'' 1-form and restores the Chern-Simons $F\w A$ term in the transformation of $G$. Bringing these results together, we have shown that 
\begin{prop}\label{psmkapranov}
    Let $M$ be a Poisson manifold and $(\mc M=T^{\ast}[1]M, Q^{\ast},\grave\nabla)$ its associated dg manifold, equipped in addition with a torsion-free connection. Then the component fields $((X^{\m}),(A_{\m}))$, $\m=1,\dots,\text{dim}\,M$ of the maps $(X,A)\in \text{Mor}(T[1]\S,\phi^{\ast}T^{\ast}[1]M)$ in the Poisson sigma model and their field strengths $(F^{\m},G_{\m})$ obey  transformation rules under the gauge symmetries of the model, which in terms of the unary and binary brackets $(b_1,b_2)$ of the Kapranov L$_{\infty}[1]$ algebra on $(\mc M,Q^{\ast},\grave\nabla)$ read 
    \bea 
    \d X^{\m}&=& \phi^{\ast}_{\times}\big(b_1(\e)^{\m}\big)\,, \\[4pt] 
\d A_{\m}&=&\dd\big(\phi^{\ast}_{\times}(\epsilon_{\m})\big) +\phi^{\ast}_{\times}\big(b_1(\epsilon)_{\m}\big)\,, \\[4pt] 
\d F^{\m} &=& -\phi^{\ast}_{\times}\big(b_2(F,\e)^{\m}\big)\,, \\[4pt]  
\d G_{\m} &=&-\phi^{\ast}_{\times}\big(b_2(F+G,\e)_{\m}\big)\,.
    \eea  
    Here $b_1$ is given by the tangent lift of the homological vector field on $\mc M$ and $b_2$ by the opposite of the Atiyah 1-cocycle for $\grave\nabla$. Moreover, $\phi^{\ast}_{\times}$ is the pull-back of the map from $T[1]\S$ to $T[1]\S\times \mc M$ introduced in \eqref{phitimes} and $F, G, \e$ are vector fields on $T[1]\S\times \mc M$ such that $F=F^{\m}\frac{\partial}{\partial{x^{\m}}}$, $G=G_{\m}\frac{\partial}{\partial a_{\m}}$ and $\e=\e_{\m}(\s^m)\frac{\partial}{\partial a_{\m}}$\,, $\s^m$ being the bosonic coordinates on $T[1]\S$ and $(x^{\m},a_{\m})$ coordinates on $\mc M$.
\end{prop}
\begin{rmk} 
This relation of field theory to a Kapranov L$_{\infty}[1]$ algebra is different than the L$_{\infty}$ algebra structure of perturbative field theories presented in \cite{Hohm:2017pnh}. First, observe that the gauge transformation of the fields is controlled by the unary bracket here, whereas it is the binary bracket that does this in \cite{Hohm:2017pnh}. Similarly, all transformations of the field strengths are controlled by the binary bracket given by the Atiyah cocycle, whereas in the picture of \cite{Hohm:2017pnh} there are also higher brackets appearing in these transformations. 
\end{rmk} 
For completeness, we briefly revisit for the simple case of the Poisson sigma model some general properties of gauge theory. Denote the field strengths collectively as $ F^{\a}=(F^{\m},G_{\m})$. Then it is a matter of simple observation to write 
\be  
{ F}^{\a}=\dd \phi^{\a}-Q^{\ast\,\a}\,,
\ee  
where $Q^{\ast\,\a}$ is understood as a pull-back. To be more precise, if we denote as $\phi$ the map from the space of fields to the target space and by $\phi^{\ast}$ the pull-back map, then 
\be  { F}^{\a} = \dd\circ\phi^{\ast}(x^{\a})-\phi^{\ast}\circ Q^{\ast}(x^{\a})\,.\ee 
This prompts the definition of the following operator \cite{Grutzmann:2014hkn}, 
\be  
{\cal F}=\dd\circ \phi^{\ast}-\phi^{\ast}\circ Q^{\ast}\,,
\ee  
such that $ F^{\a}=\mc F(x^{\a})$. The Bianchi identities \eqref{bipsm1} and \eqref{bipsm2} can be collectively written 
 \be  
\dd { F}^{\a}=- \dd Q^{\ast\,\a}\,,
 \ee  
 and in more illustrative terms, using the fundamental fact that $(Q^{\ast})^2=0$, as 
 \be  
\dd\circ {\cal F} = - {\cal F}\circ Q^{\ast}\,.
 \ee  
 Although here we confirmed this in the simple case of the Poisson sigma model, it is much more general and it holds for higher gauge theories, as shown in \cite{Grutzmann:2014hkn}. We shall return to this in Section \ref{sec6}.

\subsection{Dirac sigma models}
\label{sec53}

One of the useful aspects of the geometric approaches in Sections \ref{sec51} and \ref{sec52} is that they go beyond the vanilla AKSZ construction. Although the 2D AKSZ sigma model is the Poisson sigma model, general topological and nontopological 2D sigma models are not captured by the AKSZ formalism. Restricting to the topological case, which is better understood at the moment,{\footnote{See \cite{Grigoriev:2020xec,Grigoriev:2022zlq} for recent studies of AKSZ in nontopological settings.}} these include Wess-Zumino-Witten-Poisson sigma models \cite{Klimcik:2001vg} with an underlying twisted Poisson geometry \cite{Severa:2001qm} and more generally Dirac sigma models \cite{Kotov:2004wz}, where the target space is a Dirac manifold \cite{Courant}. In general no AKSZ construction is known for such theories. Nevertheless their BV action was determined by traditional methods recently in Ref. \cite{Chatzistavrakidis:2022wdd}. 
We are not going to repeat here the technical analysis of \cite{Chatzistavrakidis:2022wdd}. To make our point, we will essentially reconstruct via deformation the BV-BRST differential of the general topological Dirac sigma model and emphasize its relation to the basic curvature tensor(s) of suitable $E$-connection(s). 

There are essentially two equivalent ways of thinking about 2D sigma models with a Dirac manifold as target space. In the original approach they are interpolations between $G/G$ Wess-Zumino-Witten models (i.e. completely gauged ones, which are topological,) and Wess-Zumino-Poisson sigma models. In the approach of gauge theory, first studied in \cite{Salnikov:2013pwa} and in a more general context  in \cite{CDJS}, they constitute the result of gauging singular foliations. 
The action functional of Dirac sigma models is 
\be  
S_{\text{DSM}}=\int_{\S_2}\left(\frac 12 \,g_{\m\n}(X)(\dd X^{\m}-V^{\m})\w\ast (\dd X^{\m}-V^{\m})+W_{\m}\w (\dd X^{\m}-\frac 12 V^{\m})\right)+\int_{\S_3}X^{\ast}H\,,
\ee  
where $H$ is a 3-form, $X$ is the sigma model map from $\S_2$ to the target space and $\S_2=\partial\S_3$, such that the model does not depend on the choice of $\S_3$.{\footnote{Under the usual conditions for existence of the extension, which requires
that the homology class $[X(\S_3)] \in H_3(M)$ vanishes and independence on the choice of $\S_3$, which
requires that the 3-form defines an integer cohomology class $[H]/2\pi \in H_3(M, \Z)$ \cite{Figueroa-OFarrill:2005vws}.}} The two fields $V^{\m}$ and $W_{\m}$ are 1-forms on $\S_2$ with values in the pull-back tangent and cotangent bundles of the target space respectively. For more details on the model, we refer the reader to the original papers. For our present purposes it suffices to say that every gauging of a foliation associated to the action of an almost Lie algebroid on the target space $M$ is obtained by constraining this action such that the fields $V^{\m}$ and $W_{\m}$ span a Dirac structure of the $H$-twisted standard (exact) Courant algebroid over $M$.   This is established via the relations 
\be 
V^{\m}=X^{\ast}(\rho_a{}^{\m})A^{a}\,,\qquad W_{\m}=X^{\ast}(\theta_{a\m})A^{a}\,,
\ee  
where now $A^{a}$ is a 1-form gauge field valued in $X^{\ast}E$, with $E$ being the Lie algebroid associated with the Dirac structure,{\footnote{We restrict our analysis to maximal subbundles here, which give rise to topological models.}} $\rho_{a}{}^{\m}$ are the components of the anchor map and $\theta_{a\m}$ are the components of a map from $E$ to $T^{\ast}M$ that together with $\rho$ generate the isotropic and involutive image of $E$ in $\widehat{E}=TM\oplus T^{\ast}M$ \cite{CDJS}. Then the action of the Dirac sigma model takes its form as a gauge theory,
\be  
S'_{\text{DSM}}=\int_{\S_2} \left(\frac 12\, g_{\m\n}(X) F^{\m}\w\ast\, F^{\n}+A^a\w\theta_a(X)+\frac 12 \iota_{\rho_a}\theta_b(X)A^a\w A^b\right)+\int_{{\Sigma_3}}X^{\ast}H\,, 
\ee  
with $F^{\m}=\dd X^{\m}-\rho_a{}^{\m}(X)A^a$. One can see directly how to obtain the Poisson sigma model: 
\be  
S_{\text{PSM}}=S'_{\text{DSM}}|_{E=T^{\ast}M, \, g_{\m\n}=0, \, \rho=\Pi^{\sharp}, \, \theta=\text{id}, \, H=0}\,.
\ee  
Relaxing the choice $H=0$, the result is the Wess-Zumino-Poisson sigma model. It is interesting that such a simple change in the choices leads from a model whose BV action is found via AKSZ to one where AKSZ is not applicable. It is our main goal here to argue that this can be used to approach the problem of finding solutions to the classical master equation from a different angle.

To address this, we look at the BRST transformations. The fields are $X^{\m}$ and $A^{a}$ together with the ghost $\e^{a}$ of the gauge symmetry. They BRST transform as \cite{CDJS}
\begin{eqnarray}
Q_{\text{\tiny{BRST}}}X^{\m} &=&\rho_a{}^{\m}(X) \e^a\,,\\[4pt]
Q_{\text{\tiny{BRST}}}A^a &=&\dd \e^a+\tensor{C}{^a_b_c}(X)A^b\e^c+\tensor{\omega}{^a_{b\m}}(X)\e^bF^{\m}+\tensor{\phi}{^a_{b{\m}}}(X)\e^b\ast F^{\m}\,,\\[4pt]
Q_{\text{\tiny{BRST}}}\e^a &=&-\frac{1}{2}\tensor{C}{^a_b_c}(X)\e^b\e^c\,,
\end{eqnarray} 
where $C_{ab}^{c}$ are the structure functions of the Lie bracket on $E$, $\omega^{a}{}_{b\m}$ are coefficients of the canonical $E$-connection $\mathbullet\nabla^{\E}\equiv \nabla^{\omega}$ on $E$ induced by an ordinary one $\nabla$ on it (in general with torsion which depends on the 3-form $H$, see below) and $\phi^{a}_{b\m}$ are the components of an endomorphism valued 1-form $\phi\in \G(T^{\ast}M\otimes \text{End}(E))$.{\footnote{Note that these BRST transformations are already target space covariant unlike the starting point we took for the Poisson sigma model in the previous section. The end result is of course independent of whether one starts with the one or with the other. }} 
The endomorphism $\phi$ is a new element in the Dirac sigma models that did not exist in the standard Poisson sigma model; it is due to the general option in 2D sigma models of forming an 1-form via $\dd X$ of via the Hodge dual $\ast \dd X$. This introduces an additional freedom, and recalling that the difference of two connections is an endomorphism valued 1-form like $\phi$ we can define two $E$-connections on $E$ as 
\be   
\mathbullet\nabla^{\pm}=\nabla^{\omega}\pm \phi\,,
\ee 
induced by two ordinary affine connections on $M$ with torsion tensors
\be 
\Theta^{\pm}=\langle \rho_{{\cal G}_{\pm}}, H\rangle\,,
\ee 
where $\rho_{{\cal G}_{\pm}}={\cal G}_{\pm}^{-1}{}^{a}\otimes \rho_a \in\G(TM\otimes TM)$ and we defined 
\be \label{gop}
\mc G_{\pm}=\theta\pm \r^{\ast} \in \G(T^{\ast}M\otimes E^{\ast})\,,
\ee  
with $\r^{\ast}=\iota_{\rho_a}g\otimes e^{a}$ and $\theta=\theta_a\otimes e^{a}$ when  a local basis $e^{a}$ of $E^{\ast}$ is considered. 
Let us moreover denote the $E$-torsion  tensors for these two induced $E$-connections as $T^{\pm}$ and the basic curvature for the ordinary connections acting on sections of $E$ as $S^{\pm}$, given by the standard definitions of Section \ref{sec22}. Then the square of the BRST operator on the fields reads
\bea    
&& Q_{\text{\tiny{BRST}}}^2 X^{\m}=0\,,\qquad Q_{\text{\tiny{BRST}}}^2\e^{a}=0\,, \\[4pt]  && Q_{\text{\tiny{BRST}}}^2A^{a}=\frac{1}{2}\tensor{S}{^a_b_c_\m}c^bc^cF^{\m}+\frac{1}{2}\tensor{\widetilde{S}}{^a_b_c_\m}c^bc^c\ast F^{\m}\,,
\eea 
where the components appearing in the second line are those of the tensors 
\bea \label{S}
{S} = \frac{1}{2}\left({S}{^{+}}+{S}{^{-}}\right)\,,
\quad 
{\widetilde{S}}= \frac{1}{2}\left({S}{^{+}}-{S}{^{-}}\right)\,.
\eea 
As usual, this reflects the openness of the gauge algebra, since for the topological Dirac sigma models $F^{\m}=0$ is one of the equations of motion.

Let us now once again ask for an operator $Q_{\text{\tiny{BV}}}$ which is a deformation of the above that squares to zero without the need to impose the equations of motion. To achieve this we 
enlarge the space of fields by the antifields $A^{\dagger}_{a}$, $X^{\dagger}_{\m}$ and $\e^{\dagger}_{a}$
of form degrees 1, 2, 2 and ghost degrees $-1$, $-1$, $-2$ respectively. 
Furthermore, the BV-BRST formalism prompts us to require that the antifield $A^{\dagger}_{a}$ transforms to the field equation $F^{\m}$. Unlike the Poisson sigma model where this was rather obvious, in the present general case the correct action of the BV-BRST differential on it is 
\be  
Q_{\text{\tiny{BV}}}A^{\dagger}_{a}=-(\theta_{a\mu}-\r_a{}^\nu g_{\nu\mu}\ast) F^{\m}+(C^b_{ac}-\r_a{}^\m \o_{c\m}^b)A^{\dagger}_{b}\epsilon^c-\r_a{}^\m \phi_{c\m}^b\ast A^{\dagger}_{b}\epsilon^c
\ee  
Based on the above and following the same steps as the analysis of the Poisson sigma model, the final, basis-independent form of the BV-BRST differential on the fields is found to be 
\begin{tcolorbox}[ams nodisplayskip, ams align] 
\label{qbvpsm1b}&Q_{\text{\tiny{BV}}}X=\rho(\e)\,, \\[4pt] 
&Q_{\text{\tiny{BV}}}\e= T^{\E}(\e\w\e)\,,\\[4pt] 
\label{qbvpsm3b} &Q_{\text{\tiny{BV}}}A={\DD}\e-T^{\E}(A,\e)+\langle\phi,\ast F\rangle(\e)-\frac 12 \bigg(S^{+}(\e\w\e)\,\mc G^{-1}_{-}(A^{\dagger}_+)- S^{-}(\e\w\e)\,\mc G^{-1}_{+}(A^{\dagger}_{-})\bigg)\,, \\[4pt] 
\label{qbvpsm4b}&Q_{\text{\tiny{BV}}}A^{\dagger}=-\mc F-T^{\E}(A^{\dagger},\e)-\langle\iota_{\rho}\phi(\ast A),\e\rangle\,,
\end{tcolorbox}
where $\DD$ is the spacetime and target space covariant differential induced by the connection on $E$ with torsion and it depends on $H$ and $\mc F:=\langle \theta-\iota_{\rho}g\ast,F\rangle$. Moreover we defined the ``light-cone'' antifields $A^{\dagger}_{\pm}=A^{\dagger}\pm \ast A^{\dagger}$. As before, one should complement these with the action on the remaining two antifields, which we shall not present here. 
One may easily see that for the choices that lead from the Dirac  to the Poisson 
sigma model, the correct expressions are obtained. The advantage now is that this is still true for the Wess-Zumino-Poisson sigma model in presence of the 3-form $H$, even though in that case there is no known AKSZ construction. What is more this addresses the BV-BRST differential for the general topological Dirac sigma model, whose BV action was found in \cite{Chatzistavrakidis:2022wdd}. 

\section{Higher gauge theory} 
\label{sec6}

In this section, we are going to consider a higher gauge theory following closely Ref.  \cite{Grutzmann:2014hkn}. Rather than considering an action, we assume that we have a set of fields with general field strengths and gauge transformations that satisfy general compatibility conditions as in \eqref{ideal1} and \eqref{ideal2}. Then, following the same steps as in Section \ref{sec5}, we shall determine the role of the Kapranov L$_{\infty}[1]$ algebra, at least of its lower brackets, in such a gauge system. 

As in Section \ref{sec51} we consider a collective of fields $\phi^{\a}$ on the spacetime $T[1]\S$. For example $\phi^{\a}=(X^{\m},A_{\m})$ for the Poisson sigma model. Now being as general as possible,{\footnote{Or rather almost as general as possible; we neglect the possibility of higher spin fields and mixed symmetry tensor gauge fields here, which in the spirit of this approach could be treated with the bi- or multi-differential graded formalism developed in Ref. \cite{Chatzistavrakidis:2019len}.}}  these fields can be degree $0$ functions on $T[1]\S$ (scalars) $X^{\m}$, degree 1 functions (1-forms) $A^{a_1}=A^{a_1}_{m}\theta^{m}$, degree 2 functions (2-forms) $B^{a_2}=\frac 12 B^{a_2}_{mn}\theta^{m}\theta^{n}$, and so on up to $p$-forms $P^{a_p}$. Note that the (higher) color indices $a_1, a_2,\dots a_p$ are generically different and they take values in different sets. A general gauge system has a space of fields that is a dg manifold, see e.g. \cite{costello}. Let us denote the target dg manifold as $\mc M$ and assign to it coordinates of various degrees denoted collectively as $x^{\a}$ with $x^{\m}$ being those of degree 0. These coordinates include degree $1$ $a^{a_1}$, degree $2$ $b^{a_2}$ and so on up to degree-$p$ $p^{a_p}$. 
We also consider a degree-preserving map $\phi$ from the source dg manifold to the target $\mc M$, such that its pull-back  returns the fields, 
\be  
\phi^{\a}=\phi^{\ast}(x^{\a})\,. 
\ee  
Unless there is a cause of confusion, we will most of the times ignore the pull-back when it is easily understood from the context, but we shall retain it otherwise. Finally, that ${\mc M}$ is a dg manifold also means that it has a degree $1$ homological vector field   $Q\in \G(T\mc M)$. This can be expanded  as 
\be  
Q=Q^{\a}\frac\partial{\partial x^{\a}}\,.
\ee  As discussed in \cite{Grutzmann:2014hkn}, $Q^{2}=0$ yields conditions that in general correspond to L$_{\infty}$ algebroids. The setting so far is a simple yet vast generalization (both in fields and in spacetime dimension) of the 2D examples in Section \ref{sec5}, where we only had scalars and 1-forms. 

Following the general analysis of \cite{Grutzmann:2014hkn}, we assign{\footnote{This is an assumption, since the transformations may have a more general form. Nevertheless, this assumption scans a sizable part of the space of higher gauge theories.}} the following infinitesimal gauge transformations to the fields of the theory: 
\be  
\d \phi^{\a}=\dd \e^{\a}+\e^{\b}(\s)\m^{\a}_{\b}(\phi)\,, 
\ee   
where $\e^{\a}$ is a host of spacetime dependent gauge parameters and $\m^{a}_{\b}$ are unknown functions of the fields. To keep track of signs, we declare the degree of $\phi^{\a}$ to be $|\a|$, in which case the degree of $\e^{\a}$ is $|\a|-1$ and the degree of $\m^{\a}_{\b}$ is $|\a|-|\b|+1$. Moreover, define the field strengths according to 
\be  
F^{\a}=\dd\phi^{\a}-Q^{\a}(\phi)\,.
\ee  
As already shown in \cite{Grutzmann:2014hkn}, using $Q^2=0$ one obtains  
\be  
\d F^{\a}= -(-1)^{|\beta|}\e^{\b} F^{\g}\partial_{\g}\m^{\a}_{\b}\quad \Leftrightarrow \quad \m^{\a}_\b=\partial_{\b}Q^{\a}\,.
\ee
The field strength transforms covariantly---as it should---and moreover with the coefficient appearing above if and only if the unknown functions are fixed as shown. 
This means that 
 \begin{tcolorbox}[ams nodisplayskip, ams align]    
\d \phi^{\a}&= \dd\e^{\a}+\e^{\b}\partial_{\b}Q^{\a}\,, \\[4pt]
\d F^{\a}&= -(-1)^{|\beta|}\e^{\beta}F^{\g}\partial_\g\partial_\b Q^{\a}\,,
\end{tcolorbox} 
i.e. the transformations are controlled by the first and second derivatives of the components of the homological vector field, which we already know that they are associated to the tangent lift and the Atiyah cocycle. Indeed, the Atiyah cocycle is simply defined as before; we consider a torsion-free connection on the graded manifold ${\mc M}$ and examine its compatibility with the homological vector field on it. The results obtained in section \ref{sec41} go through with the reinterpretation of the index $\a$ as in the present section. This sets the stage to obtain a new, covariant perspective on the above transformations in terms of the associated Kapranov  L$_{\infty}[1]$ algebra. 
\begin{prop} \label{hgtatiyah}
For a higher gauge theory of maps $\phi=(\phi^{\a}):T[1]\S\to \mc M$ as described above and field strengths given as $F=(F^{\a})=(\dd \phi^{\a}- \phi^{\ast}(Q^{\a}))$ where $Q$ is the homological vector field on $\mc M$, the  transformations of $\phi^{\a}$ and $F^{\a}$ under the gauge symmetries of the theory are 
\begin{tcolorbox}[ams nodisplayskip, ams align] 
\d \phi^{\a}&= \dd \big(\phi^{\ast}_{\times} (\e^{\a})\big)+\phi^{\ast}_{\times}\big(b_1(\e)^{\a}\big)\,, \\[4pt] 
\d F^{\a}&= - \phi^{\ast}_{\times}\big(b_2(F,\e)^{\a}\big)\,,
\end{tcolorbox} 
where $b_1=\mc L_{Q}$ is the Lie derivative along the vector field $Q$ and $b_2=-\text{At}$ is the Atiyah cocycle for a connection without torsion on $\mc M$. Here $\phi_{\times}: T[1]\S\to T[1]\S\times \mc M$ is the extension of $\phi$ by the identity on the first factor and $F, \e$ are vector fields on $T[1]\S\times\mc M$.
\end{prop} 
We conclude with the following important observation.{\footnote{We thank R. Szabo for prompting us to calculate this.}} Consider the commutator of two gauge transformations. This is found to be 
\be  
[\d_1,\d_2]\phi^{\a}=\d_{12}\phi^{\a}+(-1)^{|\g|}\e_2^{\b}\e_1^{\g}F^{\d}\partial_\d\partial_\g\partial_\b Q^\a\,,
\ee  
where the mixed gauge parameter is $\e_{12}^{\a}=(-1)^{|\b|+1}\e_2^{\b}\e_1^\g \partial_\g\partial_\b Q^\a$ and we have only used an identity following from $Q^{2}=0$.{\footnote{From $Q^2=0$ we get $Q^\d\partial_\d Q^\a=0$ and consequently $\partial_\g\partial_\b(Q^\d\partial_\d Q^\a)=0$, from which we obtain
\be   \partial_\g Q^\d\partial_\d\partial_\b Q^\a+(-1)^{|\b|}\partial_\g\partial_\b Q^\d\partial_\d Q^\a+(-1)^{|\g(\b+1)|}Q^\d\partial_\d\partial_\b\partial_\g Q^\a+(-1)^{|\b+\g+\b\g|}\partial_\b Q^\d\partial_\d\partial_\g Q^\a=0\,,\ee 
which is precisely the combination of terms in the calculation of the commutator of gauge transformations.}} We observe that 
\be  
\e_{12}=\text{At}(\e_1,\e_2)= -b_2(\e_1,\e_2)\,.
\ee  
Moreover, employing the definitions introduced earlier we see that 
\bea   
(-1)^{|\g|}\e_2^{\b}\e_1^{\g}F^{\d}\partial_\d\partial_\g\partial_\b Q^\a=-F^{\d}\partial_\d\text{At}(\e_1,\e_2)^{\a}=-\grave\nabla_{F}\text{At}(\e_1,\e_2)^{\a}=b_3(F,\e_1,\e_2)^{\a}\,.
\eea  
Collecting these considerations, the  commutator of gauge transformations is 
\begin{tcolorbox}[ams nodisplayskip, ams align]   
[\d_1,\d_2]\phi^{\a}=\d_{-b_2(\e_1,\e_2)}\phi^{\a}+b_3(F,\e_1,\e_2)^{\a}\,.
\end{tcolorbox}
Higher operations on the gauge transformations, such as the Jacobiator, produce the higher brackets of the Kapranov L$_{\infty}[1]$ algebra. Note though that this is the one with $\grave{R}=0$; it would be interesting to see whether more general higher gauge theories that the ones considered here would use the full power of Kapranov L$_{\infty}[1]$ algebras where the curvature of the connection on $\mc M$ is nonvanishing. 

\section{Discussion} \label{sec7} 

Since our results were already summarized in the introduction, let us briefly revisit those we consider as the important ones, their limitations and with an outlook toward open questions. The first result of this paper is the definition of a basic curvature tensor for connections on Courant algebroids and its relation to the Gualtieri torsion and to the homological vector field on the QP2 manifold. This was in fact the original motivation behind this work, based on the combination of the two observations that $(\a)$ the basic curvature tensor for Lie algebroids is the coefficient in front of the 4-fermion terms in the target space covariant formulation of 2D topological sigma models and $(\b)$ the naive definition of the basic curvature tensor fails to remain a tensor when considered for connections on Courant algebroids, similarly to the $E$-torsion.
This already suggests that one should expect the basic curvature tensor for Courant algebroid connections to appear in the target space covariant formulation of 3D Courant sigma models. The latter combine Chern-Simons and BF theory in three dimensions to a topological field theory of an AKSZ type \cite{Ikeda:2002wh,Roytenberg:2006qz}. However,  manifestly target space covariant BV and BFV formulations are still lacking. Here we showed that the basic curvature for a connection on a Courant algebroid appears in the homological vector field for the supergeometric definition of a Courant algebroid as a QP2 manifold. The basic curvature is the coefficient of the degree $(3-2)$ term of the homological vector field as appears in Eq. \eqref{Qcourcov}. This already tells us that it has to play its part in the covariant description of the Courant sigma model, in particular when the BV-BRST differential is written in covariant form. This task was completed in an accompanying paper with N. Ikeda \cite{CIJ}. 

The second main result of the present paper is the explanation of how the Atiyah cocycle of a dg manifold equipped with a connection appears in higher gauge theory, at least for connections without torsion, which anyway always exist. From a covariant viewpoint the structure of gauge transformations knows about the Kapranov  L$_{\infty}[1]$ algebra, at least of its lower brackets, the unary bracket (Lie derivative with respect to the homological vector field) and the binary bracket (opposite of the Atiyah cocycle).  We showed that for a higher gauge theory that contains a tower of differential forms obeying certain compatibility conditions \cite{Grutzmann:2014hkn}, the gauge transformation of the fields is given by the unary bracket and the gauge transformation of the field strengths is given by the binary bracket. Moreover two gauge transformations close up to the ternary bracket. 

This result comes together with a restriction on the form of gauge transformations for the fields and on the definition of their strengths. As in Ref. \cite{Grutzmann:2014hkn} we did not consider a dependence on lower-degree field strengths for any of the two. This gives rise to a restricted set of higher gauge theories where only the highest form field can obey Yang-Mills type field equations whereas the lower-degree ones are of topological type. Although this class of theories has  its  merit, going beyond this to theories where all fields obey Yang-Mills type field equations is more demanding, even in the most simple cases \cite{Baez:2005qu,Baez:2004in,Baez:2010ya}. It would be interesting to explore whether previous suggestions on this matter such as the enhanced Leibniz algebras of \cite{Strobl:2016aph} find a realization in terms of the specific L$_{\infty}[1]$ algebras we considered in this paper. Moreover, it would be interesting to investigate the role of the higher brackets in gauge theory, which are presumably associated to consistency conditions of the gauge system such as Bianchi/Noether identities.  

The notion of the basic curvature should admit a generalization to higher structures as well. This refers to cases where higher brackets appear and their compatibility with a suitable higher connection is studied. From a field-theoretical perspective such higher analogs of the basic curvature have appeared in local coordinate form in the context of twisted R-Poisson sigma models \cite{Chatzistavrakidis:2021nom,Ikeda:2021rir} and they were presented in \cite{talk}. From a geometric perspective, they were studied in \cite{Jotz}. We plan to report the details of such constructions in future work.
Finally, in the present paper we have not considered models coupled to matter. In this respect, it would be interesting to explore whether there exists some relation to the L$_{\infty}$ structure of correlators found in \cite{Gerasimenko:2021sxj,Skvortsov:2022abz} with regard to Chern-Simons-Matter theories and their fermion/boson dualities that generalize particle/vortex duality in 2+1 dimensions \cite{Seiberg:2016gmd}. 

\paragraph{Acknowledgements.} We would like to thank P. Batakidis, M. Grigoriev, C. Hull, N. Ikeda, B. Jurco, S. Lavau, R. Minasian, D. Roytenberg, P. Schupp, E. Skvortsov, T. Strobl, R. Szabo and Z. \v{S}koda for useful discussions. We are particularly thankful to D. Roytenberg for bringing to our attention the potential relevance of the Atiyah class in the context of our work during its early stages and to N. Ikeda for collaboration in a closely related project \cite{CIJ}. This work was supported by the Croatian Science Foundation project IP-2019-04-4168 ``Symmetries for Quantum Gravity''.  We would like to acknowledge the Mainz Institute for Theoretical Physics (MITP) of the DFG Cluster of Excellence ``PRISMA'' (Project ID 39083149) for enabling us to complete a significant portion of this work. We would also like to thank the Erwin Schroedinger International Institute for Mathematics and Physics, where the ideas of the present paper started taking shape during the programme ``Higher Structures and Field Theory''.


\appendix 

\section{Cotangent Lie algebroid}
\label{appa}

In this appendix we collect a few useful definitions and formulas regarding the cotangent bundle of a Poisson manifold and its Lie algebroid structure. 
Consider the cotangent bundle $T^{\ast}M$ of a Poisson manifold $M$ together  with the musical isomorphism 
\bea  
 \Pi^{\sharp}:T^{\ast}M &\to& TM \nn\\ 
 \eta&\mapsto &\Pi^{\sharp}(\eta)= \Pi^{\m\n}\eta_{\m}\mf{e}_{\n}\,, 
\eea  
where by convention the contraction is on the first slot.
The Koszul-Schouten bracket is 
\be 
[\eta,\eta']_{\text{KS}}={\cal L}_{\Pi^{\sharp}(\eta)}\eta'-{\cal L}_{\Pi^{\sharp}(\eta')}\eta-\dd\left(\Pi(\eta,\eta')\right)\,,
\ee 
where $\Pi$ is the Poisson structure. Alternative expressions for the bracket include 
\bea 
[\eta,\eta']_{\text{KS}} &=& {\cal L}_{\Pi^{\sharp}(\eta)}\eta'-\iota_{\Pi^{\sharp}(\eta')}\dd \eta \nn \\[4pt] 
&=& \iota_{\Pi^{\sharp}(\eta)}\dd \eta'-\iota_{\Pi^{\sharp}(\eta')}\dd \eta+ \dd \left(\Pi(\eta,\eta')\right)\,.
\eea 
In fact the last one is the most useful for computations, while on the contrary the middle one---which is actually most often presented,---is less motivated because it does not generalize to higher forms, unlike the other two which have proper generalizations.  In local coordinates, we obtain
\bea 
 [\eta,\eta']_{\text{KS}}=  \big(\partial_{\rho}\Pi^{\m\n}\eta_{\m}\eta'_{\n}-\Pi^{\m\n}\left(\eta_{\n}\partial_{\m}\eta'_{\rho}-\eta'_{\n}\partial_{\m}\eta_{\rho}\right)\big)\mf{e}^{\rho}\,,
\eea 
for a local basis $\mf{e}^{\rho}$. Together with the Koszul-Schouten bracket on its sections, the anchored bundle $(T^{\ast}M,\Pi^{\sharp})$ has the structure of a Lie algebroid. 

The canonical $E$-connection in the present case is 
\be 
\mathbullet\nabla^{\E}_{\eta}\,\eta'= \Pi^{\m\n}\eta_{\m}(\partial_{\n}\eta'_{\rho}-\Gamma_{\n\rho}^{\s}\eta'_{\s})\mf e^{\rho}\,.
\ee 
On the other hand, an arbitrary $E$-connection is 
\be 
\nabla^{\E}_{\eta}\,\eta'=  \eta_{\m}( \Pi^{\m\n}\partial_{\n}\eta'_{\rho}+\o^{\m\s}_{\rho}\eta'_{\s}) \mf e^{\rho}\,.
\ee 
The $E$-torsion for the induced connection is found to be 
\be 
\mathbullet{T}^\E(\eta,\eta')=-\mathring\nabla\Pi (\eta,\eta') = -\, \eta_{\m}\eta'_{\n}\mathring\nabla_{\rho}\Pi^{\m\n}\mf e^{\rho}\,.
\ee 
The same formula holds also when the affine connection on $M$ has torsion. This requires the modification of the Koszul-Schouten bracket by the twist induced by this torsion. 

The $E$-curvature can be also easily computed: 
\be 
R^\E(\eta,\eta')\eta''=-\langle \eta'',R(\Pi^{\sharp}(\eta),\Pi^{\sharp}(\eta'))\rangle=\Pi^{\k\l}\Pi^{\m\n}\eta_{\l}\eta'_{\n}\eta''_{\s}R^{\s}{}_{\rho\k\m}\mf e^{\rho}\,,
\ee 
where $R$ is the ordinary curvature of $M$.  
Some additional useful formulas are: 
\bea 
\rho(\nabla_{X}\eta)&=&\Pi^{\sharp}(\nabla_{X}\eta)= \Pi^{\m\n}X^{\s}(\partial_{\s}\eta_{\n}-\eta_{\tau}\Gamma^{\tau}_{\s\n})\mf e_{\m}\,, \\[4pt] 
[\rho(\eta),X]&=&\left(\Pi^{\m\n}\eta_{\n}\partial_{\m} X^{\rho}-X^{\m}\partial_{\m}(\Pi^{\rho\n}\eta_{\n})\right)\mf e_{\rho}\,, 
\\[4pt] 
\overline{\nabla}_{\eta}X&=&\left(\Pi^{\rho\n}\eta_{\n}\partial_{\rho}X^{\m}-\Pi^{\m\n}X^{\s}\eta_{\tau}\G^{\tau}_{\s\n}-X^{\rho}\partial_{\rho}\Pi^{\m\n}\eta_{\n}\right)\mf e_{\rho}\,.
\eea 
Finally, one finds that the basic curvature is 
\be 
S^{\E}(\eta,\eta')X= \eta_{\k}\eta'_{\l}X^{\m}\left(-\nabla_{\m}\nabla_{\n}\Pi^{\k\l}+\Pi^{\rho\k}R^{\l}{}_{\n\rho\m}-\Pi^{\rho\l}R^{\k}{}_{\n\rho\m}\right)\mf e^{\n}\,. \label{Sstar}
\ee 

\section{An \texorpdfstring{$\widehat{E}$}{e}-curvature for Courant algebroids} 
\label{appb}

To provide one possible answer to the question of defining an $\widehat{E}$-curvature tensor for an arbitrary Courant algebroid connection in our approach, 
we  invoke  Eq. \eqref{REvsSE} for Lie algebroids. Recall that this equation can be seen as an alternative definition of the curvature tensor for Lie algebroids instead of Eq. \eqref{rlie}, once the basic curvature is defined. For Lie algebroids, these two equations are one and the same thing. When one passes to Courant algebroids though, as already mentioned Eq. \eqref{rlie} does not represent a tensor. However, the analog Eq. \eqref{REvsSE} does, as long as it is suitably expressed in terms of the generalized torsion tensor ${\cal T}^{\Eh}$ and the generalized basic curvature tensor ${\cal S}^{\Eh}$---since the difference of connections $\phi$ remains a tensor in an obvious way. 

According to this, the remaining task is to write down the analog of Eq. \eqref{REvsSE} for Courant algebroids. One challenging aspect is that unlike the naive torsion that takes two entries, the generalized torsion as defined in \eqref{GTE} or \eqref{GTEd} takes three.{\footnote{This could be addressed alternatively by defining suitably the generalized torsion as a map too.}} Therefore, one should find a way to express the various terms, especially the quadratic one in the torsion, in a suitable way. If one wishes to avoid considering an explicit basis, an option is to consider a tensor with six entries. Although this cannot be the analog of the 4-entry curvature  tensor, we shall see shortly that the latter can be obtained as a type of ``Ricci-like'' tensor by taking the trace.
To follow a clear path toward the definition, first we express Eq. \eqref{REvsSE} manifestly in terms of $\phi$ and $T^{\E}$ as 
\begin{align} 
R^{\E}(e_1,e_2)e_3=& - \, S^{\textit{\tiny{E}}}(e_1,e_2)\rho(e_3)+2\nabla^{\E}_{e_{[1}}\phi(e_{|3|},e_{2]}) -2\nabla^{\E}_{e_{[1}}T^{\E}(e_{|3|},e_{2]}) \nn\\[4pt]  
&+  \phi(e_3,T^{\E}(e_{1},e_{2}))-T^{\E}(e_3,T^{\E}(e_{1},e_{2}))+2\phi({\phi}(e_3,e_{[1}),e_{2]})\nn\\[4pt] 
&-2T^{\E}({\phi}(e_3,e_{[1}),e_{2]})-2\phi({T^{\E}}(e_3,e_{[1}),e_{2]})+2T^{\E}({T^{\E}}(e_3,e_{[1}),e_{2]})\,.
\end{align}
This is valid for Lie algebroids. Without further ado, for Courant algebroids we define 
\bea 
 \widehat{\mc R}^{\Eh}(e_1,e_2,e_3,e_4,e_5,e_6)&=& \biggl(- \mc S^{\Eh}(e_1,e_2,e_4)\rho(e_3) \nn\\[4pt] &&- 2\nabla^{\Eh}_{e_{[1}}\mc T^{\Eh}(e_{|3|},e_{2]},e_4)-2\mc T^{\Eh}(\phi(e_3,e_{[1}),e_{2]},e_4) \nn\\[4pt] &&+  \langle 2\nabla^{\Eh}_{e_{[1}}\phi(e_{|3|},e_{2]}) +2\phi(\phi(e_3,e_{[1}),e_{2]}),e_4\rangle_{\Eh} \biggl)\langle e_5,e_6\rangle_{\Eh} \nn\\[4pt]
&& +\mc T^{\Eh}(e_1,e_2,e_5)\left(\langle\phi(e_3,e_6),e_4\rangle-\mc T^{\Eh}(e_3,e_6,e_4)\right)\nn\\[4pt] 
&& - 2\mc T^{\Eh}(e_3,e_{[1},e_{|5})\left(\langle \phi(e_{6|},e_{2]}),e_4\rangle-\mc T^{\Eh}(e_{6|},e_{2]},e_4)\right)\,.
\label{REvsSE for CA and 6 entries}\eea 
As alluded to above, a 4-entry/4-index tensor can be obtained in a Ricci-like fashion by choosing two of the entries to be dual sections, namely their symmetric bilinear form being the identity. We choose $e_5$ and $e_6$ to play this role, $\langle e_5,e_6\rangle=\text{id}$. They can be chosen for instance to be two dual basis sections, $\mf e^{a}$ and $\mf e_{a}$. 
In this way, the following 4-entry tensor is obtained, which we denote as a trace, 
\bea 
\text{tr}\,\widehat{\mc R}^{\Eh}(e_1,e_2,e_3,e_4)&=& 
- \mc S^{\Eh}(e_1,e_2,e_4)\rho(e_3) \nn\\[4pt] &&- 2\nabla^{\Eh}_{e_{[1}}\mc T^{\Eh}(e_{|3|},e_{2]},e_4)-2\mc T^{\Eh}(\phi(e_3,e_{[1}),e_{2]},e_4) \nn\\[4pt] &&+  \langle 2\nabla^{\Eh}_{e_{[1}}\phi(e_{|3|},e_{2]}) +2\phi(\phi(e_3,e_{[1}),e_{2]}),e_4\rangle_{\Eh}  \nn\\[4pt]
&& +\mc T^{\Eh}(e_1,e_2,\mf e_a)\left(\langle\phi(e_3,\mf e^{a}),e_4\rangle-\mc T^{\Eh}(e_3,\mf e^{a},e_4)\right)\nn\\[4pt] 
&& - 2\mc T^{\Eh}(e_3,e_{[1},\mf e_a)\left(\langle \phi(\mf e^{a},e_{2]}),e_4\rangle-\mc T^{\Eh}(\mf e^{a},e_{2]},e_4)\right)\,.
\label{REvsSE for CA and 4 entries}
\eea 
That this $\widehat{E}$-curvature-like tensor on a Courant algebroid is tensorial does not even require a proof. It is obvious from its definition in terms of quantities that are all tensors in all their entries. The same is obviously true for its following descendant, antisymmetric in both pairs of entries $(e_1,e_2)$ and $(e_3,e_4)$, which we readily define and denote with a calligraphic font without a hat,
\bea 
\mc R^{\Eh}(e_1,e_2,e_3,e_4)&:=& \frac 12 \, \text{tr}\,\widehat{\mc R}^{\Eh}(e_1,e_2,e_3,e_4)- \frac 12 \, \text{tr}\,\widehat{\mc R}^{\Eh}(e_2,e_1,e_3,e_4) \nn\\[4pt] && +\, \frac 12\, \text{tr}\,\widehat{\mc R}^{\Eh}(e_2,e_1,e_4,e_3) - \frac 12\, \text{tr}\,\widehat{\mc R}^{\Eh}(e_1,e_2,e_4,e_3)  
\,.\,\,\,\,\,
\label{REvsSE for CA and 4 entries and right symmetries}
\eea 
It would be interesting to examine the relation of this tensor to what was suggested in \cite{Hohm:2012mf}, see also \cite{Jurco:2016emw}, which follows a different approach in defining a Riemann tensor for a Courant algebroid. 

\section{Algebraic Bianchi identity for the \texorpdfstring{$\rm E$}{E}-curvature}
\label{appc}

 \begin{prop} The algebraic Bianchi identity for the $E$-curvature reads
\be \text{Cycl}\big(R^\E(e_1,e_2)e_3\big)=\text{Cycl}\bigg(\nabla^\E_{e_1}T^\E(e_2,e_3)+T^\E\left(T^\E(e_1,e_2),e_3\right)\bigg)\,.\ee 
 \end{prop} 
 \paragraph{Proof.} By definition of the $E$-curvature and $E$-torsion tensors, we obtain (in the following we refrain from explicitly using the notation $\nabla^\E$ and $T^\E$ to avoid clutter; all covariant derivatives are $E$-covariant and all torsions are $E$-torsions):
 \bea  
\text{Cycl}\left(R^\E(e_1,e_2)e_3\right)&=& \text{Cycl}\left(\nabla_{e_1}\nabla_{e_2}e_3-\nabla_{e_2}\nabla_{e_1}e_3-\nabla_{[e_1,e_2]_{\E}}e_3\right) \nn\\[4pt] &=& \text{Cycl}\left(\nabla_{e_1}\left(T(e_2,e_3)\right)+\nabla_{e_1}[e_2,e_3]_{\E}-\nabla_{[e_2,e_3]_{\E}}e_1\right) \nn\\[4pt] 
 &=& \text{Cycl}\left(\nabla_{e_1}\left(T(e_2,e_3)\right)+\, T(e_1,[e_2,e_3]_{\E})-[e_1,[e_2,e_3]_{\E}]_{\E}\right)\,. \nn 
 \eea 
Due to the Jacobi identity for the Lie bracket on $E$, the last term drops out and using the Leibniz rule and linearity we remain with 
\be 
\text{Cycl}\left(R^\E(e_1,e_2)e_3\right)=\text{Cycl}\left(\nabla_{e_1}T (e_2,e_3)+ T(\nabla_{e_1}e_2,e_3)+ T(e_3,\nabla_{e_2}e_1)+ T(e_3,[e_1,e_2]_{\E})\right)\,, \nn 
\ee 
which proves the assertion by inspection, using the antisymmetry of the $E$-torsion.

\section{Further aspects of \texorpdfstring{$\rm E$}{E}-connections}
\label{appd} 

 In Section \ref{sec42} we found that there is a one-parameter family of identifications for connection coefficients that allow us to define an $E$-connection on a Lie algebroid $E$ from a connection on $E[1]$. As discussed there, the value $\k=1$ is the prominent one for our purposes and it corresponds to the canonical induced $E$-connection, however we briefly discuss here some aspects pertaining other values. 
Three values of $\k$ are of special interest, as in the following table where we present the image of the $E$-connection and of its coefficients in each case.
\begin{center}	\begin{tabular}{ |c||c|c| }
		\hline
		\multirow{3}{0.5em}{$\k$} &  & \\ & $\iota(\nabla^{\E}_e e')$ & $\o^c_{ab}$  \\ && \\
		\hline  
		\multirow{3}{0.5em}{0} & &\\ & $\grave\nabla_{\iota(e)}\mc Q\iota(e')$ & $\rho_{b}{}^{\m}\grave\o_{a\m}^{c}+C_{ab}^c$ \\ & & \\
		\hline 
			\multirow{3}{0.5em}{1} &&\\ & $\grave\nabla_{\mc Q\iota(e)}\iota(e')$& $\rho_{a}{}^{\m}\grave\o_{\m b}^{c}$
   \\ &&\\ \hline 
	 	\multirow{3}{0.5em}{$\frac 12$} &&\\ &  $\frac 12 \grave\nabla_{\iota(e)}\mc Q\iota(e')+\frac 12 \grave\nabla_{\mc Q\iota(e)}\iota(e')$& $\frac 12 (\rho_{a}{}^{\m}\grave\o_{\m b}^{c}+\rho_{b}{}^{\m}\grave\o_{a\m}^{c}+C_{ab}^c)$ \\&& \\ \hline 
	\end{tabular} 
 \end{center}
Regarding the $E$-torsion of the connection $\nabla^{\E}$, this is mapped under $\iota$ to 
  \bea 
    \iota(T^{\E}(e,e'))^{(\k)}&=&\k \bigg(\grave\nabla_{\mc Q\iota(e)}\iota(e')-\grave\nabla_{\mc Q\iota(e')}\iota(e)\bigg)+(1-\k)\bigg(\grave\nabla_{\iota(e)}\mc Q\iota(e')-\grave\nabla_{\iota(e')}\mc Q\iota(e)\bigg)- \nn\\[4pt] 
    && -\, [\mc Q\iota(e),\iota(e')] \nn\\[4pt] &=& \k\bigg(\grave\nabla_{\mc Q\iota(e)}\iota(e')-\grave\nabla_{\mc Q\iota(e')}\iota(e)-[\mc Q \iota(e),\iota(e')]\bigg) + \nn\\[4pt]
    && +\, (1-\k)\bigg(\grave\nabla_{\iota(e)}\mc Q\iota(e')-\grave\nabla_{\iota(e')}\mc Q\iota(e)-[\iota(e),\mc Q\iota(e')]\bigg)\,,
    \label{iotaTE2}
    \eea 
    where we used the identity 
    \be [\mc Q\iota(e),\iota(e')]=[\iota(e),\mc Q\iota(e')]\,.\ee 
 Then we observe that if we consider the case $\kappa=1/2$ and  since $|\iota(e)|=-1$ we can rewrite \eqref{iotaTE} as 
\be \iota(T^{\E}(e,e'))^{(\frac 12)}= \frac 12 \,\bigg(\grave T\big(\mc Q\iota(e),\iota(e')\big)-\,\grave T\big(\mc Q\iota(e'),\iota(e)\big)\bigg)
\,.\ee
Thus we see that for this value the $E$-torsion is obtained from the ordinary torsion on the dg manifold without further restrictions. This also means that a torsion-free connection on $\mc M$ gives an $E$-torsion-free connection on $E$.

Going beyond  the  case of $\k=1/2$ we can obtain several further  relations among the two sides, the Lie algebroid side and the dg manifold. For instance,  
\bea 
\iota(T^{\E}(e,e'))^{(\k)}&=&\iota(T^{\E}(e,e'))^{(0)}-\k\bigg({\text{At}}(\iota(e),\iota(e'))-\text{At}(\iota(e'),\iota(e))\bigg)\,,\label{iotaTEkappa1}
\eea 
or, alternatively,
\bea 
\iota(T^{\E}(e,e'))^{(\k)}&=&\iota(T^{\E}(e,e'))^{(1)}+(1-\k)\bigg({\text{At}}(\iota(e),\iota(e'))-\text{At}(\iota(e'),\iota(e))\bigg)\,.\label{iotaTEkappa2}
\eea 
These also mean that 
\be \text{At}(\iota(e),\iota(e'))-\text{At}(\iota(e'),\iota(e))=\iota(T^{\E}(e,e'))^{(0)}-\iota(T^{\E}(e,e'))^{(1)}\,.\ee 
We observe that in the image of the map $\iota$ the symmetric and antisymmetric parts of the Atiyah cocycle are controlled by the torsion tensors $\grave{T}$ and $T^{\E}$ respectively. 
This means in particular that 
\bea 
T^{\E}=0 \quad &\Rightarrow& \quad \text{At}(\iota(e),\iota(e'))=\text{At}(\iota(e'),\iota(e))\,, \\[4pt]
\grave{T}=0 \quad &\Rightarrow & \quad \text{At}(\iota(e),\iota(e'))=-\text{At}(\iota(e'),\iota(e))\,.
\eea 

These statements are not if and only if as presented. The second statement has a clear generalization, evident from \eqref{atiyahantisymm}. The Atiyah 1-cocycle is graded symmetric (which for functions of degree $-1$ means antisymmetric) if and only if the torsion $\grave{T}$ is $\mc Q$-closed, or equivalently if and only if it is invariant under the flow of the homological vector field $Q$.  The first statement cannot be generalized in the same way because it refers strictly to vector fields in the image of the map $\iota$, since it connects the two sides referring to the Lie algebroid and the dg manifold, unlike the second statement that refers only to the dg manifold. One should also be cautious of the fact that in general $\grave{T}=0$ does not imply that $T^{\E}=0$ or vice versa.  

The above allow us to also express the Atiyah cocycle for vector fields in the image of $\iota$ differently, in particular
\be  
\k \text{At}(\iota(e),\iota(e'))=\grave\nabla_{\iota(e)}\mc Q\iota(e') - \iota(\nabla^{\E}_{e}e')^{(\k)}\,.
\ee  
This also holds of course for the value $\k=0$, in which case we find
\be  
\iota(\nabla^{\E}_{e}e')^{(0)}=\grave\nabla_{\iota(e)}\mc Q\iota(e')\,.
\ee  
It is also interesting to explain what is special about the value $\k=1/2$. From the above definitions it follows that 
\bea 
(1-2\k)\text{At}(\iota(e),\iota(e'))=\iota(\nabla^{\E}_{e}e')^{(\k)}-\k\grave\nabla_{\iota(e)}\mc Q\iota(e')-(1-\k)\grave\nabla_{\mc Q\iota(e)}\iota(e')\,.
\eea 
We observe that for $\k=1/2$, it holds that 
\be \iota(\nabla^{\E}_{e}e')^{(\frac 12)}=\frac 12 \grave\nabla_{\iota(e)}\mc Q\iota(e')+\frac 12\grave\nabla_{\mc Q\iota(e)}\iota(e')\,, \ee 
independently of the Atiyah 1-cocycle, as expected from \eqref{iotanablaEk}. Note that at the critical value, the connection  coefficients are related through 
\be 
\omega_{ab}^{c}=\frac 12 \big(\rho_{a}{}^{\m}\grave{\omega}_{\m b}^{c}+\rho_{b}{}^{\m}\grave{\omega}^c_{a\m}+C_{ab}^c\big)\,.
\ee 
The same logic persists at the level of the torsion tensors. There exists a direct relation between $T^{\E}$ and $\grave{T}$ in the case of $\k=1/2$, given in \eqref{iotaTE}, and for any other value of $\k$ such a relation is obstructed by the Atiyah cocycle.

\end{document}